%% file: main.tex
\documentclass[11pt]{article}
\usepackage{graphicx, yk} 
\usepackage[margin=2.5cm]{geometry}
\usepackage{float}
\usepackage{amsmath, amsfonts, framed}
\usepackage{amssymb}
\usepackage{tikz}
\usepackage{multirow}
\usepackage{hyperref}       
\usepackage{textcmds}
\usepackage{longtable}
\usepackage{array}
\usepackage{dsfont}
\usepackage{algorithm}
\usepackage{adjustbox}
\usepackage{circuitikz}
\usepackage{booktabs}
\usepackage[normalem]{ulem}
\usepackage{algpseudocode}
\usepackage[bottom]{footmisc}
\hypersetup{colorlinks,linkcolor={red},citecolor={blue},urlcolor={red}}  
\tikzset{every picture/.style={line width=0.75pt}} 
\usepackage{natbib}
\setcitestyle{author year,open={(},close={)}} 

\usepackage{hyperref}       
\usepackage{url}            
\usepackage{booktabs}       
\usepackage{amsfonts}       
\usepackage{nicefrac}       
\usepackage{microtype}      
\usepackage{xcolor}         
\usepackage{float}
\usepackage{amsmath, amsfonts, framed}
\usepackage{amssymb}
\usepackage{tikz}
\usepackage{hyperref}       
\usepackage{adjustbox}
\usepackage{textcmds}
\usepackage{algorithm}
\usepackage{pifont}
\usepackage{multicol}
\usepackage{multirow}

\usepackage{enumitem}
\usepackage{mathrsfs}
\newlist{todolist}{itemize}{2}
\setlist[todolist]{label=$\square$}

\newcommand{\blf}{\boldsymbol{f}}

\usepackage{algpseudocode}

\allowdisplaybreaks
\usepackage{adjustbox}
\usepackage{circuitikz}
\usetikzlibrary{arrows.meta, positioning}
\usepackage{booktabs}
\usepackage[normalem]{ulem}
\usepackage{algpseudocode}
\hypersetup{colorlinks,linkcolor={red},citecolor={blue},urlcolor={red}}  
\tikzset{every picture/.style={line width=0.75pt}}

\newcommand{\independent}{\perp\!\!\!\!\perp}

\newcommand\blfootnote[1]{
  \begingroup
  \renewcommand\thefootnote{}\footnote{#1}
  \addtocounter{footnote}{-1}
  \endgroup
}

\providecommand{\keywords}[1]
{
  \small	
  \textbf{\textit{Keywords---}} #1
}

\title{\bf MediEncoder: Nonlinear Representation Learning for High-Dimensional Causal Mediation Analysis}
\author{Shi Bo, Debarghya Mukherjee$^\dagger$, and AmirEmad Ghassami$^\dagger$  \\
Department of Mathematics and Statistics, Boston University\vspace{7mm}}

\date{}

\begin{document}

\maketitle 

\begin{abstract}
Causal mediation analysis decomposes a treatment effect into indirect pathways through mediators and direct pathways not operating through them. Modern biomedical studies often involve high-dimensional covariates and mediators that are noisy proxies for lower-dimensional latent biological processes. Existing methods typically rely on sparsity, linear factor models, or ignore the connection among variables in the learned representations, which can be restrictive when measurements are nonlinear and covariate and mediator factors are structurally dependent. We propose \texttt{MediEncoder}, a representation-learning framework for nonlinear high-dimensional mediation analysis. MediEncoder jointly learns low-dimensional covariate and mediator representations using a coupled encoder--decoder architecture with a cross-factor network that links treatment and covariate representations to mediator representations. The learned features are then used in a cross-fitted efficient influence function-based estimator of natural direct and indirect effects. The resulting estimator is multiply robust and asymptotically normal under suitable regularity conditions. Simulations show that MediEncoder improves estimation accuracy over competing dimension-reduction approaches, and an application to Alzheimer's Disease Neuroimaging Initiative data illustrates its utility in high-dimensional biomedical causal mediation analysis.
\blfootnote{$^\dagger$ Corresponding authors.}
\end{abstract}

\input{Intro_revised}
\input{problem_setup_revised}

\input{Method_revised}

\input{Theory_DM}

\input{simulation_ablation_revised}

\input{Conclusion}

\bibliographystyle{apalike}
\bibliography{references}

\newpage
\appendix 
\input{Appendix}

\end{document}

%% file: Intro_revised.tex
\section{Introduction}
\label{section:1}

Causal mediation analysis seeks to explain how a treatment affects an outcome by decomposing the total effect into pathways operating through intermediate variables, or mediators \citep{baron1986moderator,pearl2001direct,imai2010identification,vanderweele2014mediation}. In the potential-outcomes framework \citep{rubin1974estimating}, this amounts to studying counterfactual means whose contrasts yield the natural indirect effect, natural direct effect, and total effect. Classical methods typically assume that the relevant covariates and mediators are directly observed and low-dimensional. Modern biomedical studies often violate both assumptions: the causal mechanisms of interest are latent, while the available measurements are noisy, high-dimensional proxies. This setting arises across genomics and multi-omics applications. High-dimensional molecular measurements, such as metabolomic or multi-omics profiles, are often indirect views of lower-dimensional biological states \citep{worheide2021multi,rahnenfuhrer2023statistical}. For example, oxidative stress has been proposed as a pathway linking depression to Alzheimer's disease \citep{livingston2024dementia,buccellato2021role}, but reactive oxygen species are short-lived and difficult to measure directly \citep{katerji2019approaches}. Researchers therefore rely on downstream molecular signatures, including DNA damage and DNA methylation changes \citep{pandya2013antioxidants,black2015depression,miranda2000role}. Such measurements may be nonlinear functions of latent biological processes \citep{johnson2017non,vershinina2021disentangling}; moreover, upstream covariate factors, treatments, and mediator factors can interact in ways that complicate causal identification and estimation \citep{vanderweele2015explanation}. These features motivate a mediation framework that can handle latent structure, high dimensionality, nonlinear measurement maps, and treatment--covariate--mediator interactions simultaneously.

Existing high-dimensional mediation methods address only parts of this problem. Sparse and variable-selection approaches can screen many candidate mediators \citep{zhang2022high,zhao2021pathway,bo2024debiased,jones2025causal}, but they usually treat the observed variables as the mediators and do not model latent structure. Dimension-reduction methods based on PCA or linear factor models provide low-dimensional summaries \citep{derkach2019high, fan2025factor}, but linear measurement maps can be restrictive in complex biological systems \citep{angermueller2016deep}. Autoencoders and related deep models can learn nonlinear representations \citep{hinton2006reducing,goodfellow2016deep}, yet applying them separately to covariates and mediators ignores the structural dependence between the two latent systems. Deep mediation estimators such as \citep{xu2022deepmed} improve robustness to nuisance model misspecification, but do not explicitly target nonlinear high-dimensional latent-factor settings. Variational autoencoder approaches for indirectly observed mediators \citep{jiang2023causal} model latent mediators, but do not jointly accommodate latent nonlinear covariate structure, structural dependence between covariate and mediator representations, and interaction effects; see Appendix~\ref{appdendix:related} for further discussion. Overall, existing methods address individual aspects of the problem but do not simultaneously address all the aforementioned challenges. The key missing ingredient is a framework that learns nonlinear covariate and mediator representations jointly, aligns them with the latent mediation structure, and then uses these representations in a statistically valid causal estimator.

To bridge this gap, we propose \texttt{MediEncoder}, a nonlinear representation-learning framework for high-dimensional causal mediation analysis. The method jointly learns low-dimensional representations of covariates and mediators through a structured encoder--decoder architecture. A cross-factor network links the treatment and covariate representation to the mediator representation, encouraging the learned features to respect the causal structure of the latent mediation problem. Exact recovery of the original latent factors is not required: it suffices to learn representations that preserve the information needed for the mediation functional, and our theory formalizes invariance under left-invertible transformations of the latent factors.

Our contributions are fourfold. \textbf{(i)} First, we introduce a nonlinear latent-factor framework for mediation analysis with high-dimensional covariates and mediators, avoiding linearity and sparsity requirements. \textbf{(ii)} Second, we develop MediEncoder, a coupled representation-learning algorithm that uses a cross-factor network to capture structural dependence between covariate and mediator representations. \textbf{(iii)} Third, we develop a modular asymptotic theory for the resulting cross-fitted efficient influence-function estimator. The theory formalizes that exact factor recovery is unnecessary: it suffices for the learned representation to approximate a left-invertible transformation of the latent factors, and the resulting inference remains valid under high-level factor-recovery and nuisance-rate conditions. \textbf{(iv)} Fourth, we demonstrate the empirical performance of MediEncoder in extensive simulations and in a biomedical application using data from the Alzheimer’s Disease Neuroimaging Initiative (ADNI).

%% file: problem_setup_revised.tex
\section{Model Description} \label{section:2}

We consider a binary treatment (or exposure) $A\in\{0,1\}$ and a real-valued outcome
$Y\in\mathbb R$. Let $\boldsymbol f_X\in\mathbb R^{\bar p}$ denote a vector of
pre-treatment latent covariate factors and let $\boldsymbol f_M\in\mathbb R^{\bar q}$ denote a vector of latent mediator factors. The covariate factors may confound the treatment--mediator, treatment--outcome, and mediator--outcome relationships, whereas the mediator factors are the mechanisms through which part of the effect of $A$ on $Y$ may be transmitted. We use boldface notation for random latent vectors and write $f_X$ and $f_M$ for their generic values.

A compact structural representation of the latent mediation model is
\begin{equation}
\label{eq:factor}
\begin{aligned}
\boldsymbol f_M
&=(1-A)\{\mu_0^*(\boldsymbol f_X)+\boldsymbol u_{XM}\}
  +A\{\mu_1^*(\boldsymbol f_X)+\boldsymbol u'_{XM}\},\\
Y
&=(1-A)\{\mu_0(\boldsymbol f_X,\boldsymbol f_M)+\epsilon'\}
  +A\{\mu_1(\boldsymbol f_X,\boldsymbol f_M)+\epsilon\},
\end{aligned}
\end{equation}
where $\mu_0^*,\mu_1^*:\mathbb R^{\bar p}\to\mathbb R^{\bar q}$ and
$\mu_0,\mu_1:\mathbb R^{\bar p+\bar q}\to\mathbb R$ are unknown structural functions and $\epsilon$, $\epsilon'$, $\boldsymbol u_{XM}\in \mathbb{R}^{\bar q}$, and $\boldsymbol u_{XM}'\in \mathbb{R}^{\bar q}$ are centered noise terms.

\begin{figure}[t]
    \centering
\scalebox{0.7}{
\tikzset{every picture/.style={line width=0.75pt}} 

\begin{tikzpicture}[x=0.75pt,y=0.75pt,yscale=-1,xscale=1]

\draw    (205.09,133.32) -- (172.55,72.73) ;
\draw [shift={(171.6,70.97)}, rotate = 61.76] [color={rgb, 255:red, 0; green, 0; blue, 0 }  ][line width=0.75]    (10.93,-3.29) .. controls (6.95,-1.4) and (3.31,-0.3) .. (0,0) .. controls (3.31,0.3) and (6.95,1.4) .. (10.93,3.29)   ;
\draw    (222.09,128.89) -- (256.6,69.16) ;
\draw [shift={(257.6,67.42)}, rotate = 120.01] [color={rgb, 255:red, 0; green, 0; blue, 0 }  ][line width=0.75]    (10.93,-3.29) .. controls (6.95,-1.4) and (3.31,-0.3) .. (0,0) .. controls (3.31,0.3) and (6.95,1.4) .. (10.93,3.29)   ;
\draw    (217.09,131.55) -- (348.79,70.04) ;
\draw [shift={(350.6,69.2)}, rotate = 154.97] [color={rgb, 255:red, 0; green, 0; blue, 0 }  ][line width=0.75]    (10.93,-3.29) .. controls (6.95,-1.4) and (3.31,-0.3) .. (0,0) .. controls (3.31,0.3) and (6.95,1.4) .. (10.93,3.29)   ;
\draw    (243.09,131.55) -- (459.67,70.62) ;
\draw [shift={(461.6,70.08)}, rotate = 164.29] [color={rgb, 255:red, 0; green, 0; blue, 0 }  ][line width=0.75]    (10.93,-3.29) .. controls (6.95,-1.4) and (3.31,-0.3) .. (0,0) .. controls (3.31,0.3) and (6.95,1.4) .. (10.93,3.29)   ;
\draw    (281.6,131.18) -- (180.35,74.6) ;
\draw [shift={(178.6,73.62)}, rotate = 29.2] [color={rgb, 255:red, 0; green, 0; blue, 0 }  ][line width=0.75]    (10.93,-3.29) .. controls (6.95,-1.4) and (3.31,-0.3) .. (0,0) .. controls (3.31,0.3) and (6.95,1.4) .. (10.93,3.29)   ;
\draw    (299.09,127.12) -- (263.67,70.89) ;
\draw [shift={(262.6,69.2)}, rotate = 57.79] [color={rgb, 255:red, 0; green, 0; blue, 0 }  ][line width=0.75]    (10.93,-3.29) .. controls (6.95,-1.4) and (3.31,-0.3) .. (0,0) .. controls (3.31,0.3) and (6.95,1.4) .. (10.93,3.29)   ;
\draw    (315.09,128.01) -- (353.44,74.36) ;
\draw [shift={(354.6,72.74)}, rotate = 125.56] [color={rgb, 255:red, 0; green, 0; blue, 0 }  ][line width=0.75]    (10.93,-3.29) .. controls (6.95,-1.4) and (3.31,-0.3) .. (0,0) .. controls (3.31,0.3) and (6.95,1.4) .. (10.93,3.29)   ;
\draw    (320.6,129.41) -- (464.74,73.46) ;
\draw [shift={(466.6,72.74)}, rotate = 158.79] [color={rgb, 255:red, 0; green, 0; blue, 0 }  ][line width=0.75]    (10.93,-3.29) .. controls (6.95,-1.4) and (3.31,-0.3) .. (0,0) .. controls (3.31,0.3) and (6.95,1.4) .. (10.93,3.29)   ;
\draw    (404.09,131.55) -- (195.53,74.15) ;
\draw [shift={(193.6,73.62)}, rotate = 15.39] [color={rgb, 255:red, 0; green, 0; blue, 0 }  ][line width=0.75]    (10.93,-3.29) .. controls (6.95,-1.4) and (3.31,-0.3) .. (0,0) .. controls (3.31,0.3) and (6.95,1.4) .. (10.93,3.29)   ;
\draw    (414.09,133.32) -- (275.43,72.65) ;
\draw [shift={(273.6,71.85)}, rotate = 23.63] [color={rgb, 255:red, 0; green, 0; blue, 0 }  ][line width=0.75]    (10.93,-3.29) .. controls (6.95,-1.4) and (3.31,-0.3) .. (0,0) .. controls (3.31,0.3) and (6.95,1.4) .. (10.93,3.29)   ;
\draw    (414.09,133.32) -- (364.86,72.52) ;
\draw [shift={(363.6,70.97)}, rotate = 51] [color={rgb, 255:red, 0; green, 0; blue, 0 }  ][line width=0.75]    (10.93,-3.29) .. controls (6.95,-1.4) and (3.31,-0.3) .. (0,0) .. controls (3.31,0.3) and (6.95,1.4) .. (10.93,3.29)   ;
\draw    (430.09,131.55) -- (472.42,73.47) ;
\draw [shift={(473.6,71.85)}, rotate = 126.09] [color={rgb, 255:red, 0; green, 0; blue, 0 }  ][line width=0.75]    (10.93,-3.29) .. controls (6.95,-1.4) and (3.31,-0.3) .. (0,0) .. controls (3.31,0.3) and (6.95,1.4) .. (10.93,3.29)   ;
\draw    (126.85,193.26) -- (520.85,188.85) ;
\draw [shift={(522.85,188.83)}, rotate = 179.36] [color={rgb, 255:red, 0; green, 0; blue, 0 }  ][line width=0.75]    (10.93,-3.29) .. controls (6.95,-1.4) and (3.31,-0.3) .. (0,0) .. controls (3.31,0.3) and (6.95,1.4) .. (10.93,3.29)   ;
\draw  [fill={rgb, 255:red, 155; green, 155; blue, 155 }  ,fill opacity=0.6 ] (146.05,145.35) .. controls (146.05,136.54) and (224.2,129.41) .. (320.6,129.41) .. controls (417,129.41) and (495.15,136.54) .. (495.15,145.35) .. controls (495.15,154.15) and (417,161.29) .. (320.6,161.29) .. controls (224.2,161.29) and (146.05,154.15) .. (146.05,145.35) -- cycle ;
\draw    (205.07,245.81) -- (171.15,300.47) ;
\draw [shift={(170.09,302.17)}, rotate = 301.83] [color={rgb, 255:red, 0; green, 0; blue, 0 }  ][line width=0.75]    (10.93,-3.29) .. controls (6.95,-1.4) and (3.31,-0.3) .. (0,0) .. controls (3.31,0.3) and (6.95,1.4) .. (10.93,3.29)   ;
\draw    (222.29,250.01) -- (256.4,304.6) ;
\draw [shift={(257.46,306.3)}, rotate = 238] [color={rgb, 255:red, 0; green, 0; blue, 0 }  ][line width=0.75]    (10.93,-3.29) .. controls (6.95,-1.4) and (3.31,-0.3) .. (0,0) .. controls (3.31,0.3) and (6.95,1.4) .. (10.93,3.29)   ;
\draw    (217.24,247.54) -- (350.18,304.87) ;
\draw [shift={(352.02,305.66)}, rotate = 203.33] [color={rgb, 255:red, 0; green, 0; blue, 0 }  ][line width=0.75]    (10.93,-3.29) .. controls (6.95,-1.4) and (3.31,-0.3) .. (0,0) .. controls (3.31,0.3) and (6.95,1.4) .. (10.93,3.29)   ;
\draw    (243.67,247.82) -- (462.93,305.52) ;
\draw [shift={(464.87,306.03)}, rotate = 194.74] [color={rgb, 255:red, 0; green, 0; blue, 0 }  ][line width=0.75]    (10.93,-3.29) .. controls (6.95,-1.4) and (3.31,-0.3) .. (0,0) .. controls (3.31,0.3) and (6.95,1.4) .. (10.93,3.29)   ;
\draw    (282.81,248.56) -- (179.05,298.95) ;
\draw [shift={(177.25,299.82)}, rotate = 334.1] [color={rgb, 255:red, 0; green, 0; blue, 0 }  ][line width=0.75]    (10.93,-3.29) .. controls (6.95,-1.4) and (3.31,-0.3) .. (0,0) .. controls (3.31,0.3) and (6.95,1.4) .. (10.93,3.29)   ;
\draw    (300.53,252.44) -- (263.74,303.12) ;
\draw [shift={(262.57,304.74)}, rotate = 305.97] [color={rgb, 255:red, 0; green, 0; blue, 0 }  ][line width=0.75]    (10.93,-3.29) .. controls (6.95,-1.4) and (3.31,-0.3) .. (0,0) .. controls (3.31,0.3) and (6.95,1.4) .. (10.93,3.29)   ;
\draw    (316.81,251.8) -- (354.91,300.91) ;
\draw [shift={(356.14,302.49)}, rotate = 232.19] [color={rgb, 255:red, 0; green, 0; blue, 0 }  ][line width=0.75]    (10.93,-3.29) .. controls (6.95,-1.4) and (3.31,-0.3) .. (0,0) .. controls (3.31,0.3) and (6.95,1.4) .. (10.93,3.29)   ;
\draw    (322.43,250.58) -- (468.11,302.99) ;
\draw [shift={(469.99,303.67)}, rotate = 199.79] [color={rgb, 255:red, 0; green, 0; blue, 0 }  ][line width=0.75]    (10.93,-3.29) .. controls (6.95,-1.4) and (3.31,-0.3) .. (0,0) .. controls (3.31,0.3) and (6.95,1.4) .. (10.93,3.29)   ;
\draw    (407.33,249.52) -- (194.44,299.53) ;
\draw [shift={(192.49,299.98)}, rotate = 346.78] [color={rgb, 255:red, 0; green, 0; blue, 0 }  ][line width=0.75]    (10.93,-3.29) .. controls (6.95,-1.4) and (3.31,-0.3) .. (0,0) .. controls (3.31,0.3) and (6.95,1.4) .. (10.93,3.29)   ;
\draw    (417.52,248.01) -- (275.66,301.73) ;
\draw [shift={(273.79,302.44)}, rotate = 339.26] [color={rgb, 255:red, 0; green, 0; blue, 0 }  ][line width=0.75]    (10.93,-3.29) .. controls (6.95,-1.4) and (3.31,-0.3) .. (0,0) .. controls (3.31,0.3) and (6.95,1.4) .. (10.93,3.29)   ;
\draw    (417.52,248.01) -- (366.62,302.73) ;
\draw [shift={(365.26,304.19)}, rotate = 312.93] [color={rgb, 255:red, 0; green, 0; blue, 0 }  ][line width=0.75]    (10.93,-3.29) .. controls (6.95,-1.4) and (3.31,-0.3) .. (0,0) .. controls (3.31,0.3) and (6.95,1.4) .. (10.93,3.29)   ;
\draw    (425.09,246.41) -- (475.76,303.05) ;
\draw [shift={(477.09,304.55)}, rotate = 228.19] [color={rgb, 255:red, 0; green, 0; blue, 0 }  ][line width=0.75]    (10.93,-3.29) .. controls (6.95,-1.4) and (3.31,-0.3) .. (0,0) .. controls (3.31,0.3) and (6.95,1.4) .. (10.93,3.29)   ;
\draw  [fill={rgb, 255:red, 155; green, 155; blue, 155 }  ,fill opacity=0.6 ] (146.44,235.21) .. controls (146.39,227.2) and (224.49,220.39) .. (320.89,219.98) .. controls (417.29,219.58) and (495.47,225.74) .. (495.53,233.74) .. controls (495.58,241.75) and (417.47,248.56) .. (321.08,248.97) .. controls (224.68,249.37) and (146.49,243.21) .. (146.44,235.21) -- cycle ;
\draw [color={rgb, 255:red, 0; green, 0; blue, 0 }  ,draw opacity=1 ]   (320.6,161.29) -- (320.88,217.98) ;
\draw [shift={(320.89,219.98)}, rotate = 269.72] [color={rgb, 255:red, 0; green, 0; blue, 0 }  ,draw opacity=1 ][line width=0.75]    (15.3,-4.61) .. controls (9.73,-1.96) and (4.63,-0.42) .. (0,0) .. controls (4.63,0.42) and (9.73,1.96) .. (15.3,4.61)   ;
\draw [color={rgb, 255:red, 0; green, 0; blue, 0 }  ,draw opacity=1 ]   (213.09,157.82) -- (130.97,187.15) ;
\draw [shift={(129.09,187.82)}, rotate = 340.35] [color={rgb, 255:red, 0; green, 0; blue, 0 }  ,draw opacity=1 ][line width=0.75]    (15.3,-4.61) .. controls (9.73,-1.96) and (4.63,-0.42) .. (0,0) .. controls (4.63,0.42) and (9.73,1.96) .. (15.3,4.61)   ;
\draw [color={rgb, 255:red, 0; green, 0; blue, 0 }  ,draw opacity=1 ]   (132.09,196.82) -- (212.18,221.23) ;
\draw [shift={(214.09,221.82)}, rotate = 196.96] [color={rgb, 255:red, 0; green, 0; blue, 0 }  ,draw opacity=1 ][line width=0.75]    (15.3,-4.61) .. controls (9.73,-1.96) and (4.63,-0.42) .. (0,0) .. controls (4.63,0.42) and (9.73,1.96) .. (15.3,4.61)   ;
\draw [color={rgb, 255:red, 0; green, 0; blue, 0 }  ,draw opacity=1 ]   (427.09,157.82) -- (512.19,185.21) ;
\draw [shift={(514.09,185.82)}, rotate = 197.84] [color={rgb, 255:red, 0; green, 0; blue, 0 }  ,draw opacity=1 ][line width=0.75]    (15.3,-4.61) .. controls (9.73,-1.96) and (4.63,-0.42) .. (0,0) .. controls (4.63,0.42) and (9.73,1.96) .. (15.3,4.61)   ;
\draw [color={rgb, 255:red, 0; green, 0; blue, 0 }  ,draw opacity=1 ]   (429.09,222.82) -- (513.2,193.48) ;
\draw [shift={(515.09,192.82)}, rotate = 160.77] [color={rgb, 255:red, 0; green, 0; blue, 0 }  ,draw opacity=1 ][line width=0.75]    (15.3,-4.61) .. controls (9.73,-1.96) and (4.63,-0.42) .. (0,0) .. controls (4.63,0.42) and (9.73,1.96) .. (15.3,4.61)   ;

\draw (206,134.56) node [anchor=north west][inner sep=0.75pt]    {$f_{X_{1}}$};
\draw (288,134.56) node [anchor=north west][inner sep=0.75pt]    {$f_{X_{2}}$};
\draw (348,136.51) node [anchor=north west][inner sep=0.75pt]    {$\cdots $};
\draw (409,135.45) node [anchor=north west][inner sep=0.75pt]    {$f_{X_{\overline{p}}}$};
\draw (154,46.96) node [anchor=north west][inner sep=0.75pt]    {$X_{1}$};
\draw (246,46.96) node [anchor=north west][inner sep=0.75pt]    {$X_{2}$};
\draw (343,47.84) node [anchor=north west][inner sep=0.75pt]    {$X_{3}$};
\draw (466,49.61) node [anchor=north west][inner sep=0.75pt]    {$X_{p}$};
\draw (399,49.73) node [anchor=north west][inner sep=0.75pt]    {$\cdots $};
\draw (106,183.35) node [anchor=north west][inner sep=0.75pt]    {$A$};
\draw (534,180.69) node [anchor=north west][inner sep=0.75pt]    {$Y$};
\draw (206.33,224.34) node [anchor=north west][inner sep=0.75pt]  [rotate=-359.7]  {$f_{M_{1}}$};
\draw (288.33,224) node [anchor=north west][inner sep=0.75pt]  [rotate=-359.7]  {$f_{M_{2}}$};
\draw (350.33,225.64) node [anchor=north west][inner sep=0.75pt]  [rotate=-359.7]  {$\cdots $};
\draw (409.33,224.29) node [anchor=north west][inner sep=0.75pt]  [rotate=-359.7]  {$f_{M_{\overline{q}}}$};
\draw (158.83,308.74) node [anchor=north west][inner sep=0.75pt]  [rotate=-359.64]  {$M_{1}$};
\draw (250.83,308.28) node [anchor=north west][inner sep=0.75pt]  [rotate=-359.64]  {$M_{2}$};
\draw (347.83,308.59) node [anchor=north west][inner sep=0.75pt]  [rotate=-359.64]  {$M_{3}$};
\draw (470.84,309.58) node [anchor=north west][inner sep=0.75pt]  [rotate=-359.64]  {$M_{q}$};
\draw (403.84,310.12) node [anchor=north west][inner sep=0.75pt]  [rotate=-359.64]  {$\cdots $};

\end{tikzpicture}
}
    \caption{Causal relationship in the presence of latent factors $\mathbf{f}_X$ and $\mathbf{f}_M$}
    \vspace{-5mm}
    \label{fig:DAG}
\end{figure}
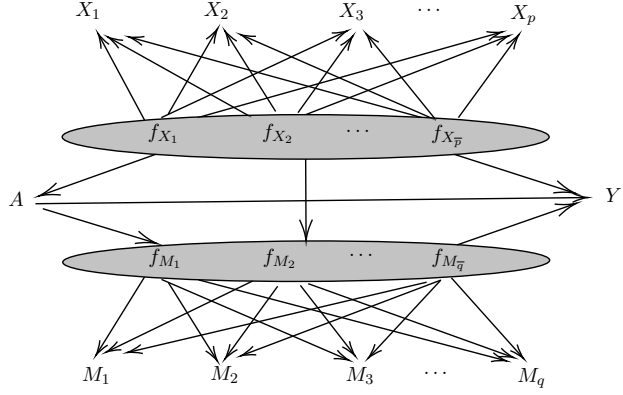

In the applications motivating this work, $\boldsymbol f_X$ and $\boldsymbol f_M$ are not directly observed. Instead, we observe high-dimensional measurements
$X\in\mathbb R^p$ and $M\in\mathbb R^q$, with $p\gg \bar p$ and $q\gg \bar q$, that serve as noisy proxies for the latent factors; see Figure~\ref{fig:DAG}. We assume the measurement model
\begin{equation}\label{eq:factors}
    X=\phi_X(\boldsymbol f_X)+\boldsymbol u_X,
    \qquad
    M=\phi_M(\boldsymbol f_M)+\boldsymbol u_M,
\end{equation}
where $\phi_X:\mathbb R^{\bar p}\to\mathbb R^p$ and
$\phi_M:\mathbb R^{\bar q}\to\mathbb R^q$ are unknown loading functions, and
$\boldsymbol u_X$ and $\boldsymbol u_M$ are centered idiosyncratic errors. The classical linear factor model is a special case of \eqref{eq:factors}, obtained by taking
$\phi_X(f_X)=\Gamma_X f_X$ and $\phi_M(f_M)=\Gamma_M f_M$ for loading matrices
$\Gamma_X\in\mathbb R^{p\times \bar p}$ and
$\Gamma_M\in\mathbb R^{q\times \bar q}$.

We next define the causal estimands at the latent-factor level. Let
$\boldsymbol f_M^{(a)}$ denote the mediator factors that would be observed if the treatment were set to $a$, and let $Y^{(a,f_M)}$ denote the potential outcome if the treatment were set to $a$ and the mediator factors were set to the value $f_M$. We use the shorthand
$Y^{(a)}=Y^{(a,\boldsymbol f_M^{(a)})}$. For treatment level $1$ relative to $0$, the total effect decomposes as
\begin{equation}
\textstyle
\label{eq:NDENIE}
\underbrace{\mathbb E\{Y^{(1)}-Y^{(0)}\}}_{\text{Total effect}}
=
\underbrace{\mathbb E\!
\left[ Y^{(1,\boldsymbol f_M^{(1)})}-Y^{(1,\boldsymbol f_M^{(0)})}\right]}_{\text{Natural indirect effect}}+
\underbrace{\mathbb E\!
\left[ Y^{(1,\boldsymbol f_M^{(0)})}-Y^{(0,\boldsymbol f_M^{(0)})}\right]}_{\text{Natural direct effect}} .
\end{equation}
Identification of parameters of the form $\theta(a,a')=\mathbb E\{Y^{(a,\boldsymbol f_M^{(a')})}\}$ for any $a,a'\in\{0,1\}$ yields identifying the natural indirect effect, natural direct effect, and total effect through \eqref{eq:NDENIE}. Specifically, $\mathrm{NIE}=\theta(1,1)-\theta(1,0)$,
    $\mathrm{NDE}=\theta(1,0)-\theta(0,0)$, and
    $\mathrm{TE}=\theta(1,1)-\theta(0,0)$.
Hence, in what follows, we focus on the cross-world mean
\begin{equation*}
    \theta_0:=\theta(1,0)
    =\mathbb E\!\bigl[Y^{(1,\boldsymbol f_M^{(0)})}\bigr].
\end{equation*}

To identify $\theta_0$, we impose the following standard causal assumptions.

\begin{assumption}[Consistency]\label{ass:consistency}
For all $a\in\{0,1\}$ and all mediator values $f_M$, if $A=a$ and
$\boldsymbol f_M=f_M$, then $Y=Y^{(a,f_M)}$. In addition, if $A=a$, then
$\boldsymbol f_M=\boldsymbol f_M^{(a)}$.
\end{assumption}

\begin{assumption}[Positivity]\label{ass:positivity}
For every value $f_X$ in the support of $\boldsymbol f_X$,
$0<\Pr(A=1\mid \boldsymbol f_X=f_X)<1$. Moreover, the conditional density of the mediator is positive, i.e., $0 < p({f}_M\mid A=a, \boldsymbol{f}_X={f}_X)$ for all $A$, $f_X$, and $f_M$.
\end{assumption}

\begin{assumption}[Sequential exchangeability]\label{assumption:Sequential}
Let $U\independent V\mid W$ denote conditional independence of $U$ and $V$ given
$W$. For all $a,a'\in\{0,1\}$ and all mediator values $f_M$,

$~~~~~1.~\{Y^{(a,f_M)},\boldsymbol f_M^{(a)}\}\independent A\mid \boldsymbol f_X;
\qquad
2.~Y^{(a,f_M)}\independent \boldsymbol f_M^{(a')}\mid \boldsymbol f_X.$
\end{assumption}

The first part rules out unmeasured confounding of the treatment--outcome and treatment--mediator relationships after conditioning on the latent covariate factors. The second part is the usual cross-world condition for natural direct and indirect effects; it rules out unmeasured mediator--outcome confounding not captured by $\boldsymbol f_X$ and excludes treatment-induced mediator--outcome confounders.

Under Assumptions \ref{ass:consistency}--\ref{assumption:Sequential}, the mediation functional is identified by the mediation formula \citep{pearl2001direct,imai2010identification}. Let
\(
    \mu_1(f_X,f_M):=\mathbb E\{Y\mid A=1,\boldsymbol f_X=f_X,\boldsymbol f_M=f_M\}.
\)
Then
\begin{equation}\label{theta:functional}
\begin{aligned}
\theta_0
&=\mathbb E\!\left[
    \mathbb E\!\left\{
        \mathbb E\left(Y\mid A=1,\boldsymbol f_X,\boldsymbol f_M\right)
        \mid A=0,\boldsymbol f_X
    \right\}
\right].
\end{aligned}
\end{equation}

A direct plug-in estimator based on \eqref{theta:functional} can be sensitive to nuisance-model misspecification. Following the semiparametric theory of \cite{tchetgen2012semiparametric}, we instead use the efficient influence function-based estimator of $\theta_0$. Define the nuisance functions
\begin{equation}
\textstyle
\label{eq:nuisance}
\begin{aligned}
\pi_1(f_X)
&:=\Pr(A=1\mid \boldsymbol f_X=f_X),
& \mu_1(f_X,f_M)
&:=\mathbb E\{Y\mid A=1,\boldsymbol f_X=f_X,\boldsymbol f_M=f_M\},
\\
\pi_2(f_X,f_M)
&:=\frac{p(f_M\mid A=0, f_X)}{p(f_M\mid A=1, f_X)},
&
\mu_{10}(f_X)
&:=\mathbb E\{\mu_1(\boldsymbol f_X,\boldsymbol f_M)
    \mid A=0,\boldsymbol f_X=f_X\}.
\end{aligned}
\end{equation}
With these definitions, $\theta_0=\mathbb E\{\mu_{10}(\boldsymbol f_X)\}$, and the efficient influence function is
\begin{equation}
\label{eq:EIF}
\textstyle
\begin{aligned}
\psi_0(O)
&=\mu_{10}(\boldsymbol f_X)-\theta_0
+\frac{A}{\pi_1(\boldsymbol f_X)}\pi_2(\boldsymbol f_X,\boldsymbol f_M)
    \{Y-\mu_1(\boldsymbol f_X,\boldsymbol f_M)\}\\
&\quad+
\frac{1-A}{1-\pi_1(\boldsymbol f_X)}
    \{\mu_1(\boldsymbol f_X,\boldsymbol f_M)-\mu_{10}(\boldsymbol f_X)\} \,,
\end{aligned}
\end{equation}
where $O = (Y, A, \blf_X, \blf_M)$. 
In the oracle setting where $(\boldsymbol f_X,\boldsymbol f_M)$ are observed, replacing the nuisance functions in \eqref{eq:EIF} by estimates and solving the corresponding estimating equation gives
\begin{equation}
\label{eq:IF}
\textstyle
\begin{aligned}
\hat\theta^{\mathrm{IF}}_0
=\mathbb P_n\bigg[&\hat\mu_{10}(\boldsymbol f_X)
+\frac{A}{\hat\pi_1(\boldsymbol f_X)}\hat\pi_2(\boldsymbol f_X,\boldsymbol f_M)
    \{Y-\hat\mu_1(\boldsymbol f_X,\boldsymbol f_M)\}\\
&\quad+
\frac{1-A}{1-\hat\pi_1(\boldsymbol f_X)}
    \{\hat\mu_1(\boldsymbol f_X,\boldsymbol f_M)-\hat\mu_{10}(\boldsymbol f_X)\}
\bigg],
\end{aligned}
\end{equation}
where $\mathbb P_n$ denotes the empirical average. However, in reality, we only observe $(X, M)$, (ultra)-high-dimensional proxies of $(\blf_X, \blf_M)$, and the latent factors in this oracle estimator are replaced by learned representations, as elaborated in the next section.

%% file: Method_revised.tex
\section{Proposed Method: MediEncoder}\label{section: method}

\label{section:method}

\begin{figure}[t]
    \centering
     \includegraphics[width=1.0\linewidth]{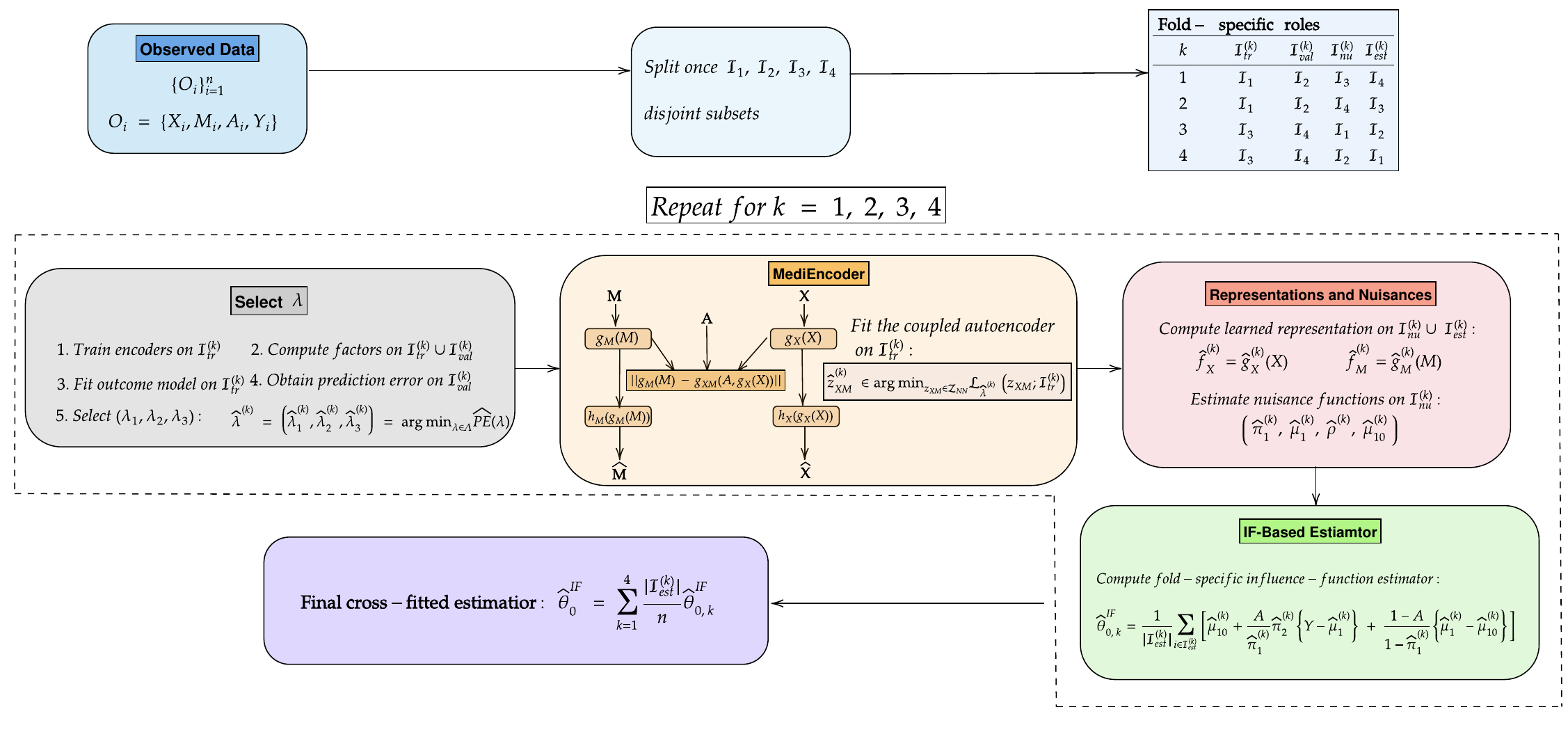}
    \caption{Schematic overview of MediEncoder. The observed data are split into disjoint subsets for representation learning, tuning, nuisance estimation, and final evaluation.}
    \label{fig:MediEncoder}
\end{figure}

We now describe MediEncoder, a cross-fitted procedure for estimating the mediation functional
\(\theta_0=\theta(1,0)=\mathbb E\{Y^{(1,\boldsymbol f_M^{(0)})}\}\). The method has two main components. First, it learns low-dimensional nonlinear representations of the high-dimensional covariates and mediators through a coupled autoencoder. Second, it uses these learned representations to estimate the nuisance functions in the efficient-influence function estimator from Section~\ref{section:2}.

\paragraph{Coupled representation learning.}
Let
\(
    z_{XM}=(g_X,g_M,g_{XM},h_X,h_M)\in\mathcal Z_{\mathrm{NN}},
\)
where \(g_X:\mathbb R^p\to\mathbb R^{\tilde p}\) and \(h_X:\mathbb R^{\tilde p}\to\mathbb R^p\) are the covariate encoder and decoder, \(g_M:\mathbb R^q\to\mathbb R^{\tilde q}\) and \(h_M:\mathbb R^{\tilde q}\to\mathbb R^q\) are the mediator encoder and decoder, and
\(g_{XM}:\{0,1\}\times\mathbb R^{\tilde p}\to\mathbb R^{\tilde q}\) maps the treatment and covariate representation to the mediator-representation space. For a training index set \(\mathcal I\), define the empirical loss
\begin{equation}
\textstyle
\label{eq:medi_loss}
\begin{aligned}
\mathcal L_{\lambda}(z_{XM};\mathcal I)
=\frac{1}{|\mathcal I|}\sum_{i\in\mathcal I}
\Big[&
\lambda_1\|X_i-h_X(g_X(X_i))\|_2^2
+\lambda_2\|M_i-h_M(g_M(M_i))\|_2^2 \\
&+\lambda_3\|g_M(M_i)-g_{XM}(A_i,g_X(X_i))\|_2^2
\Big],
\end{aligned}
\end{equation}
where \(\lambda=(\lambda_1,\lambda_2,\lambda_3)\) consists of nonnegative tuning parameters. The first two terms are reconstruction losses for \(X\) and \(M\). The third term couples the two representations by encouraging the mediator representation to be predictable from the treatment and the covariate representation, matching the structural dependence of the latent mediator factors on \((A,\boldsymbol f_X)\). 
Unlike a standard multi-view autoencoder, the coupling is \textit{directional and treatment-aware}: the mediator representation is encouraged to be predictable from $(A,g_X(X))$, obeying the latent structural equation for $\boldsymbol f_M$.
After representation learning, only the encoders \(\hat g_X\) and \(\hat g_M\) are used in the influence function-based estimator; the decoders and \(g_{XM}\) act as regularizers during training.

The working dimensions \((\tilde p,\tilde q)\) need not equal the true latent dimensions \((\bar p,\bar q)\). The theoretical results in Section~\ref{sec:theory} require the learned representation to approximate a sufficiently informative, left-invertible transformation of the latent factors. Thus, exact recovery of the original factors is not required, but the chosen representation dimensions must be large enough to retain the information relevant for the mediation functional.

\paragraph{Cross-fitted estimation.}
Algorithm~\ref{Algo:MediEncoder} gives the full procedure for estimating \(\theta_0\). The data are split into four disjoint subsets. In each fold, these subsets play four distinct roles: representation training, tuning, nuisance estimation, and final evaluation. This separation ensures that the summands used to evaluate the influence function-based estimator are not used to train either the representation or the nuisance functions. The particular rotations below are one convenient choice; any rotations with the same disjointness property can be used.

The estimator in \eqref{eq:fold_estimator_method} is the empirical analogue of the efficient influence function representation in Section~\ref{section:2}, with the latent factors replaced by fold-specific learned representations. The auxiliary propensity score
\(\rho(f_X,f_M)=\Pr(A=1\mid \boldsymbol f_X=f_X,\boldsymbol f_M=f_M)\) is used only to estimate the mediator density ratio \(\pi_2\). Indeed,
\begin{equation}
    \begin{aligned}\label{eq:pi2_ratio}
         \frac{p(f_M\mid A=0,f_X)}{p(f_M\mid A=1,f_X)}
    =
    \frac{\Pr(A=0\mid f_X,f_M)}{\Pr(A=1\mid f_X,f_M)}
    \frac{\Pr(A=1\mid f_X)}{\Pr(A=0\mid f_X)},
    \end{aligned}
\end{equation}
which yields \eqref{eq:pi2_bayes_method} after substituting \(\rho\) and \(\pi_1\).

The same construction estimates any mediation functional
\(\theta(a,a')=\mathbb E\{Y^{(a,\boldsymbol f_M^{(a')})}\}\). Consequently,
\[
    \widehat{\mathrm{NIE}}=\hat\theta^{\mathrm{IF}}(1,1)-\hat\theta^{\mathrm{IF}}(1,0),\qquad
    \widehat{\mathrm{NDE}}=\hat\theta^{\mathrm{IF}}(1,0)-\hat\theta^{\mathrm{IF}}(0,0),\qquad
    \widehat{\mathrm{TE}}=\hat\theta^{\mathrm{IF}}(1,1)-\hat\theta^{\mathrm{IF}}(0,0).
\]
Section~\ref{sec:theory} establishes conditions under which \(\hat\theta_0^{\mathrm{IF}}\) is \(\sqrt n\)-consistent and asymptotically normal.

\begin{remark}[Capacity of the cross-factor network \(g_{XM}\)]
\label{rem:gxm_capacity}
The cross-factor network \(g_{XM}\) should have moderate capacity. If it is overly expressive, it can approximate an arbitrary mapping from \((A,g_X(X))\) to \(g_M(M)\), driving the alignment loss in \eqref{eq:medi_loss} close to zero regardless of the quality of \(g_X\) and \(g_M\). In that case, \(g_{XM}\) absorbs discrepancies between the two representations, and the alignment term no longer enforces a meaningful structural relationship. Conversely, if \(g_{XM}\) is too restrictive, it may fail to capture the dependence of the mediator factors on the treatment and covariate factors. In practice, we use a lower- or moderate-capacity architecture for \(g_{XM}\), relative to the encoders and decoders, so that the alignment term couples the representations without allowing the cross-factor network to overfit.
\end{remark}

\begin{algorithm}[H]
\caption{\texttt{MediEncoder}: Estimating \(\theta_0\) with cross-fitting}
\label{Algo:MediEncoder}
\small
\begin{algorithmic}[1]
\Require Data \(\{(X_i,A_i,M_i,Y_i)\}_{i=1}^n\); representation dimensions \((\tilde p,\tilde q)\); tuning grid \(\Lambda\)
\Ensure Cross-fitted estimator \(\hat\theta_0^{\mathrm{IF}}\)

\State Split \(\{1,\ldots,n\}\) into four disjoint subsets \(\mathcal I_1,\ldots,\mathcal I_4\), each of size approximately \(n/4\). Use the following fold assignments, where the four columns denote the representation-training, validation, nuisance-estimation, and final-evaluation sets:
\vspace{-2mm}
    \[
    \begin{array}{c|cccc}
    k & \mathcal I_{\mathrm{tr}}^{(k)} & \mathcal I_{\mathrm{val}}^{(k)} & \mathcal I_{\mathrm{nu}}^{(k)} & \mathcal I_{\mathrm{est}}^{(k)} \\\hline
    1 & \mathcal I_1 & \mathcal I_2 & \mathcal I_3 & \mathcal I_4 \\
    2 & \mathcal I_1 & \mathcal I_2 & \mathcal I_4 & \mathcal I_3 \\
    3 & \mathcal I_3 & \mathcal I_4 & \mathcal I_1 & \mathcal I_2 \\
    4 & \mathcal I_3 & \mathcal I_4 & \mathcal I_2 & \mathcal I_1
    \end{array}
    \]
\vspace{-3mm}
\For{\(k=1,\ldots,4\)}
    \State Set \(\mathcal I_{\mathrm{tr}}^{(k)},\mathcal I_{\mathrm{val}}^{(k)},\mathcal I_{\mathrm{nu}}^{(k)},\mathcal I_{\mathrm{est}}^{(k)}\) according to the \(k\)-th row above.
    \State Select \(\hat\lambda^{(k)}=(\hat\lambda_1^{(k)},\hat\lambda_2^{(k)},\hat\lambda_3^{(k)})\in\Lambda\) using the training and validation sets \((\mathcal I_{\mathrm{tr}}^{(k)},\mathcal I_{\mathrm{val}}^{(k)})\) via Algorithm~\ref{Algo:TuneLambdaMediEncoder}.

    \State Fit the coupled autoencoder on \(\mathcal I_{\mathrm{tr}}^{(k)}\):
    \begin{equation}\label{eq:medi_train_fold}
    \widehat z_{XM}^{(k)}
    =
    (\hat g_X^{(k)},\hat g_M^{(k)},\hat g_{XM}^{(k)},\hat h_X^{(k)},\hat h_M^{(k)})
    \in
    \arg\min_{z_{XM}\in\mathcal Z_{\mathrm{NN}}}
    \mathcal L_{\hat\lambda^{(k)}}(z_{XM};\mathcal I_{\mathrm{tr}}^{(k)}).
    \end{equation}
\vspace{-3mm}
    \State Compute learned representations for all \(i\in\mathcal I_{\mathrm{nu}}^{(k)}\cup\mathcal I_{\mathrm{est}}^{(k)}\):
    \[
        \widehat{\boldsymbol f}_{X,i}^{(k)}=\hat g_X^{(k)}(X_i),
        \qquad
        \widehat{\boldsymbol f}_{M,i}^{(k)}=\hat g_M^{(k)}(M_i).
    \]

    \State Estimate nuisance functions via $\hat\pi_1^{(k)}$, $\hat\mu_1^{(k)}$, $\hat\mu_{10}^{(k)}$, and $\hat \rho^{(k)}$ on \(\mathcal I_{\mathrm{nu}}^{(k)}\) using the learners provided in Appendix \ref{app:alg}. Set $\hat \pi_2^{(k)}$ as per Equation \eqref{eq:pi2_ratio}. 

    \State Compute the fold-specific influence-function estimator:
    \vspace{-2mm}
    \begin{equation}\label{eq:fold_estimator_method}
    \begin{aligned}
    \hat\theta^{\mathrm{IF}}_{0,k}
    =
    \frac{1}{|\mathcal I_{\mathrm{est}}^{(k)}|}
    \sum_{i\in\mathcal I_{\mathrm{est}}^{(k)}}
    \Bigg[
    &\hat\mu_{10}^{(k)}(\widehat{\boldsymbol f}_{X,i}^{(k)})
    +\frac{A_i}{\hat\pi_1^{(k)}(\widehat{\boldsymbol f}_{X,i}^{(k)})}
    \hat\pi_2^{(k)}(\widehat{\boldsymbol f}_{X,i}^{(k)},\widehat{\boldsymbol f}_{M,i}^{(k)})
    \Big\{Y_i-\hat\mu_1^{(k)}(\widehat{\boldsymbol f}_{X,i}^{(k)},\widehat{\boldsymbol f}_{M,i}^{(k)})\Big\}\\
    &+\frac{1-A_i}{1-\hat\pi_1^{(k)}(\widehat{\boldsymbol f}_{X,i}^{(k)})}
    \Big\{\hat\mu_1^{(k)}(\widehat{\boldsymbol f}_{X,i}^{(k)},\widehat{\boldsymbol f}_{M,i}^{(k)})
    -\hat\mu_{10}^{(k)}(\widehat{\boldsymbol f}_{X,i}^{(k)})\Big\}
    \Bigg].
    \end{aligned}
    \vspace{-3mm}
    \end{equation}
\EndFor

\State Return the weighted cross-fitted estimator
\vspace{-3mm}
\[
    \hat\theta_0^{\mathrm{IF}}
    =
    \sum_{k=1}^4
    \frac{|\mathcal I_{\mathrm{est}}^{(k)}|}{n}
    \hat\theta^{\mathrm{IF}}_{0,k}.
\]
\end{algorithmic}
\end{algorithm}

%% file: Theory_DM.tex
\section{Theoretical Analysis \label{sec:theory}}

In this section, we present theoretical results establishing the $\sqrt n$-consistency and asymptotic normality of the proposed estimator $\hat\theta^{\rm IF}_0$. As is clear from Algorithm~\ref{Algo:MediEncoder}, the estimation procedure relies on two key ingredients: (i) constructing $(\hat{\boldsymbol f}_X,\hat{\boldsymbol f}_M)$, a surrogate of the latent factors $(\boldsymbol f_X,\boldsymbol f_M)$, from the high-dimensional observations $(X,M)$; and (ii) estimating the nuisance functions that enter the influence function representation by regressing either $Y$ or $A$ on these learned representations. Moreover, as discussed in the Section~\ref{section:1}, we do not assume that the true factor dimensions are known. Instead, we only require an upper bound on them, and our theoretical framework is flexible enough to accommodate this setting. We impose the following regularity condition on the data-generating process, in addition to the standard causal identification assumptions in Assumptions~\ref{ass:consistency}--\ref{assumption:Sequential}.

\begin{assumption}
\label{assm:dgp_1}
There exists a constant $C<\infty$ such that
$
\mathbb E[(Y-\mu_1(\boldsymbol f_X,\boldsymbol f_M))^2
\mid A=1,\boldsymbol f_X,\boldsymbol f_M]\le C
$
and
$
\mathbb E[
(\mu_1(\boldsymbol f_X,\boldsymbol f_M)-\mu_{10}(\boldsymbol f_X))^2
\mid A=0,\boldsymbol f_X]\le C.
$
\end{assumption}

Assumption~\ref{assm:dgp_1} is mild. It requires the outcome noise and the relevant conditional mean contrast to have finite conditional second moments, which is a basic regularity requirement for establishing a central limit theorem. 
We further have the following requirement on the estimated factor surrogates.

\begin{assumption}
    \label{assm:factor}
    The estimated factors $\hat {\boldsymbol f}=(\hat{\boldsymbol f}_X ,\hat{\boldsymbol f}_M)$ satisfy: 
\begin{equation}
\label{eq:factor_rate_left_inverse}
\textstyle
\|\hat{\boldsymbol f} - \nu(\boldsymbol f)\|_{L_2(P)} = O_p(\delta_{n, f}), \qquad \delta_{n, f} \downarrow 0 \text{ as }n \uparrow \infty, 
\end{equation}
for some $\nu: \mathbb R^{\bar p + \bar q} \mapsto \mathbb R^{\tilde p + \tilde q}$, where $\nu = (\nu_X, \nu_M)$ admits a Lipschitz left inverse $\nu^{-1}: \mathbb R^{\tilde p + \tilde q} \mapsto \mathbb R^{\bar p + \bar q}$.
\end{assumption}

In Appendix~\ref{app:prop_factor_rec} (Proposition \ref{prop:factor_recovery_sufficient}), we give a sufficient condition showing that Equation \eqref{eq:factor_rate_left_inverse} follows from three interpretable ingredients: (i) the existence of oracle encoders that recover a left-invertible transformation of the latent factors up to denoising error; (ii) a local stability, or quadratic-growth, condition for the population MediEncoder objective around the oracle representation; and (iii) a small excess-risk bound for the fitted MediEncoder. This result shows that the factor condition is not tied to a particular algorithm, but can be verified whenever the representation learning objective is stable, and the learned encoder has small population excess risk. 
In linear factor models with pervasive factors, diversified projection provides one such example \cite{fan2013large,fan2023factor,fan2022learning}. For nonlinear factor models, recent autoencoder theory \cite{xiu2024deep} provides related recovery guarantees up to functional transformations. Translating these guarantees into \eqref{eq:factor_rate_left_inverse} requires additional stability conditions on the learned bottleneck map; we discuss this connection in Appendix~\ref{app:prop_factor_rec}.

We now present our first main result which shows that if the estimated factor surrogates and the nuisance estimators converge at appropriate rates, then the final estimator $\hat\theta^{\rm IF}$ is $\sqrt n$-consistent and asymptotically normal.

\begin{theorem}
\label{thm:main}
Suppose Assumptions \ref{ass:consistency}-\ref{assm:factor} hold 
and the nuisance estimators from Line~7 of Algorithm~\ref{Algo:MediEncoder} are Lipschitz continuous, and $(\pi_1, \hat \pi_1, \rho, \hat \rho)$ are uniformly bounded away from $0$ and $1$, and 
\begin{equation}
\begin{aligned}
\|\hat\pi_1-\pi_1\circ \nu_X^{-1}\|_{L_2(P)} &= O_p(\delta_{n, \pi_1}),
\qquad &
\|\hat\mu_1-\mu_1\circ \nu^{-1}\|_{L_2(P)} &= O_p(\delta_{n, \mu_1}),
\\
\|\hat\mu_{10}-\mu_{10}\circ \nu_X^{-1}\|_{L_2(P)} &= O_p(\delta_{n, \mu_{10}}),
\qquad &
\|\hat\rho-\rho\circ \nu^{-1}\|_{L_2(P)} &= O_p(\delta_{n, \rho}) \,.
\end{aligned}
\end{equation}
for some rates $\{\delta_{n,\eta}:\eta\in\Xi\}\downarrow 0$ as $n\uparrow\infty$, that satisfy: 
\begin{equation}
\begin{aligned}
    \label{eq:rate_cond}
&\sqrt n\,
(\delta_{n,\pi_1}+\delta_{n,\rho}+\delta_{n,f}+n^{-1/2})
(\delta_{n,\mu_1}+\delta_{n,f}+n^{-1/2})
=o(1),\\
&\sqrt n\,
(\delta_{n,\pi_1}+\delta_{n,f}+n^{-1/2})
(\delta_{n,\mu_{10}}+\delta_{n,f}+n^{-1/2})
=o(1).
\end{aligned}
\end{equation}
Then, the estimator $\hat\theta_0^{\mathrm{IF}}$ satisfies 
\(
\sqrt n\bigl(\hat\theta_0^{\mathrm{IF}}-\theta_0\bigr) \overset{\mathscr{L}}{\implies} \mathcal N(0, \sigma^2_{\rm eff}),
\)
where $\sigma^2_{\rm eff}$ is the semiparametrically efficient variance of the oracle latent-factor model. A sufficient condition for Equation \eqref{eq:rate_cond} is $\max\{
\delta_{n,\pi_1},
\delta_{n,\rho},
\delta_{n,\mu_1},
\delta_{n,\mu_{10}},
\delta_{n,f}\} =
o(n^{-1/4}).
$
\end{theorem}

\begin{remark}
It is important to note that the transformation $\nu$ is an unknown population-level map and is not estimated by our procedure. It is used only to formalize the class of latent representations under which the mediation functional remains estimable at root-$n$ rate. Thus, our theory does not require recovering the true factors themselves; it only requires the learned representation to be close to some left-invertible transformation of them.
\end{remark}

Theorem~\ref{thm:main} only requires weak rate requirements on the learned representation and nuisance functions irrespective of what algorithm is being used. We now verify these rates under a compositional smoothness assumption for DNN-based estimators as used in Algorithm \ref{Algo:MediEncoder}. However, the nuisances are estimated using learned representation $\widehat{\blf}$, but their population targets are functions of $\blf$ or equivalently $\nu(\blf)$ as $\nu$ is left-invertible; hence, the factor learning error $\delta_{n, f}$ contributes to the nuisance parameter estimation. We start with a definition of the compositional function class, namely Hierarchical Compositional Model (HCM): 

\begin{definition}[Hierarchical Composition Model]
Let $\mathcal P\subset [1,\infty)\times\mathbb N$. For $\ell=1$, define a collection $\mathcal H(d,1,\mathcal P)$ consisting of functions $h(x)=g(x_{\pi(1)},\ldots,x_{\pi(t)})$, where $g:\mathbb R^t\to\mathbb R$ is $(\beta,C)$ Holder-smooth for some $(\beta,t)\in\mathcal P$. For $\ell>1$, $\mathcal H(d,\ell,\mathcal P)$ is defined recursively by $h(x)=g(h_1(x),\ldots,h_t(x))$, where $g:\mathbb R^t\to\mathbb R$ is $(\beta,C)$ H\"older-smooth for some
$(\beta,t)\in\mathcal P$, and each $h_j\in\mathcal H(d,\ell-1,\mathcal P)$. The dimensional-adjusted smoothness (DAS) is defined as $\gamma^\star:=\inf_{(\beta,t)\in\mathcal P}(\beta/t)$. 
\end{definition}

HCM includes ordinary H\"older classes as well as additive, single-index, and other low-dimensional compositional models. For instance, if $\textstyle{h(x)=\sum_{j=1}^d h_j(x_j)}$ or $h(x)=g(x^\top\beta_0)$ with $\beta$-smooth univariate components or link functions, then the effective index is $\gamma^\star=\beta$ up to smooth linear and aggregation maps, whereas a generic $d$-variate H\"older function has $\gamma^\star=\beta/d$. Thus $\gamma^\star$ can be much larger than the ambient smoothness index, and DNN-based estimators adaptively achieve faster rates than those dictated by the ambient dimension. To that end, we require the following condition. 

\begin{assumption}
    \label{assm:subg}
    The mean functions $\mu_1$ and $\mu_{10}$ are uniformly bounded, and $(\eps, \eps', \bu_X, \bu_M)$ in Equation \eqref{eq:factor} are independent of $(\blf_X ,\blf_M)$ and centered sub-Gaussian random variables. 
\end{assumption}

\begin{theorem}
\label{thm:HCM_nuisance}
Suppose Assumptions \ref{ass:consistency}-\ref{assumption:Sequential}, \ref{assm:factor}, and \ref{assm:subg} hold. Define the transformed nuisances as $\eta_{\mu_1}:=\mu_1\circ\nu^{-1}, \eta_{10}:=\mu_{10}\circ\nu_X^{-1}, \eta_{\pi_1}:=\pi_1\circ\nu_X^{-1}$ and $\eta_{\rho}:=\rho\circ\nu^{-1}$. 
Assume that each of these $\eta_j$ admits a bounded Lipschitz HCM extension with DAS $\gamma_j^\star$ to a compact neighborhood containing both the transformed $(\nu(\blf))$ and learned representations ($\hat \blf$). Assume also the overlap conditions that $\pi_1$ and $\rho$ are bounded away from $0$ and $1$. Then, under a proper choice of width and depth (see Appendix \ref{Proof_of_HCM_nuisance} for details) of deep ReLU networks, the estimators obtained in Line 7 of Algorithm \ref{Algo:MediEncoder} satisfy: 
\begin{equation*}
\textstyle
\|\hat \eta_j(\hat \blf) - \eta_j(\nu(\blf))\|_{L_2(P)}^2 = O_p(\delta_{n, j}^2 ),\ \text{ with } \delta^2_{n, j}  = n^{-2\gamma^*_j/(2\gamma^*_j + 1)}(\log{n})^{a_j} + \delta_{n, f}^2,  
\end{equation*}
for $j \in \{\pi_1, \rho, \mu_1\}$, and $a_j > 0$ fixed constant. Further $\hat \eta_{10}$ satisfies: 
\begin{equation*}
\textstyle
\|\hat \eta_{10}(\hat \blf_X) -\eta_{10}(\nu_X(\blf_X))\|_{L_2(P)}^2
= O_p(\delta_{n, 10}^2), \quad \delta^2_{n, 10} = n^{-2\gamma_{10}^\star/(2\gamma_{10}^\star+1)}(\log n)^{a_{10}}
+\delta_{n,f}^2
+\delta_{n, \mu_1}^2 \,.
\end{equation*}
\end{theorem}

Theorem \ref{thm:HCM_nuisance} implies that, if $\gamma_j^\star>1/2$ and $\delta_{n,f}=o(n^{-1/4})$, then all nuisance functions are estimable at rates faster than $n^{-1/4}$. Consequently, Equation \eqref{eq:rate_cond} in Theorem~\ref{thm:main} holds, and $\textstyle{\widehat\theta^{\rm IF}}$ is $\textstyle{\sqrt{n}}$-CAN. The rate for $\mu_{10}$ additionally depends on the rate for $\mu_1$, because $\mu_{10}$ is estimated by regressing the estimated responses $\widehat\mu_1(\widehat{\blf}_X,\widehat{\blf}_M)$ on $\widehat{\blf}_X$. Our theory extends the results of \cite{fan2024factor} to nonlinear factor estimation and nuisance estimation under logistic-link losses.

%% file: simulation_ablation_revised.tex
\section{Experiments}\label{section:simu}

\subsection{Simulation and Ablation Study}

We evaluate \texttt{MediEncoder} in nonlinear high-dimensional settings generated from the latent mediation model in Section~\ref{section:2}. The target is
$\theta_0=\theta(1,0)$, estimated by the cross-fitted efficient influence function-based estimator in \eqref{eq:IF}. We compare with the baseline representation methods described in Appendix~\ref{appendix:baseline}. Across $B=500$ Monte Carlo replications, we report the empirical standard deviation (SD), root mean squared error (RMSE), average length of the nominal 95\% confidence interval, and empirical coverage. The full data-generating process is given in Appendix~\ref{appendix:simu}.

\begin{table}[t]
\centering
\small
\setlength{\tabcolsep}{5pt}
\begin{tabular}{llcccc}
\toprule
$n$ & \textbf{Estimator} & \textbf{SD} & \textbf{RMSE} & \textbf{CI Length} & \textbf{Coverage} \\
\midrule
\multirow{5}{*}{300}
& Projection   & 1.831& 2.011& 7.175& 0.951
\\
& Autoencoder  & 
1.533& 1.672& 6.011& 0.955
\\
& VAE          & 0.412& 0.423& 1.616& 0.952
\\
& IMAVAE       & 
0.442& 0.733& 1.740& 0.875
\\
& MediEncoder  & 0.366& 0.388& 1.435& 0.945
\\
\midrule
\multirow{5}{*}{1200}
& Projection   & 
0.319& 0.439& 1.251& 0.855
\\
& Autoencoder  & 
0.415& 0.465& 1.626& 0.915
\\
& VAE          & 
0.329& 0.332& 1.290& 0.935
\\
& IMAVAE       & 
0.341& 0.390& 1.337& 0.927
\\
& MediEncoder  & 0.269& 0.288& 1.056& 0.925
\\
\midrule
\multirow{5}{*}{3000}
& Projection   & 
0.262&  0.368&  1.027&  0.862
\\
& Autoencoder  & 0.331& 0.396&  1.294&  0.885\\
& VAE          & 
0.331& 0.343&  1.294& 0.930
\\
& IMAVAE       & 0.344& 0.349& 1.350&  0.947
\\
& MediEncoder  & 0.270&  0.290& 1.059& 0.955\\
\bottomrule
\end{tabular}
\caption{Representative simulation results under the nonlinear wavelet DGP with $p=2000$, $q=1000$, $\sigma_X=2$, $\sigma_M=1$, $\sigma_Y=1$, $\bar p=\bar q=5$, $\tilde p=\tilde q=7$, and $B=500$ replications.}
\label{tab:wave_2_a}
\end{table}

\begin{figure}[t]
    \centering
    \includegraphics[width=0.78\linewidth]{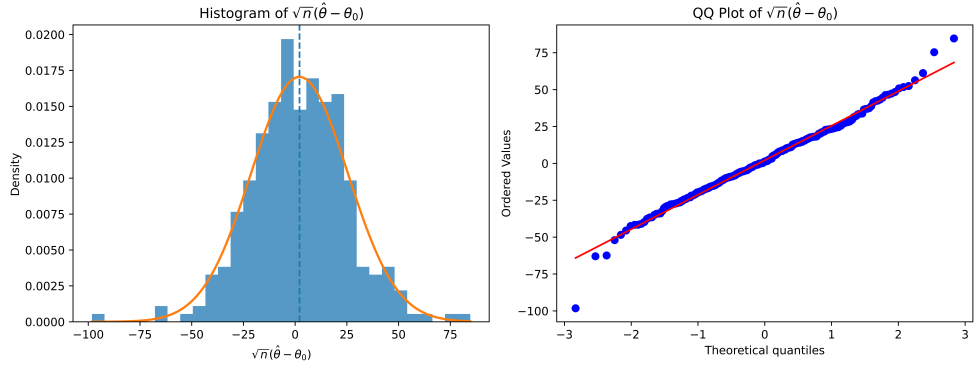}
    \caption{Histogram and Q--Q plot of $\sqrt n(\hat\theta_0^{\mathrm{IF}}-\theta_0)$ for $n=2000$ and $p+q=1000$ under the nonlinear wavelet DGP.}
    \label{fig:histqqwavelet1}
\end{figure}

\noindent\textbf{Simulation settings.}
For each replication, we generate $n\in\{100,300,800,1200,2000,3000\}$ observations and consider $p+q\in\{1000,3000,5000,10000\}$. The latent dimensions are fixed at $\bar p=\bar q=5$, while the learned representation dimensions are set to $\tilde p=\tilde q=7$. The mediator and outcome noise standard deviations are $\sigma_M=\sigma_Y=1$; the covariate noise level is set to $\sigma_X=2$ for the first two dimension settings and $\sigma_X=1.5$ for the larger settings. Observed covariates and mediators are generated from nonlinear additive Haar-wavelet loading functions with independent Gaussian idiosyncratic noise.

All representation and nuisance models are multilayer perceptrons with ReLU activations. The encoders $g_X$ and $g_M$ use two hidden layers of widths $300$ and $200$, and the decoders $h_X$ and $h_M$ use the reverse architecture. The cross-factor network $g_{XM}$ uses one hidden layer of width $50$, giving it lower capacity than the encoders and decoders. For nuisance estimation, we fit $\pi_1$, $\rho$, $\mu_1$, and $\mu_{10}$ using fully connected networks; the density ratio $\pi_2$ is then computed from $\hat\pi_1$ and $\hat\rho$ using Bayes' rule as in Algorithm~\ref{Algo:MediEncoder}. Networks are trained with Adam using initial learning rate $10^{-3}$, weight decay $10^{-3}$, a learning-rate decay factor of $0.5$ every $30$ epochs, and at most $300$ epochs. We use mean squared error for regression losses and binary cross-entropy for classification losses, with early stopping based on validation loss.

\noindent\textbf{Results.}
Table~\ref{tab:wave_2_a} reports a representative setting with $p+q=3000$; additional results are given in Appendix~\ref{appendix:res}. In this setting, \texttt{MediEncoder} has the smallest SD, RMSE, and average confidence-interval length for all displayed sample sizes, while maintaining coverage close to the nominal 95\% level. The improvement becomes clearer as $n$ increases: for example, at $n=3000$, \texttt{MediEncoder} reduces RMSE from $0.349$ for IMAVAE and $0.343 $ for VAE to $0.290$. The appendix results show the same trend for moderate and large sample sizes, although very small samples can favor simpler representation learners because deep representations are harder to train stably. Figure~\ref{fig:histqqwavelet1} shows that the scaled errors are approximately centered and have nearly linear Q--Q behavior, consistent with the asymptotic normal approximation developed in Section~\ref{sec:theory}.

\noindent\textbf{Ablation study on $\lambda_3$.}
To assess the role of the alignment loss, we set $\lambda_3=0$ and compare this ablated version with the fully tuned \texttt{MediEncoder}. The ablation results in Appendix~\ref{appendix:res} show that removing the alignment term generally increases SD, RMSE, and confidence-interval length, with the largest gap in smaller samples. This supports the use of the cross-factor alignment component for stabilizing the learned mediator representation and improving estimation efficiency, while the advantage naturally becomes less pronounced as the sample size grows.

\subsection{Real Data Application: Alzheimer's Disease Neuroimaging Initiative (ADNI)}

\begin{table}[t]
\centering
\small
\begin{tabular}{lccc}
\hline
\textbf{Effect} & \textbf{Estimate} & \textbf{Bootstrap Mean} & \textbf{Quantile CI (95\%)}\\
\hline
Direct   & 3.2163& 2.9623& (0.2268, 5.9212)\\
Indirect & 0.3527& 0.2370& (-2.1451, 3.7704)\\
Total    & 3.5691& 3.1994& (0.7321, 5.5262)\\
\hline
\end{tabular}
\caption{Estimated natural direct, natural indirect, and total effects in the ADNI analysis using the binarized threshold $\mathrm{GDS}>5$. The selected tuning parameter was $\hat\lambda=(0.5,0.2,0.3)$.}
\label{tab:combined_effects}
\end{table}

We apply \texttt{MediEncoder} to data from the Alzheimer's Disease Neuroimaging Initiative (ADNI) \citep{mueller2005alzheimer} to study whether the effect of depressive symptoms on cognitive decline is mediated by DNA methylation. The treatment $A$ is the binarized Geriatric Depression Scale (GDS), with $A=1$ indicating $\mathrm{GDS}>5$, a commonly used threshold for elevated depressive symptoms \citep{greenberg2012geriatric}. The outcome $Y$ is the Alzheimer's Disease Assessment Scale--Cognitive Subscale (ADAS-Cog), where higher scores indicate worse cognitive function \citep{cano2010adas,raghavan2013adas}. The covariates $X$ include demographic and clinical variables, and the mediators $M$ are high-dimensional DNA methylation features.

After preprocessing (Appendix~\ref{appendix:real_data}), the analysis includes $n=649$ subjects, with $\dim(X)=171$ and $\dim(M)=3{,}206$. Table~\ref{tab:combined_effects} reports the estimated natural direct effect (NDE), natural indirect effect (NIE), and total effect (TE). The total effect estimate is positive, suggesting that elevated depressive symptoms are associated, under the causal assumptions in Section~\ref{section:2}, with worse cognitive outcomes. The point estimates indicate that most of the effect operates through the direct pathway, while the estimated DNA-methylation-mediated indirect effect is positive but smaller. The bootstrap interval for the total effect excludes zero, whereas the intervals for the direct and indirect effects include zero, so the mediation finding should be interpreted cautiously.

%% file: Conclusion.tex
\section{Conclusion} \label{sec:conclusion}

We introduced \texttt{MediEncoder}, a coupled representation-learning framework for high-dimensional causal mediation analysis with nonlinear latent covariate and mediator structures. By combining a structured autoencoder with a cross-fitted efficient influence function-based estimator, the proposed method targets natural direct and indirect effects while accommodating complex high-dimensional measurements. Across simulations, \texttt{MediEncoder} achieved lower RMSE and more stable inference than competing methods, and the ablation study showed that the alignment term improves estimation accuracy and efficiency. In the ADNI application, we found a positive total effect of depressive symptoms on cognitive decline; the point estimates suggested that most of this effect operated through the direct pathway, while the estimated DNA-methylation-mediated indirect effect was smaller and more uncertain.

\textbf{Limitations and future work.} The proposed method relies on standard causal assumptions for natural mediation analysis, including no unmeasured mediator--outcome confounding. It is also currently developed for binary treatments, requires tuning representation dimensions and network hyperparameters, and the learned mediator representations may be difficult to interpret biologically. These limitations motivate future work on sensitivity analysis for unmeasured mediator--outcome confounding, extensions to continuous or multi-valued treatments, and methods for improving interpretability of the learned mediator representations.

%% file: Appendix.tex
\newpage

\begin{center}
~\\
{\bf\Large Appendix for}

{\bf\Large ``MediEncoder: Nonlinear Representation Learning for High-Dimensional Causal Mediation Analysis''}
\end{center}

\section{Related Work \label{appdendix:related}}
\textbf{High-Dimensional Mediation Analysis.}
Recent work extends causal mediation analysis to high-dimensional regimes where the number of mediators or covariates may exceed the sample size. Early approaches focus primarily on high-dimensional mediators, employing sparsity assumptions and variable selection techniques such as Lasso to estimate mediation effects \citep{zhao2016pathway, guo2022high, guo2023statistical, guo2024estimations, jones2025causal}. Subsequent work incorporates interaction effects among treatment, mediators, and covariates, allowing for more flexible modeling of complex mechanisms \citep{rakshit2024statistical}.  More recent developments consider settings where both mediators and covariates are high-dimensional and may interact, proposing estimation and inference procedures that account for their joint structure \citep{bo2024debiased}. Despite these advances, most existing methods rely on linear modeling assumptions and treat observed variables directly, which may be restrictive when the underlying mechanisms are nonlinear and only indirectly observed through noisy measurements.

\textbf{Factor Models.}
Factor models provide a natural framework for representing high-dimensional data via a lower-dimensional set of latent variables. Classical linear factor models have been widely used in statistics and econometrics for dimension reduction \citep{bai2002determining, bai2013principal, fan2013large, fan2020factor, fan2022learning}. These ideas have also been incorporated into causal inference and mediation analysis to handle high-dimensional covariates or mediators by projecting them onto latent spaces \citep{derkach2019high}. However, existing approaches predominantly rely on linear factor structures, assuming observed variables are linear combinations of latent factors. In many scientific applications, particularly in biological and genomic settings, the mapping from latent processes to observed measurements is often nonlinear. Recent work has explored nonlinear extensions of factor models using neural networks to allow more flexible representations \citep{feng2023optimal, xiu2024deep, fan2025factor}, though their integration with mediation analysis remains limited. In parallel, factor-based methods have also been developed for high-dimensional average treatment effect (ATE) estimation under doubly robust frameworks. \cite{fan2025factor} proposed a factor-informed AIPW estimator combining factor models and deep learning for high-dimensional causal inference. Their setting can be viewed as an ATE analogue of our problem, but still primarily relies on linear factor structures without considering nonlinear factor generation mechanisms.

\textbf{Autoencoders and More on Mediation Analysis.}
Deep representation learning methods, especially autoencoders, provide a flexible approach for learning nonlinear low-dimensional representations of high-dimensional data \citep{hinton2006reducing, goodfellow2016deep}. They have been widely applied in large-scale settings for nonlinear dimension reduction and feature learning \citep{lecun2015deep, khemakhem2020variational}. Variational autoencoders further introduce probabilistic latent-variable models that enable structured and generative representations \citep{kingma2013auto}. Related work on multi-view representation learning, including multi-view autoencoders and deep canonical correlation methods, aims to learn shared representations across multiple data modalities \citep{wang2015deep, guerrero2022multimodal}. \citet{xu2022deepmed} develop a deep learning-based mediation estimator that is robust to model misspecification. However, their approach learns representations independently and does not explicitly model for high-dimensionality and latent factor structure. In particular, it does not consider a shared representation capturing the dependence between covariates and mediators, which is important in high-dimensional biological systems driven by associated latent factors. \cite{jiang2023causal} propose a variational autoencoder-based mediation framework for indirectly observed mediators; however, their method does not accommodate latent and nonlinear structures in covariates, does not model the structural dependence between the latent representations of covariates and mediators, and does not allow treatment--covariate or treatment--mediator interactions. In addition, their estimation procedure may be sensitive to model misspecification.

\section{Overview of Baseline Methods} \label{appendix:baseline}

We consider several baseline methods for estimating low-dimensional latent representations (factors) from high-dimensional covariates $X$ and mediators $M$. These representations are subsequently used in downstream mediation effect estimation.

\paragraph{Diversified Projection.}
Diversified projection method \citep{fan2022learning, fan2023factor} constructs a linear surrogate of latent factors using a projection matrix $W \in \mathbb{R}^{p \times \tilde r}$ that satisfies boundedness and significance conditions. The estimated factors are obtained as
\begin{align*}
    \hat{\boldsymbol f}= p^{-1} W^\top x,
\end{align*}
which can be viewed as an affine approximation of the true latent factor $f$:
\begin{align*}
    \hat{\boldsymbol f} = H {\boldsymbol f} + \xi, \quad \text{with } H = p^{-1} W^\top B.
\end{align*}
Under mild conditions, the signal term $Hf$ dominates the noise $\xi$, allowing $\tilde f$ to serve as a consistent proxy for $f$. This approach avoids explicit estimation of the factor loading matrix and does not require precise knowledge of the factor dimension.

\paragraph{Autoencoder.} Autoencoders (AE) learn nonlinear low-dimensional representations by jointly training an encoder $g(\cdot)$ and a decoder $h(\cdot)$. Given input $x$, the latent representation is $z = g(x)$ and the reconstruction is $\hat x = h(z)$. The model is trained by minimizing the reconstruction loss:
\begin{align*}
    \min_{g,h} \; \sum_{i=1}^n \| x_i - h(g(x_i)) \|_2^2.
\end{align*}
In our setting, autoencoders are applied separately to $X$ and $M$ to obtain latent representations, which may fail to capture shared structure across modalities.

\paragraph{Variational Autoencoder.}
Variational autoencoders (VAE) extend autoencoders by introducing a probabilistic latent-variable model. The encoder defines a variational posterior $q_\phi(z|x)$, while the decoder specifies a likelihood $p_\theta(x|z)$. The model is trained by maximizing the evidence lower bound (ELBO), equivalently minimizing:
\begin{align*}
    \mathcal{L}_{\text{VAE}} 
    = \mathbb{E}_{q_\phi(z|x)}\big[\|x - \hat x\|_2^2\big]
    + \mathrm{KL}\big(q_\phi(z|x) \,\|\, p(z)\big),
\end{align*}
where $p(z)$ is a prior distribution. Similar to AE, VAEs are typically applied independently to $X$ and $M$, which may not capture their joint dependence structure.

\paragraph{IMAVAE.}
IMAVAE \citep{jiang2023causal} extends the VAE framework to mediation analysis with indirectly observed mediators by incorporating treatment and covariates as auxiliary variables. Let $u = t$ or $u = (w,t)$ denote the auxiliary input. The model jointly learns an encoder $q_\phi(z|x,u)$, decoder $p_\theta(x|z)$, and predictor $g_\gamma(z,u)$ for the outcome. The objective function is:
\begin{align*}
    \min_{\theta,\phi,\gamma} \;
    \Big\{
    \alpha \, \mathcal{L}_{\theta,\phi}(\hat x, x)
    - \beta \, \mathcal{L}_{\theta,\phi}(x,u)
    + \mathcal{L}_{\phi,S,\lambda,\gamma}(\hat y, y)
    \Big\},
\end{align*}
where $\mathcal{L}_{\theta,\phi}(x,u)$ corresponds to the ELBO term, $\mathcal{L}_{\theta,\phi}(\hat x, x)$ is the reconstruction loss, and $\mathcal{L}_{\phi,S,\lambda,\gamma}(\hat y, y)$ is the prediction loss for the outcome. This framework enables learning latent mediator representations informed by treatment and covariates, but remains tied to a specific generative model.

\section{Details of Algorithm \ref{Algo:MediEncoder}}
\label{app:alg}

\subsection{Nuisance Function Estimation}

\begin{itemize}

\item Estimate nuisance functions on \(\mathcal I_{\mathrm{nu}}^{(k)}\):
    \begin{equation}\label{eq:nuisance_opts_method}
    \begin{aligned}
    \hat\pi_1^{(k)}
    &\in
    \arg\min_{g\in\mathcal F_{\pi}}
    \sum_{i\in\mathcal I_{\mathrm{nu}}^{(k)}}
    \Big[-A_i\log g(\widehat{\boldsymbol f}_{X,i}^{(k)})
    -(1-A_i)\log\{1-g(\widehat{\boldsymbol f}_{X,i}^{(k)})\}\Big],\\
    \hat\mu_1^{(k)}
    &\in
    \arg\min_{g\in\mathcal F_{\mu}}
    \sum_{\substack{i\in\mathcal I_{\mathrm{nu}}^{(k)}\\ A_i=1}}
    \Big[Y_i-g(\widehat{\boldsymbol f}_{X,i}^{(k)},\widehat{\boldsymbol f}_{M,i}^{(k)})\Big]^2,\\
    \hat\rho^{(k)}
    &\in
    \arg\min_{g\in\mathcal F_{\rho}}
    \sum_{i\in\mathcal I_{\mathrm{nu}}^{(k)}}
    \Big[-A_i\log g(\widehat{\boldsymbol f}_{X,i}^{(k)},\widehat{\boldsymbol f}_{M,i}^{(k)})
    -(1-A_i)\log\{1-g(\widehat{\boldsymbol f}_{X,i}^{(k)},\widehat{\boldsymbol f}_{M,i}^{(k)})\}\Big],\\
    \hat\mu_{10}^{(k)}
    &\in
    \arg\min_{g\in\mathcal F_{\mu_{10}}}
    \sum_{\substack{i\in\mathcal I_{\mathrm{nu}}^{(k)}\\ A_i=0}}
    \Big[\hat\mu_1^{(k)}(\widehat{\boldsymbol f}_{X,i}^{(k)},\widehat{\boldsymbol f}_{M,i}^{(k)})
    -g(\widehat{\boldsymbol f}_{X,i}^{(k)})\Big]^2.
    \end{aligned}
    \end{equation}

    \item Estimate the density ratio \(\pi_2(f_X,f_M)=p(f_M\mid A=0,f_X)/p(f_M\mid A=1,f_X)\) by Bayes' rule:
    \begin{equation}\label{eq:pi2_bayes_method}
        \hat\pi_2^{(k)}(u,v)
        =
        \frac{1-\hat\rho^{(k)}(u,v)}{\hat\rho^{(k)}(u,v)}
        \frac{\hat\pi_1^{(k)}(u)}{1-\hat\pi_1^{(k)}(u)}.
    \end{equation}

\end{itemize}

\subsection{Tuning Algorithm}

\begin{algorithm}[t]
\caption{Tuning $\lambda=(\lambda_1,\lambda_2,\lambda_3)$ for \texttt{MediEncoder}}
\label{Algo:TuneLambdaMediEncoder}
\small
\begin{algorithmic}[1]

\Require Training indices $\mathcal I_{\mathrm{tr}}$, validation indices
$\mathcal I_{\mathrm{val}}$, tuning grid
$\Lambda\subset[0,\infty)^3\setminus\{(0,0,0)\}$, representation class
$\mathcal Z_{\mathrm{NN}}$, auxiliary outcome-regression class
$\mathcal F_{\mu}^{\mathrm{tun}}$
\Ensure Selected tuning parameter
$\hat\lambda=(\hat\lambda_1,\hat\lambda_2,\hat\lambda_3)$

\State Define the treated training and validation subsets
\[
    \mathcal I_{\mathrm{tr},1}
    :=\{i\in\mathcal I_{\mathrm{tr}}:A_i=1\},
    \qquad
    \mathcal I_{\mathrm{val},1}
    :=\{i\in\mathcal I_{\mathrm{val}}:A_i=1\}.
\]
\State Assume $|\mathcal I_{\mathrm{tr},1}|>0$ and
$|\mathcal I_{\mathrm{val},1}|>0$; otherwise, redraw the sample split.

\For{each $\lambda=(\lambda_1,\lambda_2,\lambda_3)\in\Lambda$}

\State Fit the coupled autoencoder on $\mathcal I_{\mathrm{tr}}$:
\[
\begin{aligned}
&\widehat z_{XM}^{(\lambda)}
=
\left(
\hat g_X^{(\lambda)},
\hat g_M^{(\lambda)},
\hat g_{XM}^{(\lambda)},
\hat h_X^{(\lambda)},
\hat h_M^{(\lambda)}
\right)  \\
&\in
\arg\min_{z_{XM}=(g_X,g_M,g_{XM},h_X,h_M)\in\mathcal Z_{\mathrm{NN}}}
\frac{1}{|\mathcal I_{\mathrm{tr}}|}
\sum_{i\in\mathcal I_{\mathrm{tr}}}
\Big[
\lambda_1\|X_i-h_X(g_X(X_i))\|_2^2
+\lambda_2\|M_i-h_M(g_M(M_i))\|_2^2 \\
&\hspace{12.5em}
+\lambda_3\|g_M(M_i)-g_{XM}(A_i,g_X(X_i))\|_2^2
\Big].
\end{aligned}
\]

\State Compute learned representations for all
$i\in\mathcal I_{\mathrm{tr}}\cup\mathcal I_{\mathrm{val}}$:
\[
    \widehat{\boldsymbol f}_{X,i}^{(\lambda)}
    =
    \hat g_X^{(\lambda)}(X_i),
    \qquad
    \widehat{\boldsymbol f}_{M,i}^{(\lambda)}
    =
    \hat g_M^{(\lambda)}(M_i).
\]

\State Fit an auxiliary treated-outcome regression on the training subset:
\[
\widehat m_1^{(\lambda)}
\in
\arg\min_{g\in\mathcal F_{\mu}^{\mathrm{tun}}}
\frac{1}{|\mathcal I_{\mathrm{tr},1}|}
\sum_{i\in\mathcal I_{\mathrm{tr},1}}
\left[
Y_i
-
g\!\left(
\widehat{\boldsymbol f}_{X,i}^{(\lambda)},
\widehat{\boldsymbol f}_{M,i}^{(\lambda)}
\right)
\right]^2 .
\]

\State Compute the held-out validation prediction error:
\[
\widehat{\mathrm{PE}}(\lambda)
=
\frac{1}{|\mathcal I_{\mathrm{val},1}|}
\sum_{i\in\mathcal I_{\mathrm{val},1}}
\left[
Y_i
-
\widehat m_1^{(\lambda)}
\!\left(
\widehat{\boldsymbol f}_{X,i}^{(\lambda)},
\widehat{\boldsymbol f}_{M,i}^{(\lambda)}
\right)
\right]^2 .
\]

\EndFor

\State Select any minimizer
\[
    \hat\lambda
    =
    (\hat\lambda_1,\hat\lambda_2,\hat\lambda_3)
    \in
    \arg\min_{\lambda\in\Lambda}
    \widehat{\mathrm{PE}}(\lambda).
\]

\State \Return $\hat\lambda$

\end{algorithmic}
\end{algorithm}

The auxiliary regression $\widehat m_1^{(\lambda)}$ is used only to select the
representation learning tuning parameter $\lambda$ and is not used in the final
influence function-based estimator. After $\hat\lambda$ is selected, Algorithm~\ref{Algo:MediEncoder}
refits the coupled autoencoder on the fold-specific representation-training set and estimates the
nuisance functions on a separate nuisance-estimation set.

\section{More on Simulation Study \label{appendix:simu}}
\subsection{Simulation Settings}
The data generating process is as follows:
\begin{enumerate}
\item 
We generate latent covariates \( \boldsymbol{f}_X \in \mathbb{R}^{\bar{p}} \) from 
\( \boldsymbol{f}_X \sim \mathcal U(-1, 1)^{\bar{p}} \). 
The treatment variable \(A\) is generated according to a logistic model based on \( \boldsymbol{f}_X \):
\[
\Pr(A = 1 \mid \boldsymbol{f}_X) = 
\frac{\exp(\boldsymbol{f}_X^\top \alpha)}{1 + \exp(\boldsymbol{f}_X^\top \alpha)}, 
\qquad 
A \sim \mathrm{Bernoulli}\!\left(
\frac{\exp(\boldsymbol{f}_X^\top \alpha)}{1 + \exp(\boldsymbol{f}_X^\top \alpha)}
\right),
\]
where \( \alpha \in \mathbb{R}^{\bar{p}} \) is drawn once from \( \mathcal{U}(0,2)^{\bar{p}} \).

    \item The latent mediators \( \boldsymbol{f}_M \in \mathbb{R}^{\bar{q}} \) are generated through a nonlinear transformation of \( \boldsymbol{f}_X \), the treatment indicator \(A\), and additive noise:
    \[
        \boldsymbol{f}_M 
        =
        (1 - A)\big( \boldsymbol{\delta}_0 \boldsymbol{f}_X^{\circ 2} + \mathbf u_{XM} \big)
        +
        A\big( \boldsymbol{\delta}_1 \boldsymbol{f}_X^{\circ 2} + \mathbf u_{XM}' \big),
    \quad \mathbf u_{XM},\mathbf u'_{XM} \sim \mathcal{N}(0, \Sigma_U),
    \]
    where \( \boldsymbol{f}_X^{\circ 2} \) denotes the element-wise square of \( \boldsymbol{f}_X \),
    \( \boldsymbol{\delta}_0, \boldsymbol{\delta}_1 \in \mathbb{R}^{\bar{q} \times \bar{p}} \) have entries drawn once from \( \mathcal{U}(0.5, 1.5) \), and
    \( \Sigma_U = R_U \Lambda_U R_U^\top \), where \(R_U \in \mathbb{R}^{\bar q \times \bar q}\) is an orthonormal matrix and
    \(\Lambda_U = \mathrm{diag}(\lambda_1,\ldots,\lambda_{\bar q})\) with \(\lambda_j \sim \mathcal U(1, 2)\). 

    \item The outcome variable \(Y\) is generated via a nonlinear equation:
    \begin{align*}
Y
&=
(1 - A)
\Big(
\sin(5 \boldsymbol{f}_X)^\top \beta_0
+ \log\!\big(1 + \exp(\boldsymbol{f}_X^\top \kappa_0)\big)
+ (\boldsymbol{f}_M \circ \boldsymbol{f}_X^{\circ 2})^\top \gamma_0
+ \epsilon_Y'
\Big)
\\
&\quad
+ A
\Big(
\sin(5 \boldsymbol{f}_X)^\top \beta_1
+ \log\!\big(1 + \exp(\boldsymbol{f}_X^\top \kappa_1)\big)
+ (\boldsymbol{f}_M \circ \boldsymbol{f}_X^{\circ 2})^\top \gamma_1
+ \epsilon_Y
\Big),
\end{align*}
    where
    \( \beta_0, \beta_1, \kappa_0, \kappa_1 \in \mathbb{R}^{\bar{p}} \),
    \( \gamma_0, \gamma_1 \in \mathbb{R}^{\bar{q}} \),
    each drawn once from \( \mathcal{U}(0.5, 1.5) \),
    and \( \epsilon_Y, \epsilon_Y' \sim \mathcal{N}(0, \sigma_Y^2) \).

\item We generate the observed high-dimensional covariates
\( X \in \mathbb{R}^{p} \) and mediators \( M \in \mathbb{R}^{q} \) through nonlinear factor loadings with additive noise. Specifically, for each observed coordinate \( j = 1,\dots,p \) and \( k = 1,\dots,q \), we generate
\[
    X_j
    =
    \phi_{X,j}(\boldsymbol{f}_X)
    + \boldsymbol u_{X,j},
    \qquad
    M_k
    =
    \phi_{M,k}(\boldsymbol{f}_M)
    + \boldsymbol u_{M,k},
\]
where \( \boldsymbol{f}_X \in \mathbb{R}^{\bar p} \) and \( \boldsymbol{f}_M \in \mathbb{R}^{\bar q} \) denote the latent factors.

\textbf{Nonlinear loading families.} We consider nonlinear factor loadings for
\(g_{X,j}(\cdot)\) and \(g_{M,k}(\cdot)\) in wavelet (see \ref{appendix:wavelet} for more details) based on Haar atoms, and we use an additive structure over latent coordinates:
\[
g_{X,j}(\boldsymbol{f}_X)
=
\sum_{i=1}^{\bar p}
h_{X,ji}(\boldsymbol{f}_{X_i}),
\qquad
g_{M,k}(\boldsymbol{f}_M)
=
\sum_{i=1}^{\bar q}
h_{M,ki}(\boldsymbol{f}_{M_i}),
\]
where the univariate functions \(h_{X,jm}(\cdot)\) and \(h_{M,km}(\cdot)\) are instantiated using the wavelet bases.

\item The idiosyncratic components satisfy
\[
\mathbf u_X \sim \mathcal{N}(0, \sigma_X^2 I_p),
\qquad
\mathbf u_M \sim \mathcal{N}(0, \sigma_M^2 I_q),
\]
independently of the latent factors.
\end{enumerate}

\subsubsection{Wavelet Loading\label{appendix:wavelet}} 
In the simulation experiment \ref{section:simu}, the loading functions 
\(g_{X,j}(\cdot)\) and \(g_{M,k}(\cdot)\) are constructed using nonlinear basis expansions. 
Specifically, we consider nonlinear loadings based on wavelet which follows an additive structure over latent
coordinates.

\noindent
\textbf{Haar mother wavelet.}
We adopt the standard Haar mother wavelet \( \psi : \mathbb{R} \to \mathbb{R} \), defined by
\[
\psi(t)
=
\begin{cases}
1, & 0 \le t < 1, \\
-1, & 1 \le t < 2, \\
0, & \text{otherwise}.
\end{cases}
\]
This function satisfies
\[
\int_{\mathbb{R}} \psi(t)\, dt = 0,
\qquad
\|\psi\|_{L^2} = \sqrt{2}.
\]

\textbf{Wavelet atoms.} For each wavelet atom \( \phi_\ell \), we independently sample an integer-valued
scale parameter $r_\ell \sim \mathrm{Unif}\{r_{\min},\dots,r_{\max}\}.$ To ensure that the wavelet atoms overlap with the latent support, we sample a location $t_\ell \sim \mathrm{Unif}[x_{\min}, x_{\max}],$ where \( [x_{\min}, x_{\max}] \) denotes the range of the latent variables. The corresponding translation parameter is then defined as $s_\ell = \big\lfloor 2^{r_\ell} t_\ell \big\rfloor.$ Given the sampled pair \( (r_\ell, s_\ell) \), the wavelet atom is $\psi_{r_\ell,s_\ell}(t)
=
2^{r_\ell/2}\,
\psi\!\left(2^{r_\ell} t - s_\ell\right).$

\textbf{Additive nonlinear loading.} For each latent coordinate \( i\), we select a collection of \(L\) wavelet atoms
\(
\{\phi_{\ell}(\cdot)\}_{\ell=1}^{L}
\subset
\{\psi_{r,s}(\cdot)\}.
\) The loading functions are defined as
\begin{align*}
g_{X,j}(f_X)
=
\sum_{i=1}^{\bar p}
\sum_{\ell=1}^{L}
\lambda_{X,ji\ell}^{\mathrm{wavelet}}
\, \phi_{\ell}(f_{X_i}), \qquad
g_{M,k}(f_M)
=
\sum_{i=1}^{\bar q}
\sum_{\ell=1}^{L}
\lambda_{M,ki\ell}^{\mathrm{wavelet}}
\, \phi_{\ell}(f_{M_i}),
\end{align*}
where $\lambda_{X,jm\ell}^{\mathrm{wavelet}}
\sim
\mathcal{N} \Big(0, \frac{1}{L}\Big),
\lambda_{M,km\ell}^{\mathrm{wavelet}}
\sim
\mathcal{N} \Big(0, \frac{1}{L}\Big),$ drawn once at initialization.

\section{Proof of Theorem \ref{thm:main}}
\label{appendix:proof}
For notational simplicity, we present the analysis for a single fold (e.g., $k=1$, so $|\mathcal I_4|$ is used for evaluation of $\hat \theta_0^{\mathrm{IF}}$ and write $n = |\mathcal I_4|$ for simplicity). The arguments for the remaining folds follow analogously by permuting the roles of  $\mathcal I_1,\ldots,\mathcal I_4$ (see Algorithm~\ref{Algo:MediEncoder}). 

\textbf{Decomposition of $\hat{\theta}_0^{\mathrm{IF}}-\theta_0$.}
Our goal in this step is to decompose $\hat{\theta}_0^{\mathrm{IF}}-\theta_0$ as the leading influence-function term and the remainder terms.  Although the Algorithm \ref{Algo:MediEncoder} estimates the auxiliary propensity $\rho(\boldsymbol f_X,\boldsymbol f_M)
:= \mathbb P(A=1\mid \boldsymbol f_X,\boldsymbol f_M)$ rather than estimating the mediator density ratio directly, we keep the notation \(\pi_2\) for readability in this part of the proof.  Throughout the proof, \(\pi_2\) is understood as the
Bayes-rule representation $\pi_2(\boldsymbol f_X,\boldsymbol f_M)
=
\frac{1-\rho(\boldsymbol f_X,\boldsymbol f_M)}
     {\rho(\boldsymbol f_X,\boldsymbol f_M)}
\cdot
\frac{\pi_1(\boldsymbol f_X)}
     {1-\pi_1(\boldsymbol f_X)} .$

We first claim that the estimator admits the decomposition
\begin{align*}
\hat\theta_0^{\mathrm{IF}}-\theta_0
=
\frac{1}{n}\sum_{i=1}^n \Psi_i
+
R_0+R_1+R_2+R_3+R_4,
\end{align*}
where the oracle influence-function score
\begin{align*}
\Psi_i
&:=
\mu_{10}(\boldsymbol f_{X,i})
+
\frac{A_i}{\pi_1(\boldsymbol f_{X,i})}
\pi_2(\boldsymbol f_{X,i},\boldsymbol f_{M,i})
\bigl(
Y_i-\mu_1(\boldsymbol f_{X,i},\boldsymbol f_{M,i})
\bigr)\\
&\qquad
+
\frac{1-A_i}{1-\pi_1(\boldsymbol f_{X,i})}
\bigl(
\mu_1(\boldsymbol f_{X,i},\boldsymbol f_{M,i})
-
\mu_{10}(\boldsymbol f_{X,i})
\bigr)
-\theta_0.
\end{align*}
The remainder terms collect the differences between the estimated nuisance
quantities and their population counterparts.  Specifically,
\begin{align*}
R_0
&=
\frac{1}{n}\sum_{i=1}^n
A_i
\pi_2(\boldsymbol f_{X,i},\boldsymbol f_{M,i})
\left(
\frac{1}{\hat\pi_1(\hat{\boldsymbol f}_{X,i})}
-
\frac{1}{\pi_1(\boldsymbol f_{X,i})}
\right)
\bigl(
Y_i-\mu_1(\boldsymbol f_{X,i},\boldsymbol f_{M,i})
\bigr), \\
R_1
&=
\frac{1}{n}\sum_{i=1}^n
A_i
\frac{
\hat\pi_2(\hat{\boldsymbol f}_{X,i},\hat{\boldsymbol f}_{M,i})
-
\pi_2(\boldsymbol f_{X,i},\boldsymbol f_{M,i})
}{
\hat\pi_1(\hat{\boldsymbol f}_{X,i})
}
\bigl(
Y_i-\mu_1(\boldsymbol f_{X,i},\boldsymbol f_{M,i})
\bigr), \\
R_2
&=
\frac{1}{n}\sum_{i=1}^n
\left[
\frac{1-A_i}{1-\hat\pi_1(\hat{\boldsymbol f}_{X,i})}
-
A_i
\frac{\hat\pi_2(\hat{\boldsymbol f}_{X,i},\hat{\boldsymbol f}_{M,i})}
{\hat\pi_1(\hat{\boldsymbol f}_{X,i})}
\right]
\bigl(
\hat\mu_1(\hat{\boldsymbol f}_{X,i},\hat{\boldsymbol f}_{M,i})
-
\mu_1(\boldsymbol f_{X,i},\boldsymbol f_{M,i})
\bigr), \\
R_3
&=
\frac{1}{n}\sum_{i=1}^n
(1-A_i)
\bigl(
\mu_1(\boldsymbol f_{X,i},\boldsymbol f_{M,i})
-
\mu_{10}(\boldsymbol f_{X,i})
\bigr)
\left(
\frac{1}{1-\hat\pi_1(\hat{\boldsymbol f}_{X,i})}
-
\frac{1}{1-\pi_1(\boldsymbol f_{X,i})}
\right), \\
R_4
&=
\frac{1}{n}\sum_{i=1}^n
\frac{A_i-\hat\pi_1(\hat{\boldsymbol f}_{X,i})}
{1-\hat\pi_1(\hat{\boldsymbol f}_{X,i})}
\bigl(
\hat\mu_{10}(\hat{\boldsymbol f}_{X,i})
-
\mu_{10}(\boldsymbol f_{X,i})
\bigr).
\end{align*}
For readibility, we suppress the arguments of all functions throughout the following derivation. For example, we write $\hat\pi_1 = \hat\pi_1(\hat{\boldsymbol f}_{X,i}), 
\pi_1 = \pi_1(\boldsymbol f_{X,i}), 
\hat\pi_2 = \hat\pi_2(\hat{\boldsymbol f}_{X,i},\hat{\boldsymbol f}_{M,i}), 
\pi_2 = \pi_2(\boldsymbol f_{X,i},\boldsymbol f_{M,i}),$
and similarly for $\hat\mu_1$, $\mu_1$ and $\hat \mu_{10}$, $\mu_{10}$.

We now prove this decomposition step by step. By definition of $\hat\theta_0^{\mathrm{IF}}$, we have
\begin{equation}
\begin{aligned} \label{eq:diff}
\hat\theta_0^{\mathrm{IF}}-\theta_0
&=
\frac{1}{n}\sum_{i=1}^n
\Biggl[
\hat\mu_{10}
+
\frac{A_i}{\hat\pi_1}
\hat\pi_2
\bigl(
Y_i-\hat\mu_1
\bigr) 
+
\frac{1-A_i}{1-\hat\pi_1}
\bigl(
\hat\mu_1
-
\hat\mu_{10}
\bigr)
\Biggr]
-\theta_0 .
\end{aligned}
\end{equation}
Introduce
\begin{align*}
Z_{1,i}
&:=
\frac{A_i}{\hat\pi_1}
\hat\pi_2
\bigl(
Y_i-\hat\mu_1
\bigr)
-
\frac{A_i}{\pi_1}
\pi_2
\bigl(
Y_i-\mu_1
\bigr),\\
Z_{0,i}
&:=
\frac{1-A_i}{1-\hat\pi_1}
\bigl(
\hat\mu_1-\hat\mu_{10}
\bigr)
-
\frac{1-A_i}{1-\pi_1}
\bigl(
\mu_1-\mu_{10}
\bigr).
\end{align*}

Then \eqref{eq:diff} can be rewritten as
\begin{equation}\label{eq:diff2}
\begin{aligned}
\hat\theta_0^{\mathrm{IF}}-\theta_0
&=
\frac{1}{n}\sum_{i=1}^n \Psi_i 
+
\frac{1}{n}\sum_{i=1}^n
\bigl(
\hat\mu_{10}
-
\mu_{10}
\bigr)
+
\frac{1}{n}\sum_{i=1}^n Z_{1,i}
+
\frac{1}{n}\sum_{i=1}^n Z_{0,i},
\end{aligned}
\end{equation}
where \eqref{eq:diff2} is obtained by adding and subtracting $\Psi_i$ and regrouping terms. We next expand $Z_{1,i}$ in \eqref{eq:diff2}. By direct algebra,
\allowdisplaybreaks
\begin{align*}
Z_{1,i}
&=
\frac{A_i}{\hat\pi_1}
\hat\pi_2
\bigl(
Y_i-\hat\mu_1
\bigr)
-
\frac{A_i}{\pi_1}
\pi_2
\bigl(
Y_i-\mu_1
\bigr) \\
&=
A_i
\Biggl[
\frac{\hat\pi_2}{\hat\pi_1}
\bigl(
Y_i-\hat\mu_1
\bigr)
-
\frac{\pi_2}{\pi_1}
\bigl(
Y_i-\mu_1
\bigr)
\Biggr] \\
&=
A_i
\Biggl[
\frac{\hat\pi_2}{\hat\pi_1}
\bigl(
Y_i-\hat\mu_1
\bigr)
-
\frac{\hat\pi_2}{\hat\pi_1}
\bigl(
Y_i-\mu_1
\bigr)
+
\frac{\hat\pi_2}{\hat\pi_1}
\bigl(
Y_i-\mu_1
\bigr)
-
\frac{\pi_2}{\pi_1}
\bigl(
Y_i-\mu_1
\bigr)
\Biggr] \\
&=
A_i
\Biggl[
\frac{\hat\pi_2}{\hat\pi_1}
\bigl(
\mu_1-\hat\mu_1
\bigr)
+
\left(
\frac{\hat\pi_2}{\hat\pi_1}
-
\frac{\pi_2}{\pi_1}
\right)
\bigl(
Y_i-\mu_1
\bigr)
\Biggr] \\
&=
A_i
\frac{\hat\pi_2}{\hat\pi_1}
\bigl(
\mu_1-\hat\mu_1
\bigr)
+
A_i
\left[
\frac{\hat\pi_2-\pi_2}{\hat\pi_1}
+
\pi_2
\left(
\frac{1}{\hat\pi_1}
-
\frac{1}{\pi_1}
\right)
\right]
\bigl(
Y_i-\mu_1
\bigr) \\
&=
A_i
\frac{\hat\pi_2}{\hat\pi_1}
\bigl(
\mu_1-\hat\mu_1
\bigr)
+
A_i
\frac{\hat\pi_2-\pi_2}{\hat\pi_1}
\bigl(
Y_i-\mu_1
\bigr)
+
A_i
\pi_2
\left(
\frac{1}{\hat\pi_1}
-
\frac{1}{\pi_1}
\right)
\bigl(
Y_i-\mu_1
\bigr).
\end{align*}
We next expand $Z_{0,i}$ in \eqref{eq:diff2}. Again, by direct algebra,
\begin{align*}
Z_{0,i}
&=
\hat\mu_{10}
-
\mu_{10}
+
\frac{1-A_i}{1-\hat\pi_1}
\bigl(
\hat\mu_1-\hat\mu_{10}
\bigr)
-
\frac{1-A_i}{1-\pi_1}
\bigl(
\mu_1-\mu_{10}
\bigr) \\
&=
\hat\mu_{10}
-
\mu_{10}
+
(1-A_i)
\Biggl[
\frac{\hat\mu_1-\hat\mu_{10}}{1-\hat\pi_1}
-
\frac{\mu_1-\mu_{10}}{1-\pi_1}
\Biggr] \\
&=
\hat\mu_{10}
-
\mu_{10}
+
(1-A_i)
\Biggl[
\frac{\hat\mu_1-\hat\mu_{10}}{1-\hat\pi_1}
-
\frac{\mu_1-\mu_{10}}{1-\hat\pi_1}
+
\frac{\mu_1-\mu_{10}}{1-\hat\pi_1}
-
\frac{\mu_1-\mu_{10}}{1-\pi_1}
\Biggr] \\
&=
\hat\mu_{10}
-
\mu_{10}
+
(1-A_i)
\Biggl[
\frac{
\hat\mu_1-\hat\mu_{10}-\mu_1+\mu_{10}
}{
1-\hat\pi_1
}
+
(\mu_1-\mu_{10})
\left(
\frac{1}{1-\hat\pi_1}
-
\frac{1}{1-\pi_1}
\right)
\Biggr] \\
&=
\hat\mu_{10}
-
\mu_{10}
+
(1-A_i)
\Biggl[
\frac{\hat\mu_1-\mu_1}{1-\hat\pi_1}
-
\frac{\hat\mu_{10}-\mu_{10}}{1-\hat\pi_1}
+
(\mu_1-\mu_{10})
\left(
\frac{1}{1-\hat\pi_1}
-
\frac{1}{1-\pi_1}
\right)
\Biggr] \\
&=
\hat\mu_{10}
-
\mu_{10}
-
\frac{1-A_i}{1-\hat\pi_1}
\bigl(
\hat\mu_{10}-\mu_{10}
\bigr)
+
\frac{1-A_i}{1-\hat\pi_1}
\bigl(
\hat\mu_1-\mu_1
\bigr) \\
&\qquad
+
(1-A_i)
(\mu_1-\mu_{10})
\left(
\frac{1}{1-\hat\pi_1}
-
\frac{1}{1-\pi_1}
\right) \\
&=
\left[
1-\frac{1-A_i}{1-\hat\pi_1}
\right]
(\hat\mu_{10}-\mu_{10})
+
\frac{1-A_i}{1-\hat\pi_1}
(\hat\mu_1-\mu_1) \\
&\qquad
+
(1-A_i)
(\mu_1-\mu_{10})
\left(
\frac{1}{1-\hat\pi_1}
-
\frac{1}{1-\pi_1}
\right) \\
&=
\frac{A_i-\hat\pi_1}{1-\hat\pi_1}
(\hat\mu_{10}-\mu_{10})
+
\frac{1-A_i}{1-\hat\pi_1}
(\hat\mu_1-\mu_1) \\
&\qquad
+
(1-A_i)
(\mu_1-\mu_{10})
\left(
\frac{1}{1-\hat\pi_1}
-
\frac{1}{1-\pi_1}
\right).
\end{align*}
Substituting the expansions of $Z_{1,i}$ and $Z_{0,i}$ into \eqref{eq:diff2}, we obtain
\begin{align*}
\hat\theta_0^{\mathrm{IF}}-\theta_0
&=
\frac{1}{n}\sum_{i=1}^n \Psi_i \\
&\quad
+
\frac{1}{n}\sum_{i=1}^n
A_i
\frac{\hat\pi_2}{\hat\pi_1}
(\mu_1-\hat\mu_1)
+
\frac{1}{n}\sum_{i=1}^n
A_i
\frac{\hat\pi_2-\pi_2}{\hat\pi_1}
(Y_i-\mu_1)
+
\frac{1}{n}\sum_{i=1}^n
A_i
\pi_2
\left(
\frac{1}{\hat\pi_1}-\frac{1}{\pi_1}
\right)
(Y_i-\mu_1) \\
&\quad
+
\frac{1}{n}\sum_{i=1}^n
\frac{A_i-\hat\pi_1}{1-\hat\pi_1}
(\hat\mu_{10}-\mu_{10})
+
\frac{1}{n}\sum_{i=1}^n
\frac{1-A_i}{1-\hat\pi_1}
(\hat\mu_1-\mu_1) \\
&\quad
+
\frac{1}{n}\sum_{i=1}^n
(1-A_i)
(\mu_1-\mu_{10})
\left(
\frac{1}{1-\hat\pi_1}
-
\frac{1}{1-\pi_1}
\right) \\
&=
\frac{1}{n}\sum_{i=1}^n \Psi_i \\
&\quad
+
\frac{1}{n}\sum_{i=1}^n
A_i
\pi_2
\left(
\frac{1}{\hat\pi_1}-\frac{1}{\pi_1}
\right)
(Y_i-\mu_1) \\
&\quad
+
\frac{1}{n}\sum_{i=1}^n
A_i
\frac{\hat\pi_2-\pi_2}{\hat\pi_1}
(Y_i-\mu_1) \\
&\quad
+
\frac{1}{n}\sum_{i=1}^n
\left[
\frac{1-A_i}{1-\hat\pi_1}
-
A_i
\frac{\hat\pi_2}{\hat\pi_1}
\right]
(\hat\mu_1-\mu_1) \\
&\quad
+
\frac{1}{n}\sum_{i=1}^n
(1-A_i)
(\mu_1-\mu_{10})
\left(
\frac{1}{1-\hat\pi_1}
-
\frac{1}{1-\pi_1}
\right) \\
&\quad
+
\frac{1}{n}\sum_{i=1}^n
\frac{A_i-\hat\pi_1}{1-\hat\pi_1}
(\hat\mu_{10}-\mu_{10}).
\end{align*}
Comparing with the definitions of $R_0,R_1,R_2,R_3,R_4$, we obtain
\begin{align}\label{eq:decomp}
\hat\theta_0^{\mathrm{IF}}-\theta_0
=
\frac{1}{n}\sum_{i=1}^n \Psi_i
+
R_0+R_1+R_2+R_3+R_4.
\end{align}
\textbf{Reduction to the remainder terms.} To establish the asymptotic linearity of $\hat\theta_0^{\mathrm{IF}}$, it suffices to show that $\sqrt n\,R_j = o_p(1), ~~ j=0,\ldots,4.$ Multiplying the decomposition in Equation \eqref{eq:decomp} by $\sqrt n$ yields
\begin{align*}
\sqrt n\bigl(\hat\theta_0^{\mathrm{IF}}-\theta_0\bigr)
&=
\frac{1}{\sqrt n}\sum_{i=1}^n \Psi_i
+
\sqrt n\,R_0+\sqrt n\,R_1+\sqrt n\,R_2+\sqrt n\,R_3+\sqrt n\,R_4 \\
&=: \frac{1}{\sqrt n}\sum_{i=1}^n \Psi_i+T_1+T_2+T_3+T_4+T_5.
\end{align*}
Thus, it remains to show that $T_j=o_p(1)$ for $j=1,\ldots,5$, in which case
\[
\sqrt n\bigl(\hat\theta_0^{\mathrm{IF}}-\theta_0\bigr)
=
\frac{1}{\sqrt n}\sum_{i=1}^n \Psi_i + o_p(1).
\]
We now write the remainder terms explicitly:
\begin{align*}
T_1
&=
\frac{1}{\sqrt n}\sum_{i=1}^n
A_i
\pi_2(\boldsymbol f_{X,i},\boldsymbol f_{M,i})
\left(
\frac{1}{\hat\pi_1(\hat{\boldsymbol f}_{X,i})}
-
\frac{1}{\pi_1(\boldsymbol f_{X,i})}
\right)
\bigl(
Y_i-\mu_1(\boldsymbol f_{X,i},\boldsymbol f_{M,i})
\bigr), \\
T_2
&=
\frac{1}{\sqrt n}\sum_{i=1}^n
A_i
\frac{
\hat\pi_2(\hat{\boldsymbol f}_{X,i},\hat{\boldsymbol f}_{M,i})
-
\pi_2(\boldsymbol f_{X,i},\boldsymbol f_{M,i})
}{
\hat\pi_1(\hat{\boldsymbol f}_{X,i})
}
\bigl(
Y_i-\mu_1(\boldsymbol f_{X,i},\boldsymbol f_{M,i})
\bigr), \\
T_3
&=
\frac{1}{\sqrt n}\sum_{i=1}^n
\left[
\frac{1-A_i}{1-\hat\pi_1(\hat{\boldsymbol f}_{X,i})}
-
A_i
\frac{\hat\pi_2(\hat{\boldsymbol f}_{X,i},\hat{\boldsymbol f}_{M,i})}
{\hat\pi_1(\hat{\boldsymbol f}_{X,i})}
\right]
\bigl(
\hat\mu_1(\hat{\boldsymbol f}_{X,i},\hat{\boldsymbol f}_{M,i})
-
\mu_1(\boldsymbol f_{X,i},\boldsymbol f_{M,i})
\bigr), \\
T_4
&=
\frac{1}{\sqrt n}\sum_{i=1}^n
(1-A_i)
\bigl(
\mu_1(\boldsymbol f_{X,i},\boldsymbol f_{M,i})
-
\mu_{10}(\boldsymbol f_{X,i})
\bigr)
\left(
\frac{1}{1-\hat\pi_1(\hat{\boldsymbol f}_{X,i})}
-
\frac{1}{1-\pi_1(\boldsymbol f_{X,i})}
\right), \\
T_5
&=
\frac{1}{\sqrt n}\sum_{i=1}^n
\frac{A_i-\hat\pi_1(\hat{\boldsymbol f}_{X,i})}
{1-\hat\pi_1(\hat{\boldsymbol f}_{X,i})}
\bigl(
\hat\mu_{10}(\hat{\boldsymbol f}_{X,i})
-
\mu_{10}(\boldsymbol f_{X,i})
\bigr).
\end{align*}
Before proceeding with the analysis, we define the following sigma-algebras:
\begin{align*}
\mathcal G_n^{T_1}
&:=
\sigma\!\Big(
\{\boldsymbol f_{X,i},\boldsymbol f_{M,i},
\boldsymbol u_{X,i},\boldsymbol u_{M,i}: i \in \mathcal I_4\},
\hat\pi_1
\Big), \\
\mathcal G_n^{T_2}
&:=
\sigma\!\Big(
\{\boldsymbol f_{X,i},\boldsymbol f_{M,i},
\boldsymbol u_{X,i},\boldsymbol u_{M,i} : i \in \mathcal I_4\},
\hat\pi_1,\hat\rho
\Big), \\
\mathcal G_n^{T_3}
&:=
\sigma\!\Big(
\{\boldsymbol f_{X,i},\boldsymbol f_{M,i},
\boldsymbol u_{X,i},\boldsymbol u_{M,i}: i \in \mathcal I_4\},
\hat\pi_1,\hat\rho,\hat\mu_1
\Big), \\
\mathcal G_n^{T_4} = \mathcal G_n^{T_5}
&:=
\sigma\!\Big(
\{\boldsymbol f_{X,i},\boldsymbol f_{M,i},
\boldsymbol u_{X,i},\boldsymbol u_{M,i}: i \in \mathcal I_4\},
\hat\pi_1,\hat\mu_{10}
\Big)\,.
\end{align*}
\textbf{Analysis of $T_1$.} We consider a single fold $\mathcal I_4$ for evaluation and $|\mathcal I_4|= n$, and all summations below are taken over $i\in \mathcal I_4$. We first study the conditional expectation of $T_1$: 
\begin{align*}
\mathbb E[T_1 \mid \mathcal G_n^{T_1}]
&=
\mathbb E\!\left[
\frac{1}{\sqrt n}\sum_{i\in \mathcal I_4}
A_i
\pi_2(\boldsymbol f_{X,i},\boldsymbol f_{M,i})
\left(
\frac{1}{\hat\pi_1(\hat{\boldsymbol f}_{X,i})}
-
\frac{1}{\pi_1(\boldsymbol f_{X,i})}
\right)
\bigl(
Y_i-\mu_1(\boldsymbol f_{X,i},\boldsymbol f_{M,i})
\bigr)
\;\middle|\;
\mathcal G_n^{T_1}
\right] \\
&=
\frac{1}{\sqrt n}\sum_{i\in \mathcal I_4}
\pi_2(\boldsymbol f_{X,i},\boldsymbol f_{M,i})
\left(
\frac{1}{\hat\pi_1(\hat{\boldsymbol f}_{X,i})}
-
\frac{1}{\pi_1(\boldsymbol f_{X,i})}
\right)
\mathbb E\!\left[
A_i
\bigl(
Y_i-\mu_1(\boldsymbol f_{X,i},\boldsymbol f_{M,i})
\bigr)
\;\middle|\;
\mathcal G_n^{T_1}
\right] \\
&=
\frac{1}{\sqrt n}\sum_{i\in \mathcal I_4}
\pi_2(\boldsymbol f_{X,i},\boldsymbol f_{M,i})
\left(
\frac{1}{\hat\pi_1(\hat{\boldsymbol f}_{X,i})}
-
\frac{1}{\pi_1(\boldsymbol f_{X,i})}
\right) \\
&\qquad\qquad\cdot
\mathbb E\!\left[
A_i
\mathbb E\!\left[
Y_i-\mu_1(\boldsymbol f_{X,i},\boldsymbol f_{M,i})
\;\middle|\;
A_i=1,\boldsymbol f_{X,i},\boldsymbol f_{M,i}
\right]
\;\middle|\;
\mathcal G_n^{T_1}
\right] \\
&=0.
\end{align*}
Therefore, $\mathbb E[T_1 \mid \mathcal G_n^{T_1}]
=0$.  We next study the conditional variance of $T_1$. Using conditional independence across observations given $\mathcal G_n^{T_1}$, we have:
\begin{align*}
\mathrm{Var}(T_1 \mid \mathcal G_n^{T_1})
&=
\frac{1}{n}\sum_{i\in\mathcal I_4}
\mathrm{Var}\!\Bigg(
A_i
\pi_2(\boldsymbol f_{X,i},\boldsymbol f_{M,i})
\left(
\frac{1}{\hat\pi_1(\hat{\boldsymbol f}_{X,i})}
-
\frac{1}{\pi_1(\boldsymbol f_{X,i})}
\right)
\bigl(
Y_i-\mu_1(\boldsymbol f_{X,i},\boldsymbol f_{M,i})
\bigr)
\;\Bigg|\;
\mathcal G_n^{T_1}
\Bigg) \\
&\le
\frac{1}{n}\sum_{i\in\mathcal I_4}
\pi_2(\boldsymbol f_{X,i},\boldsymbol f_{M,i})^2
\left(
\frac{1}{\hat\pi_1(\hat{\boldsymbol f}_{X,i})}
-
\frac{1}{\pi_1(\boldsymbol f_{X,i})}
\right)^2 \\
&\qquad\qquad\cdot
\mathbb E\!\left[
A_i
\bigl(
Y_i-\mu_1(\boldsymbol f_{X,i},\boldsymbol f_{M,i})
\bigr)^2
\;\middle|\;
\mathcal G_n^{T_1}
\right] \\
&\le
C
\frac{1}{n}\sum_{i\in\mathcal I_4}
\left(
\frac{1}{\hat\pi_1(\hat{\boldsymbol f}_{X,i})}
-
\frac{1}{\pi_1(\boldsymbol f_{X,i})}
\right)^2 \\
&=
C
\frac{1}{n}\sum_{i\in\mathcal I_4}
\left|
\frac{
\pi_1(\boldsymbol f_{X,i})
-
\hat\pi_1(\hat{\boldsymbol f}_{X,i})
}{
\hat\pi_1(\hat{\boldsymbol f}_{X,i})
\pi_1(\boldsymbol f_{X,i})
}
\right|^2 \\
&\le
C
\frac{1}{n}\sum_{i\in\mathcal I_4}
\bigl(
\hat\pi_1(\hat{\boldsymbol f}_{X,i})
-
\pi_1(\boldsymbol f_{X,i})
\bigr)^2 \\
&=
C
\left\|
\hat\pi_1(\hat{\boldsymbol f}_X)
-
\pi_1(\boldsymbol f_X)
\right\|_{L_2(P_n)}^2 \\
&=
C
\left\|
\hat\pi_1(\hat{\boldsymbol f}_X)
-
\pi_1\circ v_X^{-1}(\hat{\boldsymbol f}_X)
+
\pi_1\circ v_X^{-1}(\hat{\boldsymbol f}_X)
-
\pi_1(\boldsymbol f_X)
\right\|_{L_2(P_n)}^2 \\
&\le
C
\Big(
\big\|
\hat\pi_1(\hat{\boldsymbol f}_X)
-
\pi_1\circ v_X^{-1}(\hat{\boldsymbol f}_X)
\big\|_{L_2(P_n)}
+
\big\|
\pi_1\circ v_X^{-1}(\hat{\boldsymbol f}_X)
-
\pi_1(\boldsymbol f_X)
\big\|_{L_2(P_n)}
\Big)^2 \\
&\le
C
\big\|
\hat\pi_1(\hat{\boldsymbol f}_X)
-
\pi_1\circ v_X^{-1}(\hat{\boldsymbol f}_X)
\big\|_{L_2(P_n)}^2
+
C
\big\|
\pi_1\circ v_X^{-1}(\hat{\boldsymbol f}_X)
-
\pi_1(\boldsymbol f_X)
\big\|_{L_2(P_n)}^2 .
\end{align*}
The last inequality follows from Young's inequality, equivalently $(a+b)^2\le 2a^2+2b^2$. The second inequality above uses $A_i\le 1$, boundedness of $\pi_2$, $\pi_1$ and $\hat\pi_1$, and the bounded conditional second moment in Assumption~\ref{assm:dgp_1}. It remains to bound the first empirical $L_2( P_n)$ term. We first add and subtract its population counterpart:
\begin{align*}
&
\big\|
\hat\pi_1(\hat{\boldsymbol f}_X)
-
\pi_1\circ v_X^{-1}(\hat{\boldsymbol f}_X)
\big\|_{L_2( P_n)}^2 \\
&\qquad =
\big\|
\hat\pi_1
-
\pi_1\circ v_X^{-1}
\big\|_{L_2( P)}^2
+
\left\{
\big\|
\hat\pi_1(\hat{\boldsymbol f}_X)
-
\pi_1\circ v_X^{-1}(\hat{\boldsymbol f}_X)
\big\|_{L_2( P_n)}^2
-
\big\|
\hat\pi_1
-
\pi_1\circ v_X^{-1}
\big\|_{L_2( P)}^2
\right\}\\
&\qquad =
O_p(\delta_{n,\pi_1}^2+n^{-1/2})\,,
\end{align*}
where the first term follows from the population $L_2$ rate in Theorem~\ref{thm:main}, and the second term is obtained via Lemma \ref{lemma:l2-norm} applied to the empirical average over the evaluation fold. Combining the above bounds, we obtain $\mathrm{Var}(T_1\mid \mathcal G_n^{T_1})
=
O_p(\delta_{n,\pi_1}^2+n^{-1/2}+\delta_{n,f}^2)
=
o_p(1).$ Together with $\mathbb E[T_1\mid \mathcal G_n^{T_1}]=0$, Chebyshev's inequality gives, for any $\varepsilon>0$,
\begin{align*}
\mathbb P(|T_1|>\varepsilon\mid \mathcal G_n^{T_1})
\le
\frac{\mathrm{Var}(T_1\mid \mathcal G_n^{T_1})}{\varepsilon^2}
=
o_p(1).
\end{align*}
Since $0\le \mathbb P(|T_1|>\varepsilon\mid \mathcal G_n^{T_1})\le 1$, the dominated convergence argument gives $\mathbb P(|T_1|>\varepsilon)
=
\mathbb E\!\left[
\mathbb P(|T_1|>\varepsilon\mid \mathcal G_n^{T_1})
\right]
\to 0.$ Hence $T_1=o_p(1)$.

\noindent
\textbf{Analysis of $T_2$.} Recall that the density ratio $\hat \pi_2$ is represented as $\hat\pi_2(\hat{\boldsymbol f}_X,\hat{\boldsymbol f}_M)
=
\frac{1-\hat\rho(\hat{\boldsymbol f}_X,\hat{\boldsymbol f}_M)}
     {\hat\rho(\hat{\boldsymbol f}_X,\hat{\boldsymbol f}_M)}
\cdot
\frac{\hat\pi_1(\hat{\boldsymbol f}_X)}
     {1-\hat\pi_1(\hat{\boldsymbol f}_X)} .$ 
We first study the conditional expectation of $T_2$ given $\mathcal G_n^{T_2}$:
\begin{align*}
\mathbb E[T_2 \mid \mathcal G_n^{T_2}]
&=
\mathbb E\!\left[
\frac{1}{\sqrt n}\sum_{i\in \mathcal I_4}
A_i
\frac{
\hat\pi_2(\hat{\boldsymbol f}_{X,i},\hat{\boldsymbol f}_{M,i})
-
\pi_2(\boldsymbol f_{X,i},\boldsymbol f_{M,i})
}{
\hat\pi_1(\hat{\boldsymbol f}_{X,i})
}
\bigl(
Y_i-\mu_1(\boldsymbol f_{X,i},\boldsymbol f_{M,i})
\bigr)
\;\middle|\;
\mathcal G_n^{T_2}
\right] \\
&=
\frac{1}{\sqrt n}\sum_{i\in \mathcal I_4}
\frac{
\hat\pi_2(\hat{\boldsymbol f}_{X,i},\hat{\boldsymbol f}_{M,i})
-
\pi_2(\boldsymbol f_{X,i},\boldsymbol f_{M,i})
}{
\hat\pi_1(\hat{\boldsymbol f}_{X,i})
}
\mathbb E\!\left[
A_i
\bigl(
Y_i-\mu_1(\boldsymbol f_{X,i},\boldsymbol f_{M,i})
\bigr)
\;\middle|\;
\mathcal G_n^{T_2}
\right] \\
&=
\frac{1}{\sqrt n}\sum_{i\in \mathcal I_4}
\frac{
\hat\pi_2(\hat{\boldsymbol f}_{X,i},\hat{\boldsymbol f}_{M,i})
-
\pi_2(\boldsymbol f_{X,i},\boldsymbol f_{M,i})
}{
\hat\pi_1(\hat{\boldsymbol f}_{X,i})
}\\
&~~~~~~~~~~~~~~~~\cdot\mathbb E\!\left[
A_i\,
\mathbb E\!\left[
Y_i-\mu_1(\boldsymbol f_{X,i},\boldsymbol f_{M,i})
\;\middle|\;
A_i=1,\boldsymbol f_{X,i},\boldsymbol f_{M,i}
\right]
\;\middle|\;
\mathcal G_n^{T_2}
\right] \\
&=0.
\end{align*}
We next study the conditional variance of $T_2$:
\begin{align*}
\mathrm{Var}(T_2 \mid \mathcal G_n^{T_2})
&=
\frac{1}{n}\sum_{i\in \mathcal I_4}
\mathrm{Var}\!\left(
A_i
\frac{
\hat\pi_2(\hat{\boldsymbol f}_{X,i},\hat{\boldsymbol f}_{M,i})
-
\pi_2(\boldsymbol f_{X,i},\boldsymbol f_{M,i})
}{
\hat\pi_1(\hat{\boldsymbol f}_{X,i})
}
\bigl(
Y_i-\mu_1(\boldsymbol f_{X,i},\boldsymbol f_{M,i})
\bigr)
\;\middle|\;
\mathcal G_n^{T_2}
\right) \\
&\le
\frac{1}{n}\sum_{i\in \mathcal I_4}
\left(
\frac{
\hat\pi_2(\hat{\boldsymbol f}_{X,i},\hat{\boldsymbol f}_{M,i})
-
\pi_2(\boldsymbol f_{X,i},\boldsymbol f_{M,i})
}{
\hat\pi_1(\hat{\boldsymbol f}_{X,i})
}
\right)^2
\mathbb E\!\left[
A_i
\bigl(
Y_i-\mu_1(\boldsymbol f_{X,i},\boldsymbol f_{M,i})
\bigr)^2
\;\middle|\;
\mathcal G_n^{T_2}
\right] \\
&\le
C
\frac{1}{n}\sum_{i\in \mathcal I_4}
\bigl(
\hat\pi_2(\hat{\boldsymbol f}_{X,i},\hat{\boldsymbol f}_{M,i})
-
\pi_2(\boldsymbol f_{X,i},\boldsymbol f_{M,i})
\bigr)^2 .
\end{align*}
The second inequality uses $A_i\le 1$, positivity of $\hat\pi_1$, and the bounded conditional second moment in Assumption~\ref{assm:dgp_1}. Now, using the ratio representation of $\pi_2$ together with the uniform boundedness of $(\pi_1, \hat \pi_1, \rho, \hat \rho)$, we obtain
\begin{align*}
&\left|
\hat\pi_2(\hat{\boldsymbol f}_{X,i},\hat{\boldsymbol f}_{M,i})
-
\pi_2(\boldsymbol f_{X,i},\boldsymbol f_{M,i})
\right| \\
&\le
C
\Big(
\left|
\hat\rho(\hat{\boldsymbol f}_{X,i},\hat{\boldsymbol f}_{M,i})
-
\rho\circ v^{-1}(\hat{\boldsymbol f}_{X,i},\hat{\boldsymbol f}_{M,i})
\right|
+
\left|
\hat\pi_1(\hat{\boldsymbol f}_{X,i})
-
\pi_1\circ v_X^{-1}(\hat{\boldsymbol f}_{X,i})
\right| \\
&\qquad\qquad+
\left|
\rho\circ v^{-1}(\hat{\boldsymbol f}_{X,i},\hat{\boldsymbol f}_{M,i})
-
\rho(\boldsymbol f_{X,i},\boldsymbol f_{M,i})
\right|
+
\left|
\pi_1\circ v_X^{-1}(\hat{\boldsymbol f}_{X,i})
-
\pi_1(\boldsymbol f_{X,i})
\right|
\Big),
\end{align*}
where we use that the maps $x\mapsto (1-x)/x$ and $x\mapsto x/(1-x)$ are Lipschitz. Therefore, we obtain
\begin{align*}
&\frac{1}{n}\sum_{i\in \mathcal I_4}
\bigl(
\hat\pi_2(\hat{\boldsymbol f}_{X,i},\hat{\boldsymbol f}_{M,i})
-
\pi_2(\boldsymbol f_{X,i},\boldsymbol f_{M,i})
\bigr)^2 \\
&\le
C
\left\|
\hat\rho(\hat{\boldsymbol f}_X,\hat{\boldsymbol f}_M)
-
\rho\circ v^{-1}(\hat{\boldsymbol f}_X,\hat{\boldsymbol f}_M)
\right\|_{L_2( P_n)}^2
+
C
\left\|
\hat\pi_1(\hat{\boldsymbol f}_X)
-
\pi_1\circ v_X^{-1}(\hat{\boldsymbol f}_X)
\right\|_{L_2( P_n)}^2 \\
&\quad+
C
\left\|
\rho\circ v^{-1}(\hat{\boldsymbol f}_X,\hat{\boldsymbol f}_M)
-
\rho(\boldsymbol f_X,\boldsymbol f_M)
\right\|_{L_2( P_n)}^2
+
C
\left\|
\pi_1\circ v_X^{-1}(\hat{\boldsymbol f}_X)
-
\pi_1(\boldsymbol f_X)
\right\|_{L_2( P_n)}^2 \\
&=
O_p(\delta_{n,\rho}^2+n^{-1/2})
+
O_p(\delta_{n,\pi_1}^2+n^{-1/2})
+
O_p(\delta_{n,f}^2)
+
O_p(\delta_{n,f}^2) \\
&=
O_p(\delta_{n,\rho}^2+\delta_{n,\pi_1}^2+\delta_{n,f}^2+n^{-1/2}).
\end{align*}
where the inequality follows $(a+b+c+d)^2\le C(a^2+b^2+c^2+d^2)$. The first two rates follow by adding and subtracting their population $L_2( P)$ counterparts, using the nuisance rates in Theorem~\ref{thm:main}, and applying Lemma \ref{lemma:l2-norm}. The last two rates follow from the Lipschitz continuity of $\rho$ and $\pi_1$ together with the factor-recovery rate in Theorem~\ref{thm:main}. Combining the above bounds, we obtain $\mathrm{Var}(T_2\mid \mathcal G_n^{T_2})
=
O_p(\delta_{n,\rho}^2+\delta_{n,\pi_1}^2+\delta_{n,f}^2+n^{-1/2})=
o_p(1).$ Together with $\mathbb E[T_2\mid \mathcal G_n^{T_2}]=0$, Chebyshev's inequality gives, for any $\varepsilon>0$,
\begin{align*}
\mathbb P(|T_2|>\varepsilon \mid \mathcal G_n^{T_2})
\le
\frac{\mathrm{Var}(T_2\mid \mathcal G_n^{T_2})}{\varepsilon^2}
=
o_p(1).
\end{align*}
Hence $T_2=o_p(1)$ by the similar argument as in proof of $T_1$.

\textbf{Analysis of $T_3$.} Consider $T_3$, which can be expressed as
\begin{align*}
T_3
&=
\frac{1}{\sqrt n}\sum_{i\in \mathcal I_4}
\frac{
(1-A_i)
-
A_i
\frac{1-\hat\rho(\hat{\boldsymbol f}_{X,i},\hat{\boldsymbol f}_{M,i})}
     {\hat\rho(\hat{\boldsymbol f}_{X,i},\hat{\boldsymbol f}_{M,i})}
}
{1-\hat\pi_1(\hat{\boldsymbol f}_{X,i})}
\bigl(
\hat\mu_1(\hat{\boldsymbol f}_{X,i},\hat{\boldsymbol f}_{M,i})
-
\mu_1(\boldsymbol f_{X,i},\boldsymbol f_{M,i})
\bigr) \\
&=
\frac{1}{\sqrt n}\sum_{i\in \mathcal I_4}
\frac{
(1-A_i)
-
A_i
\frac{1-\rho(\boldsymbol f_{X,i},\boldsymbol f_{M,i})}
     {\rho(\boldsymbol f_{X,i},\boldsymbol f_{M,i})}
}
{1-\pi_1(\boldsymbol f_{X,i})}
\bigl(
\hat\mu_1(\hat{\boldsymbol f}_{X,i},\hat{\boldsymbol f}_{M,i})
-
\mu_1(\boldsymbol f_{X,i},\boldsymbol f_{M,i})
\bigr) \\
&\quad+
\frac{1}{\sqrt n}\sum_{i\in \mathcal I_4}
\left[
\frac{
(1-A_i)
-
A_i
\frac{1-\hat\rho(\hat{\boldsymbol f}_{X,i},\hat{\boldsymbol f}_{M,i})}
     {\hat\rho(\hat{\boldsymbol f}_{X,i},\hat{\boldsymbol f}_{M,i})}
}
{1-\hat\pi_1(\hat{\boldsymbol f}_{X,i})}
-
\frac{
(1-A_i)
-
A_i
\frac{1-\rho(\boldsymbol f_{X,i},\boldsymbol f_{M,i})}
     {\rho(\boldsymbol f_{X,i},\boldsymbol f_{M,i})}
}
{1-\pi_1(\boldsymbol f_{X,i})}
\right] \\
&\qquad\qquad\cdot
\bigl(
\hat\mu_1(\hat{\boldsymbol f}_{X,i},\hat{\boldsymbol f}_{M,i})
-
\mu_1(\boldsymbol f_{X,i},\boldsymbol f_{M,i})
\bigr) \\
&=: T_{31}+T_{32}.
\end{align*}
We first study $T_{31}$. Then
\begin{align*}
\mathbb E[T_{31}\mid \mathcal G_n^{T_3}]
&=
\frac{1}{\sqrt n}\sum_{i\in \mathcal I_4}
\frac{
\hat\mu_1(\hat{\boldsymbol f}_{X,i},\hat{\boldsymbol f}_{M,i})
-
\mu_1(\boldsymbol f_{X,i},\boldsymbol f_{M,i})
}
{1-\pi_1(\boldsymbol f_{X,i})} \\
&\qquad\qquad\cdot
\mathbb E\!\left[
(1-A_i)
-
A_i
\frac{1-\rho(\boldsymbol f_{X,i},\boldsymbol f_{M,i})}
     {\rho(\boldsymbol f_{X,i},\boldsymbol f_{M,i})}
\;\middle|\;
\mathcal G_n^{T_3}
\right] \\
&=
\frac{1}{\sqrt n}\sum_{i\in \mathcal I_4}
\frac{
\hat\mu_1(\hat{\boldsymbol f}_{X,i},\hat{\boldsymbol f}_{M,i})
-
\mu_1(\boldsymbol f_{X,i},\boldsymbol f_{M,i})
}
{1-\pi_1(\boldsymbol f_{X,i})} \\
&\qquad\qquad\cdot
\mathbb E\!\left[
\mathbb E\!\left[
(1-A_i)
-
A_i
\frac{1-\rho(\boldsymbol f_{X,i},\boldsymbol f_{M,i})}
     {\rho(\boldsymbol f_{X,i},\boldsymbol f_{M,i})}
\;\middle|\;
\boldsymbol f_{X,i},\boldsymbol f_{M,i}
\right]
\;\middle|\;
\mathcal G_n^{T_3}
\right] \\
&=
\frac{1}{\sqrt n}\sum_{i\in \mathcal I_4}
\frac{
\hat\mu_1(\hat{\boldsymbol f}_{X,i},\hat{\boldsymbol f}_{M,i})
-
\mu_1(\boldsymbol f_{X,i},\boldsymbol f_{M,i})
}
{1-\pi_1(\boldsymbol f_{X,i})} \cdot 0 \\
&=0,
\end{align*}
where
\begin{align*}
\mathbb E\!\left[
(1-A_i)
-
A_i
\frac{1-\rho(\boldsymbol f_{X,i},\boldsymbol f_{M,i})}
     {\rho(\boldsymbol f_{X,i},\boldsymbol f_{M,i})}
\;\middle|\;
\boldsymbol f_{X,i},\boldsymbol f_{M,i}
\right]&=
1-\rho(\boldsymbol f_{X,i},\boldsymbol f_{M,i})
-
\rho(\boldsymbol f_{X,i},\boldsymbol f_{M,i})
\frac{1-\rho(\boldsymbol f_{X,i},\boldsymbol f_{M,i})}
     {\rho(\boldsymbol f_{X,i},\boldsymbol f_{M,i})}\\
&=0.
\end{align*}
Therefore, $\mathbb E[T_{31}\mid \mathcal G_n^{T_3}]=0.$ Next, Given $\mathcal G_n^{T_3}$,
\begin{align*}
\mathrm{Var}(T_{31}\mid \mathcal G_n^{T_3})
&\le
C\frac{1}{n}\sum_{i\in \mathcal I_4}
\bigl(
\hat\mu_1(\hat{\boldsymbol f}_{X,i},\hat{\boldsymbol f}_{M,i})
-
\mu_1(\boldsymbol f_{X,i},\boldsymbol f_{M,i})
\bigr)^2 \\
&=
C
\left\|
\hat\mu_1(\hat{\boldsymbol f}_X,\hat{\boldsymbol f}_M)
-
\mu_1(\boldsymbol f_X,\boldsymbol f_M)
\right\|_{L_2( P_n)}^2 \\
&\le
C
\left\|
\hat\mu_1(\hat{\boldsymbol f}_X,\hat{\boldsymbol f}_M)
-
\mu_1\circ v^{-1}(\hat{\boldsymbol f}_X,\hat{\boldsymbol f}_M)
\right\|_{L_2( P_n)}^2
+
C
\left\|
\mu_1\circ v^{-1}(\hat{\boldsymbol f}_X,\hat{\boldsymbol f}_M)
-
\mu_1(\boldsymbol f_X,\boldsymbol f_M)
\right\|_{L_2( P_n)}^2 \\
&=
O_p(\delta_{n,\mu_1}^2+n^{-1/2})
+
O_p(\delta_{n,f}^2) \\
&=
O_p(\delta_{n,\mu_1}^2+\delta_{n,f}^2+n^{-1/2})
=
o_p(1).
\end{align*}
Together with $\mathbb E[T_{31}\mid \mathcal G_n^{T_3}]=0$, Chebyshev's inequality and the same dominated-convergence argument as in the proof of $T_1$ imply $T_{31}=o_p(1)$. 

We next study $T_{32}$. Recall that
\begin{align*}
T_{32}
=
\frac{1}{\sqrt n}\sum_{i\in\mathcal I_4} Z_{1,i}Z_{2,i},
\end{align*}
where
\begin{align*}
Z_{1,i}
&:=
\frac{
(1-A_i)
-
A_i
\frac{1-\hat\rho(\hat{\boldsymbol f}_{X,i},\hat{\boldsymbol f}_{M,i})}
     {\hat\rho(\hat{\boldsymbol f}_{X,i},\hat{\boldsymbol f}_{M,i})}
}
{1-\hat\pi_1(\hat{\boldsymbol f}_{X,i})}
-
\frac{
(1-A_i)
-
A_i
\frac{1-\rho(\boldsymbol f_{X,i},\boldsymbol f_{M,i})}
     {\rho(\boldsymbol f_{X,i},\boldsymbol f_{M,i})}
}
{1-\pi_1(\boldsymbol f_{X,i})},
\\
Z_{2,i}
&:=
\hat\mu_1(\hat{\boldsymbol f}_{X,i},\hat{\boldsymbol f}_{M,i})
-
\mu_1(\boldsymbol f_{X,i},\boldsymbol f_{M,i}).
\end{align*}
By Cauchy--Schwarz,
\begin{align*}
|T_{32}|
&\le
\frac{1}{\sqrt n}
\left(
\sum_{i\in\mathcal I_4}Z_{1,i}^2
\right)^{1/2}
\left(
\sum_{i\in\mathcal I_4}Z_{2,i}^2
\right)^{1/2} =
\sqrt n
\left(
\frac{1}{n}\sum_{i\in\mathcal I_4}Z_{1,i}^2
\right)^{1/2}
\left(
\frac{1}{n}\sum_{i\in\mathcal I_4}Z_{2,i}^2
\right)^{1/2}.
\end{align*}
By positivity and Lipschitz continuity of the maps $x\mapsto (1-x)/x$ and $x\mapsto 1/(1-x)$,
\begin{align*}
|Z_{1,i}|
&\le
C
\left|
\hat\rho(\hat{\boldsymbol f}_{X,i},\hat{\boldsymbol f}_{M,i})
-
\rho\circ v^{-1}(\hat{\boldsymbol f}_{X,i},\hat{\boldsymbol f}_{M,i})
\right|
+
C
\left|
\rho\circ v^{-1}(\hat{\boldsymbol f}_{X,i},\hat{\boldsymbol f}_{M,i})
-
\rho(\boldsymbol f_{X,i},\boldsymbol f_{M,i})
\right| \\
&\quad+
C
\left|
\hat\pi_1(\hat{\boldsymbol f}_{X,i})
-
\pi_1\circ v_X^{-1}(\hat{\boldsymbol f}_{X,i})
\right|
+
C
\left|
\pi_1\circ v_X^{-1}(\hat{\boldsymbol f}_{X,i})
-
\pi_1(\boldsymbol f_{X,i})
\right|.
\end{align*}
Therefore, using $(a+b+c+d)^2\le C(a^2+b^2+c^2+d^2)$, Lemma~\ref{lemma:l2-norm}, and Assumptions in Theorem \ref{thm:main}, we have:
\begin{align*}
\left(
\frac{1}{n}\sum_{i\in\mathcal I_4}Z_{1,i}^2
\right)^{1/2}
&\le
C
\left\|
\hat\rho(\hat{\boldsymbol f}_X,\hat{\boldsymbol f}_M)
-
\rho\circ v^{-1}(\hat{\boldsymbol f}_X,\hat{\boldsymbol f}_M)
\right\|_{L_2( P_n)} \\
&\quad+
C
\left\|
\rho\circ v^{-1}(\hat{\boldsymbol f}_X,\hat{\boldsymbol f}_M)
-
\rho(\boldsymbol f_X,\boldsymbol f_M)
\right\|_{L_2( P_n)} \\
&\quad+
C
\left\|
\hat\pi_1(\hat{\boldsymbol f}_X)
-
\pi_1\circ v_X^{-1}(\hat{\boldsymbol f}_X)
\right\|_{L_2( P_n)} \\
&\quad+
C
\left\|
\pi_1\circ v_X^{-1}(\hat{\boldsymbol f}_X)
-
\pi_1(\boldsymbol f_X)
\right\|_{L_2( P_n)} \\
&=
O_p(\delta_{n,\rho}+n^{-1/2})
+
O_p(\delta_{n,f})
+
O_p(\delta_{n,\pi_1}+n^{-1/2})
+
O_p(\delta_{n,f}) \\
&=
O_p(\delta_{n,\rho}+\delta_{n,\pi_1}+\delta_{n,f}+n^{-1/2})\,.
\end{align*}
Similarly,
\begin{align*}
\left(
\frac{1}{n}\sum_{i\in\mathcal I_4}Z_{2,i}^2
\right)^{1/2}
&\le
C
\left\|
\hat\mu_1(\hat{\boldsymbol f}_X,\hat{\boldsymbol f}_M)
-
\mu_1\circ v^{-1}(\hat{\boldsymbol f}_X,\hat{\boldsymbol f}_M)
\right\|_{L_2( P_n)} \\
&\quad+
C
\left\|
\mu_1\circ v^{-1}(\hat{\boldsymbol f}_X,\hat{\boldsymbol f}_M)
-
\mu_1(\boldsymbol f_X,\boldsymbol f_M)
\right\|_{L_2( P_n)} \\
&=
O_p(\delta_{n,\mu_1}+n^{-1/2})
+
O_p(\delta_{n,f}) \\
&=
O_p(\delta_{n,\mu_1}+\delta_{n,f}+n^{-1/2}),
\end{align*}
again by Lemma~\ref{lemma:l2-norm}. Combining the above bounds gives
\begin{align*}
|T_{32}|
&=
O_p\!\left[
\sqrt n
(\delta_{n,\rho}+\delta_{n,\pi_1}+\delta_{n,f}+n^{-1/2})
(\delta_{n,\mu_1}+\delta_{n,f}+n^{-1/2})
\right] \\
&=
o_p(1),
\end{align*}
Combining the results for $T_{31}$ and $T_{32}$, we conclude $T_{3}=o_p(1).$

\noindent
\textbf{Analysis of $T_4$.}
We first study the conditional expectation of $T_4$ given $\mathcal G_n^{T_4}$:
\begin{align*}
\mathbb E[T_4 \mid \mathcal G_n^{T_4}]
&=
\frac{1}{\sqrt n}\sum_{i\in\mathcal I_4}
\left(
\frac{1}{1-\hat\pi_1(\hat{\boldsymbol f}_{X,i})}
-
\frac{1}{1-\pi_1(\boldsymbol f_{X,i})}
\right) \\
&\qquad\qquad\cdot
\mathbb E\!\left[
(1-A_i)
\bigl(
\mu_1(\boldsymbol f_{X,i},\boldsymbol f_{M,i})
-
\mu_{10}(\boldsymbol f_{X,i})
\bigr)
\;\middle|\;
\mathcal G_n^{T_4}
\right] \\
&=
\frac{1}{\sqrt n}\sum_{i\in\mathcal I_4}
\left(
\frac{1}{1-\hat\pi_1(\hat{\boldsymbol f}_{X,i})}
-
\frac{1}{1-\pi_1(\boldsymbol f_{X,i})}
\right) \\
&\qquad\qquad\cdot
\mathbb E\!\left[
\mathbb E\!\left[
(1-A_i)
\bigl(
\mu_1(\boldsymbol f_{X,i},\boldsymbol f_{M,i})
-
\mu_{10}(\boldsymbol f_{X,i})
\bigr)
\;\middle|\;
A_i,\boldsymbol f_{X,i}
\right]
\;\middle|\;
\mathcal G_n^{T_4}
\right] \\
&=0.
\end{align*}
The last equality follows from the definition
$\mu_{10}(\boldsymbol f_X)
=
\mathbb E[
\mu_1(\boldsymbol f_X,\boldsymbol f_M)
\mid A=0,\boldsymbol f_X
]$, which implies
\begin{align*}
&\mathbb E\!\left[
(1-A_i)
\bigl(
\mu_1(\boldsymbol f_{X,i},\boldsymbol f_{M,i})
-
\mu_{10}(\boldsymbol f_{X,i})
\bigr)
\;\middle|\;
A_i,\boldsymbol f_{X,i}
\right] \\
&\qquad=
(1-A_i)
\mathbb E\!\left[
\mu_1(\boldsymbol f_{X,i},\boldsymbol f_{M,i})
-
\mu_{10}(\boldsymbol f_{X,i})
\;\middle|\;
A_i,\boldsymbol f_{X,i}
\right]
=0.
\end{align*}
We next study the conditional variance:
\begin{align*}
\mathrm{Var}(T_4 \mid \mathcal G_n^{T_4})
&\le
\frac{1}{n}\sum_{i\in\mathcal I_4}
\mathbb E\!\left[
(1-A_i)
\bigl(
\mu_1(\boldsymbol f_{X,i},\boldsymbol f_{M,i})
-
\mu_{10}(\boldsymbol f_{X,i})
\bigr)^2
\left(
\frac{1}{1-\hat\pi_1(\hat{\boldsymbol f}_{X,i})}
-
\frac{1}{1-\pi_1(\boldsymbol f_{X,i})}
\right)^2
\;\middle|\;
\mathcal G_n^{T_4}
\right] \\
&\le
C\frac{1}{n}\sum_{i\in\mathcal I_4}
\left(
\frac{1}{1-\hat\pi_1(\hat{\boldsymbol f}_{X,i})}
-
\frac{1}{1-\pi_1(\boldsymbol f_{X,i})}
\right)^2 \\
&\le
C
\frac{1}{n}\sum_{i\in\mathcal I_4}
\bigl(
\hat\pi_1(\hat{\boldsymbol f}_{X,i})
-
\pi_1(\boldsymbol f_{X,i})
\bigr)^2 \\
&=
C
\left\|
\hat\pi_1(\hat{\boldsymbol f}_X)
-
\pi_1(\boldsymbol f_X)
\right\|_{L_2( P_n)}^2 \\
&\le
C
\left\|
\hat\pi_1(\hat{\boldsymbol f}_X)
-
\pi_1\circ v_X^{-1}(\hat{\boldsymbol f}_X)
\right\|_{L_2( P_n)}^2
+
C
\left\|
\pi_1\circ v_X^{-1}(\hat{\boldsymbol f}_X)
-
\pi_1(\boldsymbol f_X)
\right\|_{L_2( P_n)}^2 \\
&=
O_p(\delta_{n,\pi_1}^2+n^{-1/2})
+
O_p(\delta_{n,f}^2) \\
&=
O_p(\delta_{n,\pi_1}^2+\delta_{n,f}^2+n^{-1/2})
=
o_p(1).
\end{align*}
The second inequality uses $1-A_i\le 1$, boundedness of $\pi_1$ and $\hat\pi_1$, and the bounded conditional second moment in Assumption~\ref{assm:dgp_1}. Together with $\mathbb E[T_4\mid \mathcal G_n^{T_4}]=0$, Chebyshev's inequality and the same bounded-convergence-in-probability argument as in the proof of $T_1$ imply $T_4=o_p(1)$.

\noindent
\textbf{Analysis of $T_5$.} Consider $T_5$, which can be expressed as
\begin{align*}
T_5
&=
\frac{1}{\sqrt n}\sum_{i\in \mathcal I_4}
\frac{A_i-\hat\pi_1(\hat{\boldsymbol f}_{X,i})}
{1-\hat\pi_1(\hat{\boldsymbol f}_{X,i})}
\bigl(
\hat\mu_{10}(\hat{\boldsymbol f}_{X,i})
-
\mu_{10}(\boldsymbol f_{X,i})
\bigr) \\
&=
\frac{1}{\sqrt n}\sum_{i\in \mathcal I_4}
\frac{A_i-\pi_1(\boldsymbol f_{X,i})}
{1-\pi_1(\boldsymbol f_{X,i})}
\bigl(
\hat\mu_{10}(\hat{\boldsymbol f}_{X,i})
-
\mu_{10}(\boldsymbol f_{X,i})
\bigr) \\
&\quad+
\frac{1}{\sqrt n}\sum_{i\in \mathcal I_4}
\left[
\frac{A_i-\hat\pi_1(\hat{\boldsymbol f}_{X,i})}
{1-\hat\pi_1(\hat{\boldsymbol f}_{X,i})}
-
\frac{A_i-\pi_1(\boldsymbol f_{X,i})}
{1-\pi_1(\boldsymbol f_{X,i})}
\right] \\
&\qquad\qquad\cdot
\bigl(
\hat\mu_{10}(\hat{\boldsymbol f}_{X,i})
-
\mu_{10}(\boldsymbol f_{X,i})
\bigr) \\
&=: T_{51}+T_{52}.
\end{align*}

We first study $T_{51}$. Then
\begin{align*}
\mathbb E[T_{51}\mid \mathcal G_n^{T_5}]
&=
\frac{1}{\sqrt n}\sum_{i\in \mathcal I_4}
\frac{
\hat\mu_{10}(\hat{\boldsymbol f}_{X,i})
-
\mu_{10}(\boldsymbol f_{X,i})
}
{1-\pi_1(\boldsymbol f_{X,i})}
\mathbb E\!\left[
A_i-\pi_1(\boldsymbol f_{X,i})
\;\middle|\;
\mathcal G_n^{T_5}
\right] \\
&=
\frac{1}{\sqrt n}\sum_{i\in \mathcal I_4}
\frac{
\hat\mu_{10}(\hat{\boldsymbol f}_{X,i})
-
\mu_{10}(\boldsymbol f_{X,i})
}
{1-\pi_1(\boldsymbol f_{X,i})}
\mathbb E\!\left[
\mathbb E\!\left[
A_i-\pi_1(\boldsymbol f_{X,i})
\;\middle|\;
\boldsymbol f_{X,i}
\right]
\;\middle|\;
\mathcal G_n^{T_5}
\right] \\
&=0.
\end{align*}
We next study the conditional variance:
\begin{align*}
\mathrm{Var}(T_{51}\mid \mathcal G_n^{T_5})
&\le
C\frac{1}{n}\sum_{i\in \mathcal I_4}
\bigl(
\hat\mu_{10}(\hat{\boldsymbol f}_{X,i})
-
\mu_{10}(\boldsymbol f_{X,i})
\bigr)^2 \\
&=
C
\left\|
\hat\mu_{10}(\hat{\boldsymbol f}_X)
-
\mu_{10}(\boldsymbol f_X)
\right\|_{L_2( P_n)}^2 \\
&\le
C
\left\|
\hat\mu_{10}(\hat{\boldsymbol f}_X)
-
\mu_{10}\circ v_X^{-1}(\hat{\boldsymbol f}_X)
\right\|_{L_2( P_n)}^2
+
C
\left\|
\mu_{10}\circ v_X^{-1}(\hat{\boldsymbol f}_X)
-
\mu_{10}(\boldsymbol f_X)
\right\|_{L_2( P_n)}^2 \\
&=
O_p(\delta_{n,\mu_{10}}^2+n^{-1/2})
+
O_p(\delta_{n,f}^2) \\
&=
O_p(\delta_{n,\mu_{10}}^2+\delta_{n,f}^2+n^{-1/2})
=
o_p(1).
\end{align*}
Together with $\mathbb E[T_{51}\mid \mathcal G_n^{T_5}]=0$, Chebyshev's inequality and the same bounded-convergence-in-probability argument as in the proof of $T_1$ imply $T_{51}=o_p(1)$.

We next study $T_{52}$. Let $Z_{1,i}
:=
\frac{A_i-\hat\pi_1(\hat{\boldsymbol f}_{X,i})}
{1-\hat\pi_1(\hat{\boldsymbol f}_{X,i})}
-
\frac{A_i-\pi_1(\boldsymbol f_{X,i})}
{1-\pi_1(\boldsymbol f_{X,i})},$ and $Z_{2,i}
:=
\hat\mu_{10}(\hat{\boldsymbol f}_{X,i})
-
\mu_{10}(\boldsymbol f_{X,i}).$
Then, we write
\begin{align*}
T_{52}
=
\frac{1}{\sqrt n}\sum_{i\in\mathcal I_4} Z_{1,i}Z_{2,i}.
\end{align*}
By Cauchy--Schwarz,
\begin{align*}
|T_{52}|
&\le
\frac{1}{\sqrt n}
\left(
\sum_{i\in\mathcal I_4} Z_{1,i}^2
\right)^{1/2}
\left(
\sum_{i\in\mathcal I_4} Z_{2,i}^2
\right)^{1/2} =
\sqrt n
\left(
\frac{1}{n}\sum_{i\in\mathcal I_4} Z_{1,i}^2
\right)^{1/2}
\left(
\frac{1}{n}\sum_{i\in\mathcal I_4} Z_{2,i}^2
\right)^{1/2}.
\end{align*}
By positivity and Lipschitz continuity of the map $x\mapsto (A_i-x)/(1-x)$ on the relevant range,
\begin{align*}
|Z_{1,i}|
&\le
C
\left|
\hat\pi_1(\hat{\boldsymbol f}_{X,i})
-
\pi_1(\boldsymbol f_{X,i})
\right| \\
&\le
C
\left|
\hat\pi_1(\hat{\boldsymbol f}_{X,i})
-
\pi_1\circ v_X^{-1}(\hat{\boldsymbol f}_{X,i})
\right|
+
C
\left|
\pi_1\circ v_X^{-1}(\hat{\boldsymbol f}_{X,i})
-
\pi_1(\boldsymbol f_{X,i})
\right|.
\end{align*}
Therefore,
\begin{align*}
\left(
\frac{1}{n}\sum_{i\in\mathcal I_4} Z_{1,i}^2
\right)^{1/2}
&\le
C
\left\|
\hat\pi_1(\hat{\boldsymbol f}_X)
-
\pi_1\circ v_X^{-1}(\hat{\boldsymbol f}_X)
\right\|_{L_2( P_n)}
+
C
\left\|
\pi_1\circ v_X^{-1}(\hat{\boldsymbol f}_X)
-
\pi_1(\boldsymbol f_X)
\right\|_{L_2( P_n)} \\
&=
O_p(\delta_{n,\pi_1}+n^{-1/2})
+
O_p(\delta_{n,f}) \\
&=
O_p(\delta_{n,\pi_1}+\delta_{n,f}+n^{-1/2}).
\end{align*}

Similarly,
\begin{align*}
|Z_{2,i}|
&\le
\left|
\hat\mu_{10}(\hat{\boldsymbol f}_{X,i})
-
\mu_{10}\circ v_X^{-1}(\hat{\boldsymbol f}_{X,i})
\right|
+
\left|
\mu_{10}\circ v_X^{-1}(\hat{\boldsymbol f}_{X,i})
-
\mu_{10}(\boldsymbol f_{X,i})
\right|,
\end{align*}
and hence
\begin{align*}
\left(
\frac{1}{n}\sum_{i\in\mathcal I_4} Z_{2,i}^2
\right)^{1/2}
&\le
C
\left\|
\hat\mu_{10}(\hat{\boldsymbol f}_X)
-
\mu_{10}\circ v_X^{-1}(\hat{\boldsymbol f}_X)
\right\|_{L_2(\mathbb P_n)}
+
C
\left\|
\mu_{10}\circ v_X^{-1}(\hat{\boldsymbol f}_X)
-
\mu_{10}(\boldsymbol f_X)
\right\|_{L_2(\mathbb P_n)} \\
&=
O_p(\delta_{n,\mu_{10}}+n^{-1/2})
+
O_p(\delta_{n,f}) \\
&=
O_p(\delta_{n,\mu_{10}}+\delta_{n,f}+n^{-1/2}).
\end{align*}
Combining the above bounds gives
\begin{align*}
|T_{52}|
&=
O_p\!\left[
\sqrt n
(\delta_{n,\pi_1}+\delta_{n,f}+n^{-1/2})
(\delta_{n,\mu_{10}}+\delta_{n,f}+n^{-1/2})
\right] = o_p(1).
\end{align*}
we obtain $T_{52}=o_p(1)$. Together with $T_{51}=o_p(1)$, this implies $T_5=o_p(1).$

\textbf{Asymptotics.} Combining the bounds for $T_1,\ldots,T_5$, we obtain
\begin{align*}
\sqrt n\bigl(\hat\theta_0^{\mathrm{IF}}-\theta_0\bigr)
=
\frac{1}{\sqrt n}\sum_{i=1}^n \Psi_i
+
o_p(1).
\end{align*}
Since $\mathbb E[\Psi_i]=0$ and $\mathrm{Var}(\Psi_i)=\sigma_{\mathrm{eff}}^2<\infty$ by the central limit theorem,
\begin{align*}
\frac{1}{\sqrt n}\sum_{i=1}^n \Psi_i
\overset{d}{\longrightarrow}
\mathcal N(0,\sigma_{\mathrm{eff}}^2).
\end{align*}
Applying Slutsky's theorem yields
\begin{align*}
\sqrt n\bigl(\hat\theta_0^{\mathrm{IF}}-\theta_0\bigr)
\overset{d}{\longrightarrow}
\mathcal N(0,\sigma_{\mathrm{eff}}^2).
\end{align*}

\begin{lemma}
 \label{lemma:l2-norm}
    Suppose $f: \cX \to [-B, B]$ where $\cX \subseteq \reals^d$. Let $X_1, \dots, X_n \sim P$ for some probability measure $P$ on $\cX$. Define $L_2(\bbP_n)$ and $L_2(P)$ norm of $f$ as: 
    $$
    \|f\|_n := \|f\|_{L_2(\bbP_n)} = \sqrt{\frac1n \sum_i f^2(X_i)}, \qquad \|f\|_2 = \|f\|_{L_2(P)} = \sqrt{\int f^2(x) \ dP(x)} \,.
    $$
    Then, we have: 
    $$
    \bbP\left(\left|\|f\|_n - \|f\|_2\right| > t\right)  \le \frac{B^2}{nt^2} \,.
    $$
    As a consequence, $|\|f\|_n - \|f\|_2| = O_p(n^{-1/2})$. 
\end{lemma}

\section{Proof of Lemma \ref{lemma:l2-norm}}
From definition we have: 
$$
\left|\|f\|_n - \|f\|_2\right| = \frac{\left|\|f\|^2_n - \|f\|^2_2\right|}{\|f\|_n + \|f\|_2} \le \frac{\left|\|f\|^2_n - \|f\|^2_2\right|}{\|f\|_2}
$$
As a consequence, we have: 
\begin{align*}
\bbP\left(\left|\|f\|_n - \|f\|_2\right| > t\right)  & \le \bbP\left(\left|\|f\|^2_n - \|f\|^2_2\right| > t\|f\|_2\right)  \\
& = \bbP\left(\left|\frac1n \sum_{i = 1}^n (f^2(X_i) - \bbE[f^2(X)])\right| > t\|f\|_2\right) \\
& \le \frac{\var(f^2(X))}{nt^2 \|f\|_2^2} \\
& \le \frac{\bbE(f^4(X))}{nt^2 \|f\|_2^2} \\
& \le  \frac{B^2\bbE(f^2(X))}{nt^2 \|f\|_2^2} = \frac{B^2}{nt^2} \,.
\end{align*}
The last inequality follows from the fact that $\|f\|_\infty \le B$. This completes the proof.

\section{Proof of Theorem \ref{thm:HCM_nuisance}}
\label{Proof_of_HCM_nuisance}
The proof mostly follows using a similar argument as in the proof of Theorem 1 of \cite{fan2023factor}. We highlight the key parts here. As the proof for $\hat \pi_1$ and $\hat \rho$ are analogous, we skip the latter for brevity and present the proof of the other three nuisances here.  
\subsection{Rate for \texorpdfstring{$\widehat\mu_1$}{mu1} under the transformed factor representation}
\label{subsec:mu1-rate}
Recall our notation $\blf = (\blf_X, \blf_M)$ denotes the unobserved factors, and $\hat \blf = (\hat \blf_X, \hat \blf_M)$ denotes the estimated factors, i.e. $\hat \blf_X = \hat g_X(X)$ and $\hat \blf_M = \hat g_M(M)$. By the regression model, we have: 
$$
Y_i = \mu_1(\blf_i) + \eps_i
$$
where $\bbE[\eps_i \mid \blf_i, \bu_{X, i}, \bu_{M, i}, A] = 0$ and $\eps_i$ is sub-gaussian random variable. As $\hat \blf$ is a function of $(\blf, \bu_X, \bu_X)$, and the encoder is estimated from a separate dataset, it is immediate that $\bbE[\eps \mid \hat \blf, A] = 0$. We use the notation $\eta = \mu_1 \circ \nu^{-1}$. As per Theorem \ref{thm:HCM_nuisance}, $\eta$ is bounded, Lipschitz, and moreover $\eta \in \cH(d_\eta, \ell_\eta, \cP)$ with DAS $\gamma^*_\eta$. Let $\cI_1 = \{i \in \cI_{\rm nuis}: A_i = 1\}$. As per our assumption, 
\[
\|\hat \blf- \nu(\blf)\|_{L_2(P_1)} = O_p(\delta_{n,f}),
\]
Let $\mathcal G_{\eta,n} := \cG(L_\eta,d_\eta,1,N_\eta,M_\eta,B_\eta)$ be a collection of truncated ReLU network class used to estimate $\eta_1$, where $d_\eta = \widetilde p+\widetilde q$ be the input
dimension, $L_\eta$ is the depth, $N_\eta$ is the width, $M_\eta$ is the truncation
level, and $B_\eta$ is the weight bound. We take $M_\eta$ larger than $\|\eta\|_\infty$. The estimator $\hat \eta_1$ is defined as: 
\[
\widehat\eta
\in
\arg\min_{g\in\mathcal G_{\eta,n}}
\frac{1}{n_1}
\sum_{i\in \cI_1}
\left\{
    Y_i-g(\hat \blf_i) 
\right\}^2.
\]
By Theorem~4 of \cite{fan2024factor}, since
$\eta \in \cH(d_\eta, \ell_\eta, \cP)$ and is supported on a bounded domain, there exist constants $C_1,C_2,C_3,C_4>0$, depending only on
the HCM parameters, such that for any integer $N_\mu\ge 2$, there exists a
ReLU network
\[
\eta^\dagger
\in
\mathcal G(C_1,d_\eta,1,C_2N_\eta,\infty,C_3N_\eta^{C_4})
\]
satisfying
\[
        \|\eta^\dagger-\eta\|_{\infty}
        \le
        C N_\eta^{-2\gamma_\eta^\star}.
        \tag{A.2}
\]
After truncation at level $M_\eta$, the approximation error does not increase,
provided $M_\eta>\|\eta\|_\infty$. Therefore, by enlarging constants,
we may regard $\eta^\dagger\in \mathcal G_{\eta,n}$ and retain the same approximation bound. Consequently,
\begin{equation}
\label{eq:nn_approx_mu_1}
\|\eta^\dagger-\eta\|_{\infty}^2
\le
C N_\eta^{-4\gamma_\eta^\star}.
\end{equation}
However, our estimator is evaluated at the learned representation $\hat \blf$, whereas the target is $\mu_1(\blf) = \eta(\nu(\blf))$. Thus, the relevant oracle
error is
\[
\|\eta^\dagger(\hat \blf)-\eta( \nu(\blf))\|_{L_2(P_1)}^2.
\]
Using $(a+b)^2\le 2a^2+2b^2$, we have
\[
\begin{aligned}
\mathbb E_1
    \left[
        \left\{
            \eta^\dagger(\hat \blf)-\eta(\nu(\blf))
        \right\}^2
    \right] \le
    2\mathbb E_1
    \left[
        \left\{
            \eta^\dagger(\hat \blf)-\eta(\hat \blf)
        \right\}^2
    \right]
    +
    2\mathbb E_1
    \left[
        \left\{
            \eta(\hat \blf)-\eta(\nu(\blf))
        \right\}^2
    \right].
\end{aligned}
\]
The first term is controlled by the uniform approximation bound:
\[
        \mathbb E_1
        \left[
            \left\{
                \eta^\dagger(\hat \blf)-\eta(\hat \blf)
            \right\}^2
        \right]
        \le
        \|\eta^\dagger-\eta\|_\infty^2
        \le
        C N_\eta^{-4\gamma_\eta^\star}.
\]
For the second term, since $\eta$ is Lipschitz,
\[
        \mathbb E_1
        \left[
            \left\{
                \eta(\hat \blf)-\eta(\nu(\blf))
            \right\}^2
        \right]
        \le
        L_\eta^2
        \mathbb E_1\|\hat \blf- \nu(\blf)\|_2^2
        =
        O_p(\delta_{n,f}^2).
\]
Therefore,
\begin{align}
\label{eq:approx_bound_dnn_1}
\|\eta^\dagger(\hat \blf)-\eta(\nu(\blf))\|_{L_2(P_1)}^2
=
O_p\left(
    N_\eta^{-4\gamma_\eta^\star}
    +
    \delta_{n,f}^2
\right).
\end{align}
Next, we bound the stochastic error. 
For any $g\in\mathcal G_{\eta,n}$, define $ r_g(i) := g(\hat \blf_i)-\eta(\nu(\blf_i)).$ In particular,
\[
r_{\widehat \eta}(i)
=
\widehat \eta(\hat \blf_i)-\eta(\nu(\blf_i)),
\qquad
r_{\eta^\dagger}(i)
=
\eta^\dagger(\hat \blf_i)-\eta(\nu(\blf_i)).
\]
Since $ Y_i=\eta(\nu(\blf_i))+\varepsilon_i,$ the empirical loss can be written as
\[
        \frac{1}{n_1}\sum_{i\in\cI_1}
        \{Y_i-g(\hat \blf_i)\}^2
        =
        \frac{1}{n_1}\sum_{i\in\cI_1}
        \{\varepsilon_i-r_g(i)\}^2.
\]
By the empirical-risk minimizing property of $\widehat\eta$,
\[
        \frac{1}{n_1}\sum_{i\in\cI_1}
        \{\varepsilon_i-r_{\widehat\eta_1}(i)\}^2
        \le
        \frac{1}{n_1}\sum_{i\in\cI_1}
        \{\varepsilon_i-r_{\eta^\dagger}(i)\}^2.
\]
Expanding both sides and canceling the common term
$n^{-1}\sum_i\varepsilon_i^2$, we obtain the basic inequality
\begin{align}
\label{eq:bound_1}
\|r_{\widehat\eta_1}\|_{n,1}^2 & \le \|r_{\eta^\dagger}\|_{n,1}^2 + \frac{2}{n_1}\sum_{i\in\cI_1} \varepsilon_i\left\{\widehat\eta_1(\hat \blf_i)-\eta^\dagger(\hat \blf_i)\right\}, \notag \\
& \le \|r_{\eta^\dagger}\|_{n,1}^2 +2\sup_{g\in\mathcal G_{\eta,n}} \left|\frac{1}{n_1}\sum_{i\in \cI_1} \varepsilon_i \left\{g(\hat \blf_i)-\eta^\dagger(\hat \blf_i)\right\}\right|, 
\end{align}
where
\[
\|h\|_{n,1}^2
:=
\frac{1}{n_1}\sum_{i\in \cI_1}h_i^2.
\]
Thus, it remains to control the supremum of the empirical process.  By Lemma 4 of \cite{fan2024factor}, the suprema can be bounded in terms of the pseudo-dimension of the network class. Let $V_n = \mathrm{Pdim}(\cG_{\eta, n})$ denotes the pseudo-dimension of $\cG_{\eta, n}$. Then with probability at least $1-e^{-t}$,
\[
\begin{aligned}
\left|
        \frac{1}{n_1}\sum_{i\in \cI_1}
        \varepsilon_i
        \left\{
            g(\hat \blf_i)-\eta^\dagger(\hat \blf_i)
        \right\}
    \right| \le
    C
    \left(
        \|g(\hat \blf)-\eta^\dagger(\hat \blf)\|_{n,1}
        +
    \sqrt{\frac{V_n \log{n_1}}{n_1}}
    \right)
    \sqrt{
        \frac{V_n \log{n_1}}{n_1}+\frac{t}{n}
    }
\end{aligned}
\]
simultaneously for all $g\in\mathcal G_{\eta,n}$. Applying this inequality to
$g=\widehat\eta$ gives
\[
\begin{aligned}
\left|
        \frac{1}{n_1}\sum_{i\in \cI_1}
        \varepsilon_i
        \left\{
            \widehat\eta_1(\hat \blf_i)-\eta^\dagger(\hat \blf_i)
        \right\}
    \right| & \le
    C
    \left(
        \|\widehat\eta_1(\hat \blf)-\eta^\dagger(\hat \blf)\|_{n,1}
        +
        \sqrt{\frac{V_n \log{n_1}}{n_1}}
    \right)
    \sqrt{
         \frac{V_n \log{n_1}}{n_1}+\frac{t}{n}
    } \\
    & \le  C
    \left(
          \|r_{\widehat\eta_1}\|_{n,1}
        +
        \|r_{\eta^\dagger}\|_{n,1}
        +
        \sqrt{\frac{V_n \log{n_1}}{n_1}}
    \right)
    \sqrt{
         \frac{V_n \log{n_1}}{n_1}+\frac{t}{n}
    }
\end{aligned}
\]
where the last line follows from the fact that: 
\[
\|\widehat\eta_1(\hat \blf)-\eta^\dagger(\hat \blf)\|_{n,1}
\le
\|r_{\widehat\eta_1}\|_{n,1}
+
\|r_{\eta^\dagger}\|_{n,1} \,.
\]
This, along with Equation \eqref{eq:bound_1} yields: 
\begin{align*}
\|r_{\widehat\eta_1}\|^2_{n,1} \le  \|r_{\eta^\dagger}\|^2_{n,1} +  C
\left(
\|r_{\widehat\eta_1}\|_{n,1}
+
\|r_{\eta^\dagger}\|_{n,1}
+
\sqrt{\frac{V_n \log{n_1}}{n_1}}
\right)
\sqrt{
\frac{V_n \log{n_1}}{n_1}+\frac{t}{n}
} \,.
\end{align*}
We then apply Young's inequality and change sides to obtain: 
\[
\|r_{\widehat\eta_1}\|_{n,1}^2
\le
C \left(\|r_{\eta^\dagger}\|^2_{n,1} + \frac{V_n \log{n_1}}{n_1}+\frac{t}{n}\right).
\]
Conditional on the representation-learning fold, $r_{\eta^\dagger}$ is a fixed
bounded function of the nuisance-fold observation. Therefore, by Bernstein's inequality applied to the fixed oracle
$r_{\eta^\dagger}$,
\[
\|r_{\eta^\dagger}\|_{n,1}^2
\le
C
\|r_{\eta^\dagger}\|_{L_2(P_1)}^2
+
C\frac{t}{n}
\]
with probability at least $1-e^{-t}$. This, along with the previous equation, yields with probability $\ge 1 - 2e^{-t}$: 
\begin{equation}
\label{eq:bound_2_dnn}
\|r_{\widehat\eta_1}\|_{n,1}^2
\lesssim 
\left(\|r_{\eta^\dagger}\|^2_{L_2(P_1)} + \frac{V_n \log{n_1}}{n_1}+\frac{t}{n}\right)
\end{equation}
It remains to bound $V_\mu=\operatorname{Pdim}(\mathcal G_{\mu,n})$. By Theorem 7 of \cite{bartlett2019nearly}, we have: 
$$
V_n \le C W_\eta L_\eta \log{W_\eta} \le C_1 N_\eta^2 L^2_\eta \log{(N^2_\eta L_\eta)} \,. 
$$
Choosing $L_\eta \asymp 1$ and $N_\eta \asymp n^{1/(4\gamma_\eta^* + 2)}$, we conclude from Equation \eqref{eq:approx_bound_dnn_1} and \eqref{eq:bound_2_dnn}: 
$$
\|r_{\widehat\eta_1}\|_{n,1}^2 = O_p\left( n^{-\frac{2\gamma_\eta^*}{2\gamma_\eta^* + 1}}(\log{n})^a  + \delta_{n, f}^2 \right) \,.
$$
Finally, applying the empirical-to-population norm comparison, as in
Lemma 3 and the last step of the proof of Theorem~1 of
\cite{fan2024factor}, gives the same bound for the population norm:
$$
\|r_{\widehat\eta_1}\|_{L_2(P_1)}^2 = O_p\left( n^{-\frac{2\gamma_\eta^*}{2\gamma_\eta^* + 1}}(\log{n})^a  + \delta_{n, f}^2 \right) \,.
$$
\begin{remark}
    As $n_1 \sim n$ (by positivity assumption $\bbP(A = 1) > 0$) we use $n$ instead of $n_1$ in the rate. 
\end{remark}
\subsection{Rate for \texorpdfstring{$\widehat \pi_1$}{pi1-hat} under the transformed factor representation}
\label{subsec:pi1-rate}

We now prove the rate for the treatment propensity nuisance $\pi_1(\blf_X):=\bbP(A=1\mid \blf_X).$ Define as before, the relevant nuisance parameter, 
\[
\eta := \pi_1\circ \nu^{-1},
\qquad
\eta(\nu(\blf_X))=\pi_1(\blf_X).
\]
Let $\cI_\pi=\cI_{\rm nuis}$ be the nuisance-estimation fold and write
$n=|\cI_\pi|$ for simplicity. Assume
\[
\|\hat\blf_X-\nu(\blf_X)\|_{L_2(P)}
=
O_p(\delta_{n,f}).
\tag{B.1}
\]
From our assumption in Theorem \ref{thm:main}, the propensity score is uniformly bounded away from $0$ and $1$, i.e., for some $\kappa\in(0,1/2)$,
\[
        \kappa
        \le
        \eta(\nu(\blf_X))
        \le
        1-\kappa
\]
almost surely. We restrict the neural-network estimators to take values in
$[\kappa/2,1-\kappa/2]$ by clipping. This does not affect the rate.
Suppose $\eta$ admits a bounded Lipschitz HCM extension to a compact set
containing both $\nu(\blf_X)$ and $\hat\blf_X$ with probability tending to one. Let $\eta\in \cH(d_\eta,\ell_\eta,\cP)$ with DAS $\gamma_\eta^*$, 
where $d_\eta=\widetilde p$ is the input dimension. Let
\[
\mathcal G_{\eta,n}
:=
\cG(L_\eta,d_\eta,1,N_\eta,M_\eta,B_\eta)
\]
be the clipped ReLU network class. As per Line 7 of Algorithm \ref{Algo:MediEncoder}
\[
\widehat\eta
\in
\arg\min_{g\in\mathcal G_{\eta,n}}
\frac1n\sum_{i\in\cI_\pi}
\ell(A_i,g(\hat\blf_{X,i})) := \arg\min_{g\in\mathcal G_{\eta,n}} \ \bbP_n(\ell(A, g(\hat \blf_X)) \,
\]
where
\[
\ell(a,u)
:=
-a\log u-(1-a)\log(1-u),
\qquad
u\in[\kappa/2,1-\kappa/2].
\]
As in the previous subsection, we can use the approximation theorem (e.g., Theorem 4 of \cite{fan2024factor}), there exists
\[
        \eta^\dagger
        \in
        \cG(C_1,d_\eta,1,C_2N_\eta,\infty,C_3N_\eta^{C_4})
\]
such that
\[
        \|\eta^\dagger-\eta\|_\infty
        \le
        C N_\eta^{-2\gamma_\eta^\star}.
\]
After clipping, we may take $\eta^\dagger\in\mathcal G_{\eta,n}$ and retain
\begin{equation}
\label{eq:pi1_nn_approx_shorter}
        \|\eta^\dagger-\eta\|_\infty^2
        \le
        C N_\eta^{-4\gamma_\eta^\star}.
\end{equation}
As in the proof for $\widehat\mu_1$, the relevant oracle error is
\[
        \|\eta^\dagger(\hat\blf_X)-\eta(\nu(\blf_X))\|_{L_2(P)}^2.
\]
Using the same decomposition as in the previous subsection, and the Lipschitzness of $\eta$,
\begin{align*}
\|\eta^\dagger(\hat\blf_X)-\eta(\nu(\blf_X))\|_{L_2(P)}^2  \lesssim \|\eta^\dagger-\eta\|_\infty^2 + \|\hat\blf_X-\nu(\blf_X)\|_{L_2(P)}^2 \,.
\end{align*}
Therefore,
\begin{equation}
\label{eq:pi1_oracle_shorter}
        \|\eta^\dagger(\hat\blf_X)-\eta(\nu(\blf_X))\|_{L_2(P)}^2
        =
        O_p\left(
            N_\eta^{-4\gamma_\eta^\star}
            +
            \delta_{n,f}^2
        \right).
\end{equation}
By empirical risk minimization,
\[
\bbP_n\ell(A,\widehat\eta(\hat\blf_X))
\le
\bbP_n\ell(A,\eta^\dagger(\hat\blf_X)).
\]
Adding and subtracting population risks gives
\begin{align}
\label{eq:pi1_basic_shorter}
P\{\ell(A,\widehat\eta(\hat\blf_X))
        -
        \ell(A,\eta(\nu(\blf_X)))\}
    &\le
    P\{\ell(A,\eta^\dagger(\hat\blf_X))
        -
        \ell(A,\eta(\nu(\blf_X)))\}
    \notag\\
    &\qquad\quad+
    (P-\bbP_n)
    \{\ell(A,\widehat\eta(\hat\blf_X))
        -
        \ell(A,\eta^\dagger(\hat\blf_X))\}.
\end{align}
Since $\hat\blf_X$ is a function of $(\blf_X,\bu_X)$ and $A$ is indepndent of $\bu_X$,
\[
        \bbP(A=1\mid \nu(\blf_X),\hat\blf_X)
        = \pi_1(\blf_X) = 
        \eta(\nu(\blf_X)).
\]
Hence, for any $g\in\mathcal G_{\eta,n}$,
\[
\begin{aligned}
P\{\ell(A,g(\hat\blf_X)) - \ell(A,\eta(\nu(\blf_X)))\}  =
\bbE\left[\mathrm{KL}\{\mathrm{Bern}(\eta(\nu(\blf_X))) \mid \mid \mathrm{Bern}(g(\hat\blf_X))
\}
\right].
\end{aligned}
\]
By overlap and clipping, this KL divergence is equivalent to squared error:
\[
        c_\kappa
        \{g(\hat\blf_X)-\eta(\nu(\blf_X))\}^2
        \le
        \mathrm{KL}\{\cdot,\cdot\}
        \le
        C_\kappa
        \{g(\hat\blf_X)-\eta(\nu(\blf_X))\}^2 .
\]
Therefore, applying the lower bound to $\widehat\eta$ and the upper bound to
$\eta^\dagger$, \eqref{eq:pi1_basic_shorter} yields
\begin{align}
\label{eq:pi1_kl_shorter}
    \|\widehat\eta(\hat\blf_X)-\eta(\nu(\blf_X))\|_{L_2(P)}^2
    &\lesssim
    \|\eta^\dagger(\hat\blf_X)-\eta(\nu(\blf_X))\|_{L_2(P)}^2
    \notag\\
    &\quad+
    (P-\bbP_n)
    \{\ell(A,\widehat\eta(\hat\blf_X))
        -
        \ell(A,\eta^\dagger(\hat\blf_X))\}.
\end{align}
The logistic loss is Lipschitz on $[\kappa/2,1-\kappa/2]$ since
\[
        \left|
        \partial_u\ell(a,u)
        \right|
        \le
        \frac{2}{\kappa}.
\]
Thus, by the contraction inequality and the pseudo-dimension bound used in
Lemma~4 of \cite{fan2024factor}, if
\[
        V_n:=\operatorname{Pdim}(\mathcal G_{\eta,n}),
\]
then with probability at least $1-e^{-t}$,
\[
\begin{aligned}
    &(P-\bbP_n)
    \{\ell(A,\widehat\eta(\hat\blf_X))
        -
        \ell(A,\eta^\dagger(\hat\blf_X))\}
    \\
    &\qquad\le
    \zeta
    \|\widehat\eta(\hat\blf_X)-\eta(\nu(\blf_X))\|_{L_2(P)}^2
    +
    C_\zeta
    \|\eta^\dagger(\hat\blf_X)-\eta(\nu(\blf_X))\|_{L_2(P)}^2+
    C_\zeta
    \left(
        \frac{V_n\log n}{n}
        +
        \frac{t}{n}
    \right).
\end{aligned}
\]
Choosing $\zeta$ small enough and combining this with
\eqref{eq:pi1_kl_shorter}, we obtain
\[
\begin{aligned}
    \|\widehat\eta(\hat\blf_X)-\eta(\nu(\blf_X))\|_{L_2(P)}^2
    \lesssim
    \|\eta^\dagger(\hat\blf_X)-\eta(\nu(\blf_X))\|_{L_2(P)}^2
    +
    \frac{V_n\log n}{n}
    +
    \frac{t}{n}.
\end{aligned}
\]
Using \eqref{eq:pi1_oracle_shorter},
\begin{equation}
\label{eq:pi1_pre_balance_shorter}
    \|\widehat\eta(\hat\blf_X)-\eta(\nu(\blf_X))\|_{L_2(P)}^2
    =
    O_p\left(
        N_\eta^{-4\gamma_\eta^\star}
        +
        \delta_{n,f}^2
        +
        \frac{V_n\log n}{n}
    \right).
\end{equation}

Finally, as in the proof for $\widehat\mu_1$,
\[
        V_n
        \lesssim
        \left(
            L_\eta^2N_\eta^2+d_\eta L_\eta N_\eta
        \right)
        \log(L_\eta N_\eta d_\eta).
\]
Taking $L_\eta\asymp 1$ and balancing
\[
        N_\eta^{-4\gamma_\eta^\star}
        \qquad\text{and}\qquad
        \frac{N_\eta^2\log N_\eta\log n}{n}
\]
gives
\[
        N_\eta\asymp n^{1/(4\gamma_\eta^\star+2)}
\]
up to logarithmic factors. Therefore,
\[
\|\widehat\eta(\hat\blf_X)-\eta(\nu(\blf_X))\|_{L_2(P)}^2
=
O_p\left(
    n^{-2\gamma_\eta^\star/(2\gamma_\eta^\star+1)}
    (\log n)^a
    +
    \delta_{n,f}^2
\right).
\]
Equivalently,
\[
\|\widehat\pi_1(\hat\blf_X)-\pi_1(\blf_X)\|_{L_2(P)}^2
=
O_p\left(
    n^{-2\gamma_\eta^\star/(2\gamma_\eta^\star+1)}
    (\log n)^a
    +
    \delta_{n,f}^2
\right).        
\]
\subsection{Rate for \texorpdfstring{$\widehat \mu_{10}$}{mu10-hat} under the transformed factor representation}
\label{subsec:mu10-rate}
We now prove the rate for $\mu_{10}(\blf_X) :=\bbE[\mu_1(\blf_X,\blf_M)\mid A=0,\blf_X]$. Define the relevant nuisance parameter $\eta := \mu_{10}\circ \nu_X^{-1}$ so that $\eta(\nu_X(\blf_X))=\mu_{10}(\blf_X)$. 
Let us also define
\[
        \cI_0:=\{i\in\cI_{\rm nuis}:A_i=0\},
        \qquad
        n_0:=|\cI_0|.
\]
The estimator $\widehat\eta$ is obtained by regressing the generated outcome
$\widehat\mu_1(\widehat\blf_{X,i},\widehat\blf_{M,i})$ on $\widehat\blf_{X,i}$:
\[
        \widehat\eta
        \in
        \arg\min_{g\in\mathcal G_{\eta,n}}
        \frac1{n_0}
        \sum_{i\in\cI_0}
        \left\{
            \widehat\mu_1(\widehat\blf_{X,i},\widehat\blf_{M,i})
            -
            g(\widehat\blf_{X,i})
        \right\}^2 .
\]
Assume
\[
        \|\widehat\blf_X-\nu_X(\blf_X)\|_{L_2(P_0)}
        =
        O_p(\delta_{n,f}),
\]
where $P_0$ denotes the distribution conditional on $A=0$. Assume also that
$\eta$ admits a bounded Lipschitz HCM extension to a compact set containing both
$\nu_X(\blf_X)$ and $\widehat\blf_X$ with probability tending to one, and $ \eta\in\cH(d_\eta,\ell_\eta,\cP)$ with DAS $\gamma^*_\eta$. As before, let
\[
        \mathcal G_{\eta,n}
        :=
        \cG(L_\eta,d_\eta,1,N_\eta,M_\eta,B_\eta)
\]
be the truncated ReLU class. By Theorem 4 of \cite{fan2024factor}, there exists
\[
        \eta^\dagger
        \in
        \cG(C_1,d_\eta,1,C_2N_\eta,\infty,C_3N_\eta^{C_4})
\]
such that
\[
        \|\eta^\dagger-\eta\|_\infty
        \le
        C N_\eta^{-2\gamma_\eta^\star}.
\]
After truncation, we may regard $\eta^\dagger\in\mathcal G_{\eta,n}$ and retain
\[
        \|\eta^\dagger-\eta\|_\infty^2
        \le
        C N_\eta^{-4\gamma_\eta^\star}.
\]
As in the proof of $\widehat\mu_1$, the relevant oracle error is
\[
        \|\eta^\dagger(\widehat\blf_X)-\eta(\nu_X(\blf_X))\|_{L_2(P_0)}^2.
\]
Using the same decomposition as before and Lipschitzness of $\eta$,
\begin{equation}
\label{eq:mu10_oracle_approx}
        \|\eta^\dagger(\widehat\blf_X)-\eta(\nu_X(\blf_X))\|_{L_2(P_0)}^2
        =
        O_p\left(
            N_\eta^{-4\gamma_\eta^\star}
            +
            \delta_{n,f}^2
        \right).
\end{equation}
Now, for $i\in\cI_0$, define
\begin{align*}
     \xi_i & := \mu_1(\blf_{X,i},\blf_{M,i})-\mu_{10}(\blf_{X,i}), \\
     \Delta_i &:=\widehat\mu_1(\widehat\blf_{X,i},\widehat\blf_{M,i})-\mu_1(\blf_{X,i},\blf_{M,i}).
\end{align*}
Then the generated outcome satisfies
\[
        \widehat\mu_1(\widehat\blf_{X,i},\widehat\blf_{M,i})
        =
        \eta(\nu_X(\blf_{X,i}))
        +
        \xi_i
        +
        \Delta_i.
\]
Here $\xi_i$ is the intrinsic second-stage regression noise, while $\Delta_i$
is the first-stage error from estimating $\mu_1$. By independent of $(\blf_X, \blf_M)$ with $(\bu_X, \bu_M)$, we have: 
\[
        \bbE[\xi_i\mid \blf_{X,i},\bu_{X,i},A_i=0]=0,
\]
so that, conditional on the representation-learning fold,
\[
        \bbE[\xi_i\mid \widehat\blf_{X,i},A_i=0]=0.
\]
For any $g\in\mathcal G_{\eta,n}$, define
\[
        r_g(i)
        :=
        g(\widehat\blf_{X,i})
        -
        \eta(\nu_X(\blf_{X,i})).
\]
By empirical risk minimization,
\[
        \frac1{n_0}\sum_{i\in\cI_0}
        \{\xi_i+\Delta_i-r_{\widehat\eta}(i)\}^2
        \le
        \frac1{n_0}\sum_{i\in\cI_0}
        \{\xi_i+\Delta_i-r_{\eta^\dagger}(i)\}^2 .
\]
Expanding both sides and canceling common terms gives
\begin{align}
\label{eq:mu10_basic}
\|r_{\widehat\eta}\|_{n,0}^2 \le
\|r_{\eta^\dagger}\|_{n,0}^2
+
\frac{2}{n_0}\sum_{i\in\cI_0}
\xi_i
\left\{
    \widehat\eta(\widehat\blf_{X,i})
    -
    \eta^\dagger(\widehat\blf_{X,i})
\right\} +
\frac{2}{n_0}\sum_{i\in\cI_0}
\Delta_i
\left\{
    \widehat\eta(\widehat\blf_{X,i})
    -
    \eta^\dagger(\widehat\blf_{X,i})
\right\}.
\end{align}
The first stochastic term is handled exactly as in the proof of
$\widehat\mu_1$, using Lemma~4 of \cite{fan2024factor}. The second term is new
and comes from the generated outcome.

By Cauchy's inequality and Young's inequality,
\[
\begin{aligned}
\left|\frac{2}{n_0}\sum_{i\in\cI_0}\Delta_i\left\{\widehat\eta(\widehat\blf_{X,i})-\eta^\dagger(\widehat\blf_{X,i})\right\}\right| & \le 2\|\Delta\|_{n,0}\|\widehat\eta(\widehat\blf_X)\eta^\dagger(\widehat\blf_X)\|_{n,0}\\
&\qquad \le \zeta \|r_{\widehat\eta}\|_{n,0}^2+C_\zeta \|r_{\eta^\dagger}\|_{n,0}^2+C_\zeta \|\Delta\|_{n,0}^2.
\end{aligned}
\]
Combining this with \eqref{eq:mu10_basic}, and applying the same empirical
process bound as in the proof of $\widehat\mu_1$, we obtain
\[
        \|r_{\widehat\eta}\|_{n,0}^2
        \lesssim
        \|r_{\eta^\dagger}\|_{n,0}^2
        +
        \frac{V_n\log n}{n}
        +
        \|\Delta\|_{n,0}^2
        +
        \frac{t}{n},
\]
where
\[
        V_n:=\operatorname{Pdim}(\mathcal G_{\eta,n}).
\]
By Bernstein's inequality for the fixed oracle and the already established
rate for $\widehat\mu_1$ under $P_0$,
\[
        \|\Delta\|_{n,0}^2
        =
        O_p(r_{\mu_1,0,n}^2),
\]
where
\[
        r_{\mu_1,0,n}^2
        :=
        \|\widehat\mu_1(\widehat\blf_X,\widehat\blf_M)
        -
        \mu_1(\blf_X,\blf_M)\|_{L_2(P_0)}^2 .
\]
Therefore,
\begin{equation}
\label{eq:mu10_pre_balance}
        \|\widehat\eta(\widehat\blf_X)-\eta(\nu_X(\blf_X))\|_{L_2(P_0)}^2
        =
        O_p\left(
            N_\eta^{-4\gamma_\eta^\star}
            +
            \delta_{n,f}^2
            +
            \frac{V_n\log n}{n}
            +
            r_{\mu_1,0,n}^2
        \right).
\end{equation}

Finally, as before,
\[
        V_n
        \lesssim
        \left(
            L_\eta^2N_\eta^2+d_\eta L_\eta N_\eta
        \right)
        \log(L_\eta N_\eta d_\eta).
\]
Taking $L_\eta\asymp 1$ and balancing
\[
        N_\eta^{-4\gamma_\eta^\star}
        \qquad\text{and}\qquad
        \frac{N_\eta^2\log N_\eta\log n}{n}
\]
gives
\[
        N_\eta\asymp n^{1/(4\gamma_\eta^\star+2)}
\]
up to logarithmic factors. Hence,
\[
\|\widehat\eta(\widehat\blf_X)-\eta(\nu_X(\blf_X))\|_{L_2(P_0)}^2
=
O_p\left(
n^{-2\gamma_\eta^\star/(2\gamma_\eta^\star+1)}
(\log n)^a
+
\delta_{n,f}^2
+
r_{\mu_1,0,n}^2
\right).
\]
Equivalently,
\[
\|\widehat\mu_{10}(\widehat\blf_X)-\mu_{10}(\blf_X)\|_{L_2(P_0)}^2
=
O_p\left(
n^{-2\gamma_\eta^\star/(2\gamma_\eta^\star+1)}
(\log n)^a
+
\delta_{n,f}^2
+
r_{\mu_1,0,n}^2
\right).
\]

\section{A sufficient condition for factor recovery}
\label{app:prop_factor_rec}
\begin{proposition}[Sufficient condition for factor recovery]
\label{prop:factor_recovery_sufficient}
Let
\begin{align*}
\mathcal L_\lambda(z_{XM}) & = \mathbb \bbE\left[
\lambda_1\|X-h_X(g_X(X))\|_2^2
+\lambda_2\|M-h_M(g_M(M))\|_2^2 \right. \\
& \hspace{15em}\left. 
+\lambda_3\|g_M(M)-g_{XM}(A,g_X(X))\|_2^2
\right]
\end{align*}
denote the population MediEncoder objective, where
$z=(g_X,g_M,g_{XM},h_X,h_M)\in\mathcal Z$.
Let
\[
\mathcal L_\lambda^\star:=\inf_{z\in\mathcal Z}\mathcal L_\lambda(z).
\]
Let $\mathcal V$ be a class of maps $\nu=(\nu_X,\nu_M):
\mathbb R^{\bar p+\bar q}\to \mathbb R^{\tilde p+\tilde q}$ such that each $\nu\in\mathcal V$ admits a Lipschitz left inverse on the support of
$(\boldsymbol f_X,\boldsymbol f_M)$.
Assume that the representation class contains oracle encoders
$g_X^\star$ and $g_M^\star$, together with a transformation
$\nu^\star=(\nu_X^\star,\nu_M^\star)\in\mathcal V$, such that
\[
\|g_X^\star(X)-\nu_X^\star(\boldsymbol f_X)\|_{L_2(P)}
\le \delta_{X, \rm app},
\qquad
\|g_M^\star(M)-\nu_M^\star(\boldsymbol f_M)\|_{L_2(P)}
\le \delta_{M, \rm app}.
\]
Assume further that the population objective satisfies the quotient-stability condition
\begin{equation}
\label{eq:QS}
d_{\mathcal V}^2(z)
\le
C\{\mathcal L_\lambda(z)-\mathcal L_\lambda^\star\}
+
C(\delta_{X, \rm app}^2+\delta_{M, \rm app}^2),
\end{equation}
where, for $z \in \cZ$, 
\[
d_{\mathcal V}^2(z)
:=
\inf_{\nu\in\mathcal V}
\left\{
\|g_X(X)-\nu_X(\boldsymbol f_X)\|_{L_2(P)}^2
+
\|g_M(M)-\nu_M(\boldsymbol f_M)\|_{L_2(P)}^2
\right\}.
\]
Suppose the fitted MediEncoder $\hat z=(\hat g_X,\hat g_M,\hat g_{XM},\hat h_X,\hat h_M)$ satisfies the excess-risk bound
\begin{equation}
\label{eq:curvature_mediencoder}
\mathcal L_\lambda(\hat z)-\mathcal L_\lambda^\star
=
O_p(r_n^2).
\end{equation}
Then there exists a possibly $n$-dependent transformation
$\nu_n=(\nu_{X,n},\nu_{M,n})\in\mathcal V$ such that
\[
\|\hat g_X(X)-\nu_{X,n}(\boldsymbol f_X)\|_{L_2(P)}
+
\|\hat g_M(M)-\nu_{M,n}(\boldsymbol f_M)\|_{L_2(P)}
=
O_p(r_n+\delta_{X, \rm app}+\delta_{M, \rm app}).
\]
In particular, if $r_n+\delta_{X, \rm app}+\delta_{M, \rm app}=o(n^{-1/4})$, then the factor-recovery condition
\[
\|\hat{\boldsymbol f}-\nu_n(\boldsymbol f)\|_{L_2(P)}
=
o_p(n^{-1/4})
\]
holds, where
\[
\hat{\boldsymbol f}:=(\hat g_X(X),\hat g_M(M)),
\qquad
\nu_n(\boldsymbol f):=(\nu_{X,n}(\boldsymbol f_X),\nu_{M,n}(\boldsymbol f_M)).
\]
\end{proposition}

\begin{remark}
    Condition \eqref{eq:QS} is the natural analog of a curvature condition for autoencoder-based representations. A fixed encoder is generally not identifiable from reconstruction loss, since invertible reparametrizations of the bottleneck coordinates can be absorbed by the decoder. The distance $d_{\mathcal V}$ therefore measures recovery only up to the equivalence class of Lipschitz left-invertible transformations, which is precisely the level of recovery required by Theorem~\ref{thm:main}. The terms $\delta_{X, \rm app}$ and $\delta_{M, \rm app}$ represent the approximation errors, which depend on the expressability of the underlying function class used for the encoders.
\end{remark}

\section{Proof of Proposition \ref{app:prop_factor_rec}}
By condition \eqref{eq:QS} and \eqref{eq:curvature_mediencoder}, we have: 
\[
d_{\mathcal V}^2(\hat z)
\le
C\{\mathcal L_\lambda(\hat z)-\mathcal L_\lambda^\star\}
+
C(\delta_{X, \rm app}^2+\delta_{M, \rm app}^2)
=
O_p(r_n^2+\delta_{X, \rm app}^2+\delta_{M, \rm app}^2).
\]
Therefore,
\[
d_{\mathcal V}(\hat z)
=
O_p(r_n+\delta_{X, \rm app}+\delta_{M, \rm app}).
\]
By the definition of $d_{\mathcal V}$, for every $\varepsilon>0$ there exists
$\nu_\varepsilon=(\nu_{X,\varepsilon},\nu_{M,\varepsilon})\in\mathcal V$ such that
\[
\begin{aligned}
&\|\hat g_X(X)-\nu_{X,\varepsilon}(\boldsymbol f_X)\|_{L_2(P)}^2
+
\|\hat g_M(M)-\nu_{M,\varepsilon}(\boldsymbol f_M)\|_{L_2(P)}^2  \\
&\qquad\le
d_{\mathcal V}^2(\hat z)+\varepsilon .
\end{aligned}
\]
Taking, for instance, $\varepsilon=n^{-1}$ and writing
$\nu_n:=\nu_{1/n}$ gives
\[
\|\hat g_X(X)-\nu_{X,n}(\boldsymbol f_X)\|_{L_2(P)}^2
+
\|\hat g_M(M)-\nu_{M,n}(\boldsymbol f_M)\|_{L_2(P)}^2
=
O_p(r_n^2+\delta_{X, \rm app}^2+\delta_{M, \rm app}^2).
\]
Taking square roots and using $\sqrt{x+y+z}\le \sqrt x+\sqrt y+\sqrt z$ yields
\[
\left(
\|\hat g_X(X)-\nu_{X,n}(\boldsymbol f_X)\|_{L_2(P)}^2
+
\|\hat g_M(M)-\nu_{M,n}(\boldsymbol f_M)\|_{L_2(P)}^2
\right)^{1/2}
=
O_p(r_n+\delta_{X, \rm app}+\delta_{M, \rm app}).
\]
Consequently,
\[
\|\hat g_X(X)-\nu_{X,n}(\boldsymbol f_X)\|_{L_2(P)}
+
\|\hat g_M(M)-\nu_{M,n}(\boldsymbol f_M)\|_{L_2(P)}
=
O_p(r_n+\delta_{X, \rm app}+\delta_{M, \rm app}),
\]
where the last step follows because the sum of two nonnegative terms is bounded by
$\sqrt 2$ times their Euclidean norm.

Finally, by the definition of
\[
\hat{\boldsymbol f}:=(\hat g_X(X),\hat g_M(M)),
\qquad
\nu_n(\boldsymbol f):=(\nu_{X,n}(\boldsymbol f_X),\nu_{M,n}(\boldsymbol f_M)),
\]
we obtain
\[
\|\hat{\boldsymbol f}-\nu_n(\boldsymbol f)\|_{L_2(P)}
=
O_p(r_n+\delta_{X, \rm app}+\delta_{M, \rm app}).
\]
Thus, if $r_n+\delta_{X, \rm app}+\delta_{M, \rm app}=o(n^{-1/4})$, then
\[
\|\hat{\boldsymbol f}-\nu_n(\boldsymbol f)\|_{L_2(P)}
=
o_p(n^{-1/4}),
\]
which is the desired factor-recovery condition.

\section{Additional Real Data Details} \label{appendix:real_data}

We apply our proposed method to investigate how depressive symptoms influence cognitive decline in Alzheimer’s disease, and whether this relationship is mediated by DNA methylation. Understanding this pathway is important because prior studies have established associations between depression and cognitive impairment, but the biological mechanisms—particularly epigenetic mediation—remain less clear. Here, we present the detailed data preprocessing procedure.

\paragraph{Dataset.} We conduct our analysis using data from the Alzheimer’s Disease Neuroimaging Initiative (ADNI), a longitudinal, multi-center observational study designed to validate biomarkers for Alzheimer’s disease (AD) clinical trials \citep{mueller2005alzheimer}. The dataset is accessible via \url{https://adni.loni.usc.edu/}. Our treatment variable $A$ is the Geriatric Depression Scale (GDS), a self-reported measure ranging from 0 to 15, with higher values indicating more severe depressive symptoms. We binarize GDS using the clinically standard threshold (e.g., GDS $> 5$ indicates depression). The outcome $Y$ is the Alzheimer’s Disease Assessment Scale–Cognitive Subscale (ADAS-Cog), ranging from 0 to 85, where higher scores reflect greater cognitive impairment \citep{cano2010adas, raghavan2013adas}.

\paragraph{Preprocessing.} We include clinically relevant covariates such as demographic and health-related variables, denoted by $X$. Missing values in $X$ are imputed using \texttt{mice} \citep{van2011mice}. To enhance model flexibility, we construct higher-order polynomial features and interaction terms, resulting in a covariate dimension of $\dim(X) = 171$. The mediator $M$ consists of DNA methylation measurements obtained using the Illumina HumanMethylationEPIC BeadChip array. We preprocess the methylation data using the \texttt{minfi} package \citep{aryee2014minfi}, and we select CpG sites that have been reported to be significantly associated with Alzheimer’s disease in large-scale epigenome-wide association studies (EWAS), including a meta-analysis of Braak stage and re-analysis studies of publicly available GEO datasets \citep{zhang2020epigenome, lunnon2014methylomic, battram2022ewas}.

 based on missingness and variability, retaining only those with no missing values. This yields $\dim(M) = 3{,}206$ mediators. After preprocessing, we retain $n = 649$ subjects with complete data across all variables.

\paragraph{Choice of $\tilde p$ and $\tilde q$.} We apply our proposed estimation procedure to estimate the natural indirect effect (NIE), natural direct effect (NDE), and total effect (TE). The projection dimensions $(\tilde p, \tilde q)$ are selected based on principal component analysis (PCA) where $\tilde p = 5$ and $\tilde q = 40$ number of coodinates is able to explain 80\% variance.

\newpage 
\section{More on Simulation and Ablation Study \label{appendix:res}}

All experiments were conducted in a cluster computing environment on a CPU compute node. The computational environment used Python 3.10.12 with 32 CPU cores. A typical full experimental run with 500 replications took approximately 84 hours to complete. Tables~\ref{tab:wave_1}--\ref{tab:wave_4} and Tables~\ref{tab:ab_1}--\ref{tab:ab_4} report the performance of different methods and ablation study compared with MediEncoder across varying sample sizes and covariate/mediator dimensions. The ablation study further evaluates the performance of MediEncoder after removing the alignment component in the loss function~\eqref{eq:medi_loss}.

\begin{table}[htbp]
\centering
\small
\setlength{\tabcolsep}{5pt}
\begin{tabular}{clcccc}
\toprule
$n$ & \textbf{Estimator} & \textbf{SD} & \textbf{RMSE} & \textbf{CI Length} & \textbf{Coverage} \\
\midrule

\multirow{5}{*}{100}
& Projection   & 37.471& 37.839& 146.884& 0.980
\\
& Autoencoder  & 42.495& 44.101& 166.575& 0.965
\\
& VAE          & 0.571& 0.580& 2.239& 0.965
\\
& IMAVAE       & 0.574& 1.234& 2.250& 0.637
\\
& MediEncoder  & 0.503& 0.528& 1.974& 0.963
\\

\midrule
\multirow{5}{*}{300}
& Projection   & 1.589& 1.890& 6.228& 0.930
\\
& Autoencoder  & 1.255& 1.363& 4.918& 0.950
\\
& VAE          & 0.391& 0.401& 1.533& 0.950
\\
& IMAVAE       & 0.439& 0.644& 1.759& 0.830
\\
& MediEncoder  & 0.402& 0.419& 1.575& 0.955
\\

\midrule
\multirow{5}{*}{800}
& Projection   & 0.402& 0.549& 1.577& 0.880
\\
& Autoencoder  & 0.379& 0.435& 1.487& 0.920
\\
& VAE          & 0.335& 0.335& 1.314& 0.940
\\
& IMAVAE       & 0.372& 0.411& 1.459& 0.927
\\
& MediEncoder  & 0.277& 0.299& 1.085& 0.935
\\

\midrule
\multirow{5}{*}{1200}
& Projection   & 0.280& 0.428& 1.099& 0.820
\\
& Autoencoder  & 0.383& 0.457& 1.502& 0.915
\\
& VAE          & 0.322& 0.327& 1.262& 0.925
\\
& IMAVAE       & 0.335& 0.360& 1.312& 0.950
\\
& MediEncoder  & 0.267& 0.290& 1.047& 0.940
\\

\midrule
\multirow{5}{*}{2000}
& Projection   & 0.265& 0.411& 1.040& 0.780
\\
& Autoencoder  & 0.321& 0.415& 1.257& 0.865
\\
& VAE          & 0.337& 0.338& 1.321& 0.945
\\
& IMAVAE       & 0.340& 0.349& 1.333& 0.953
\\
& MediEncoder  & 0.275& 0.299& 1.079& 0.935
\\

\midrule
\multirow{5}{*}{3000}
& Projection   & 0.269& 0.370& 1.055& 0.845
\\
& Autoencoder  & 0.328& 0.412& 1.288& 0.880
\\
& VAE          & 0.321& 0.340& 1.257& 0.925
\\
& IMAVAE       & 0.322& 0.334& 1.251& 0.943
\\
& MediEncoder  & 0.253& 0.278& 0.992& 0.945
\\

\bottomrule
\end{tabular}
\caption{Estimator performance comparison by sample size under the nonlinear wavelet DGP with $p=800$, $q=200$, $\sigma_X=2$, $\sigma_M=1$, $\sigma_Y=1$, $\bar p=\bar q=5$, $\tilde p=\tilde q=7$, and $B=500$ replications.}
\label{tab:wave_1}
\end{table}

\begin{table}[htbp]
\centering
\begin{tabular}{ccccccc}
\cmidrule(lr){1-7}& \multicolumn{2}{c}{\textbf{SD}} 
& \multicolumn{2}{c}{\textbf{RMSE}} 
& \multicolumn{2}{c}{\textbf{CI Length}} 
\\
\cmidrule(lr){1-7}& $\lambda_3=0$& Tuning
& $\lambda_3=0$& Tuning
& $\lambda_3=0$& Tuning
\\
\midrule

100  &  0.594
&  0.503&  0.596
&  0.528&  2.330
&  1.974\\
300  &  0.410
&  0.402&  0.414
&  0.419&  1.605
&  1.575\\
800  &  0.360
&  0.277&  0.359
&  0.299&  1.413
&  1.085\\
1200 &  0.285
&  0.267&  0.287
&  0.290&  1.118
&  1.047\\
2000 &  0.358
&  0.275&  0.356
&  0.299&  1.403
&  1.079\\
3000 &  0.297
&  0.253&  0.296
&  0.278&  1.165
&  0.992\\

\bottomrule
\end{tabular}
\caption{Ablation study on the alignment term ($\lambda_3$) in \texttt{MediEncoder} with $p + q = 1000$, $\sigma_X = 2$, $\sigma_M = 1$, $\sigma_Y = 1$, $B = 500$, $\bar p = \bar q = 5$, and $\tilde p = \tilde q = 7$.}
\label{tab:ab_1}
\end{table}

\begin{table}[htbp]
\centering
\small
\setlength{\tabcolsep}{5pt}
\begin{tabular}{clcccc}
\toprule
$n$ & \textbf{Estimator} & \textbf{SD} & \textbf{RMSE} & \textbf{CI Length} & \textbf{Coverage} \\
\midrule

\multirow{5}{*}{100}
& Projection   & 14.190& 14.631& 55.626& 0.942
\\
& Autoencoder  & 24.003& 24.196& 94.09& 0.945
\\
& VAE          & 0.549& 0.561& 2.153& 0.953
\\
& IMAVAE       & 0.591& 1.500& 2.325& 0.885
\\
& MediEncoder  & 0.644& 0.665& 2.525& 0.962
\\

\midrule
\multirow{5}{*}{300}
& Projection   & 1.831& 2.011& 7.175& 0.951
\\
& Autoencoder  & 1.533& 1.672& 6.011& 0.955
\\
& VAE          & 0.412& 0.423& 1.616& 0.952
\\
& IMAVAE       & 0.442& 0.733& 1.740& 0.875
\\
& MediEncoder  & 0.366& 0.388& 1.435& 0.945
\\

\midrule
\multirow{5}{*}{800}
& Projection   & 0.390& 0.537& 1.528& 0.871
\\
& Autoencoder  & 0.482& 0.572& 1.889& 0.913
\\
& VAE          & 0.341& 0.340& 1.336& 0.945
\\
& IMAVAE       & 0.389& 0.451& 1.524& 0.755
\\
& MediEncoder  & 0.278& 0.297& 1.091& 0.945
\\

\midrule
\multirow{5}{*}{1200}
& Projection   & 0.319& 0.439& 1.251& 0.855
\\
& Autoencoder  & 0.415& 0.465& 1.626& 0.915
\\
& VAE          & 0.329& 0.332& 1.290& 0.935
\\
& IMAVAE       & 0.341& 0.390& 1.337& 0.927
\\
& MediEncoder  & 0.269& 0.288& 1.056& 0.925
\\

\midrule
\multirow{5}{*}{2000}
& Projection   & 0.262& 0.410& 1.021& 0.771
\\
& Autoencoder  & 0.365& 0.469& 1.410& 0.875
\\
& VAE          & 0.343& 0.339& 1.333& 0.944
\\
& IMAVAE       & 0.357& 0.376& 1.399& 0.755
\\
& MediEncoder  & 0.283& 0.296& 1.108& 0.945
\\

\midrule
\multirow{5}{*}{3000}
& Projection   & 0.262& 0.368& 1.027& 0.862
\\
& Autoencoder  & 0.331& 0.396& 1.294& 0.885
\\
& VAE          & 0.331& 0.343& 1.294& 0.930
\\
& IMAVAE       & 0.344& 0.349& 1.350& 0.947
\\
& MediEncoder  & 0.270& 0.290& 1.059& 0.955
\\

\bottomrule
\end{tabular}
\caption{Estimator performance comparison by sample size under the nonlinear wavelet DGP with $p=2000$, $q=1000$, $\sigma_X=2$, $\sigma_M=1$, $\sigma_Y=1$, $\bar p=\bar q=5$, $\tilde p=\tilde q=7$, and $B=500$ replications.}
\label{tab:wave_2}
\end{table}

\begin{table}[htbp]
\centering
\begin{tabular}{ccccccc}
\cmidrule(lr){1-7}& \multicolumn{2}{c}{\textbf{SD}} 
& \multicolumn{2}{c}{\textbf{RMSE}} 
& \multicolumn{2}{c}{\textbf{CI Length}} 
\\
\cmidrule(lr){1-7}& $\lambda_3=0$& Tuning
& $\lambda_3=0$& Tuning
& $\lambda_3=0$& Tuning
\\
\midrule

100  &  0.544
&  0.644&  0.548
&  0.665&  2.134
&  2.525\\
300  &  0.416
&  0.366&  0.420
&  0.388&  1.629
&  1.435\\
800  &  0.371
&  0.278&  0.371
&  0.297&  1.453
&  1.091\\
1200 &  0.306
&  0.269&  0.305
&  0.288&  1.199
&  1.056\\
2000 &  0.336
&  0.283&  0.335
&  0.296&  1.318
&  1.108\\
3000 &  0.307
&  0.270&  0.306
&  0.290&  1.203
&  1.059\\

\bottomrule
\end{tabular}
\caption{Ablation study on the alignment term ($\lambda_3$) in \texttt{MediEncoder} with $p+q=3000$, $\sigma_X = 2$, $\sigma_M = 1$, $\sigma_Y = 1$, $B = 500$, $\bar p = \bar q = 5$, and $\tilde p = \tilde q = 7$.}
\label{tab:ab_2}
\end{table}

\begin{table}[htbp]
\centering
\small
\setlength{\tabcolsep}{5pt}
\begin{tabular}{clcccc}
\toprule
$n$ & \textbf{Estimator} & \textbf{SD} & \textbf{RMSE} & \textbf{CI Length} & \textbf{Coverage} \\
\midrule

\multirow{5}{*}{100}
& Projection   & 14.559& 15.367& 57.072& 0.939
\\
& Autoencoder  & 50.642& 51.505& 198.441& 0.954
\\
& VAE          & 0.734& 0.733& 2.879& 0.985
\\
& IMAVAE       & 0.760& 1.103& 2.697& 0.910
\\
& MediEncoder  & 0.579& 0.588& 2.270& 0.967
\\

\midrule
\multirow{5}{*}{300}
& Projection   & 1.308& 1.518& 5.128& 0.895
\\
& Autoencoder  & 2.729& 2.770& 10.696& 0.970
\\
& VAE          & 0.438& 0.449& 1.717& 0.954
\\
& IMAVAE       & 0.405& 0.554& 1.508& 0.858
\\
& MediEncoder  & 0.390& 0.394& 1.528& 0.967
\\

\midrule
\multirow{5}{*}{800}
& Projection   & 0.322& 0.499& 1.263& 0.781
\\
& Autoencoder  & 0.444& 0.524& 1.739& 0.914
\\
& VAE          & 0.331& 0.331& 1.299& 0.941
\\
& IMAVAE       & 0.383& 0.411& 1.500& 0.818
\\
& MediEncoder  & 0.336& 0.335& 1.319& 0.958
\\

\midrule
\multirow{5}{*}{1200}
& Projection   & 0.285& 0.412& 1.116& 0.805
\\
& Autoencoder  & 0.417& 0.489& 1.636& 0.925
\\
& VAE          & 0.341& 0.341& 1.336& 0.935
\\
& IMAVAE       & 0.330& 0.356& 1.294& 0.830
\\
& MediEncoder  & 0.291& 0.293& 1.141& 0.950
\\

\midrule
\multirow{5}{*}{2000}
& Projection   & 0.269& 0.418& 1.053& 0.784
\\
& Autoencoder  & 0.336& 0.449& 1.318& 0.870
\\
& VAE          & 0.340& 0.340& 1.333& 0.953
\\
& IMAVAE       & 0.366& 0.376& 1.435& 0.825
\\
& MediEncoder  & 0.326& 0.325& 1.278& 0.925
\\

\midrule
\multirow{5}{*}{3000}
& Projection   & 0.245& 0.350& 0.962& 0.835
\\
& Autoencoder  & 0.345& 0.427& 1.353& 0.885
\\
& VAE          & 0.334& 0.345& 1.310& 0.944
\\
& IMAVAE       & 0.341& 0.343& 1.337& 0.865
\\
& MediEncoder  & 0.270& 0.280& 1.059& 0.975
\\

\bottomrule
\end{tabular}
\caption{Estimator performance comparison by sample size under the nonlinear wavelet DGP with $p=2500$, $q=2500$, $\sigma_X=1.5$, $\sigma_M=1$, $\sigma_Y=1$, $\bar p=\bar q=5$, $\tilde p=\tilde q=7$, and $B=500$ replications.}
\label{tab:wave_3}
\end{table}

\begin{table}[htbp]
\centering
\begin{tabular}{ccccccc}
\cmidrule(lr){1-7}& \multicolumn{2}{c}{\textbf{SD}} 
& \multicolumn{2}{c}{\textbf{RMSE}} 
& \multicolumn{2}{c}{\textbf{CI Length}} 
\\
\cmidrule(lr){1-7}& $\lambda_3=0$& Tuning
& $\lambda_3=0$& Tuning
& $\lambda_3=0$& Tuning
\\
\midrule

100  &  0.722
&  0.579
&  0.720
&  0.588
&  2.831
&  2.270
\\
300  &  
0.414
&  0.390
&  0.414
&  0.394
&  1.624
&  1.528
\\
800  &  0.353
&  0.336
&  0.360
&  0.335
&  1.382
&  1.319
\\
1200 &  
0.328
&  0.291&  0.329
&  0.293&  1.285
&  1.141\\
2000 &  0.368
&  0.326
&  0.386
&  0.325
&  1.441
&  1.278
\\
3000 &  
0.340
&  0.270&  0.341&  0.280&  1.331
&  1.059\\

\bottomrule
\end{tabular}
\caption{Ablation study on the alignment term ($\lambda_3$) in \texttt{MediEncoder} with $p+q = 5000$, $\sigma_X = 1.5$, $\sigma_M = 1$, $\sigma_Y = 1$, $B = 500$, $\bar p = \bar q = 5$, and $\tilde p = \tilde q = 7$.}
\label{tab:ab_3}
\end{table}

\begin{table}[htbp]
\centering
\small
\setlength{\tabcolsep}{5pt}
\begin{tabular}{clcccc}
\toprule
$n$ & \textbf{Estimator} & \textbf{SD} & \textbf{RMSE} & \textbf{CI Length} & \textbf{Coverage} \\
\midrule

\multirow{5}{*}{100}
& Projection   & 15.814& 16.352& 61.992& 0.950
\\
& Autoencoder  & 59.019& 59.821& 231.348& 0.954
\\
& VAE          & 1.227& 1.228& 4.810& 0.966
\\
& IMAVAE       & 0.537& 1.248& 2.104& 0.892
\\
& MediEncoder  & 0.565& 0.572& 2.214& 0.934
\\

\midrule
\multirow{5}{*}{300}
& Projection   & 2.736& 2.945& 10.725& 0.985
\\
& Autoencoder  & 2.213& 2.275& 8.675& 0.955
\\
& VAE          & 0.414& 0.414& 1.621& 0.945
\\
& IMAVAE       & 0.380& 0.568& 1.490& 0.880
\\
& MediEncoder  & 0.389& 0.406& 1.525& 0.973
\\

\midrule
\multirow{5}{*}{800}
& Projection   & 0.401& 0.570& 1.572& 0.847
\\
& Autoencoder  & 0.509& 0.600& 1.995& 0.912
\\
& VAE          & 0.329& 0.349& 1.290& 0.925
\\
& IMAVAE       & 0.398& 0.443& 1.561& 0.795
\\
& MediEncoder  & 0.358& 0.358& 1.402& 0.966
\\

\midrule
\multirow{5}{*}{1200}
& Projection   & 0.310& 0.424& 1.216& 0.854
\\
& Autoencoder  & 0.437& 0.493& 1.711& 0.912
\\
& VAE          & 0.330& 0.330& 1.292& 0.940
\\
& IMAVAE       & 0.339& 0.366& 1.328& 0.900
\\
& MediEncoder  & 0.322& 0.323& 1.537& 0.952
\\

\midrule
\multirow{5}{*}{2000}
& Projection   & 0.281& 0.417& 1.100& 0.820
\\
& Autoencoder  & 0.342& 0.431& 1.342& 0.902
\\
& VAE          & 0.332& 0.338& 1.301& 0.955
\\
& IMAVAE       & 0.380& 0.387& 1.491& 0.861
\\
& MediEncoder  & 0.324& 0.325& 1.269& 0.958
\\

\midrule
\multirow{5}{*}{3000}
& Projection   & 0.255& 0.351& 0.999& 0.841
\\
& Autoencoder  & 0.352& 0.422& 1.378& 0.897
\\
& VAE          & 0.336& 0.337& 1.317& 0.930
\\
& IMAVAE       & 0.367& 0.367& 1.439& 0.837
\\
& MediEncoder  & 0.277& 0.277& 1.085& 0.950
\\

\bottomrule
\end{tabular}
\caption{Estimator performance comparison by sample size under the nonlinear wavelet DGP with $p=5000$, $q=5000$, $\sigma_X=1.5$, $\sigma_M=1$, $\sigma_Y=1$, $\bar p=\bar q=5$, $\tilde p=\tilde q=7$, and $B=500$ replications.}
\label{tab:wave_4}
\end{table}

\begin{table}[htbp]
\centering
\begin{tabular}{ccccccc}
\cmidrule(lr){1-7}& \multicolumn{2}{c}{\textbf{SD}} 
& \multicolumn{2}{c}{\textbf{RMSE}} 
& \multicolumn{2}{c}{\textbf{CI Length}} 
\\
\cmidrule(lr){1-7}& $\lambda_3=0$& Tuning
& $\lambda_3=0$& Tuning
& $\lambda_3=0$& Tuning
\\
\midrule

100  &  0.666
&  0.565&  0.663
&  0.572&  2.610
&  2.214\\
300  &  0.440
&  0.389&  0.442
&  0.406&  1.725
&  1.525\\
800  &  0.372
&  0.358&  0.391
&  0.358&  1.458
&  1.402\\
1200 &  0.357
&  0.322 &  0.363
&  0.323 &  1.401
&  1.537\\
2000 &  0.392
&  0.324 &  0.408
&  0.325&  1.536
&  1.269 \\
3000 &  0.343
&  0.277&  0.352
&  0.277&  1.345
&  1.085 \\

\bottomrule
\end{tabular}
\caption{Ablation study on the alignment term ($\lambda_3$) in \texttt{MediEncoder} with $p+q = 10000$, $\sigma_X = 1.5$, $\sigma_M = 1$, $\sigma_Y = 1$, $B = 500$, $\bar p = \bar q = 5$, and $\tilde p = \tilde q = 7$.}
\label{tab:ab_4}
\end{table}